\documentclass{aa}
\usepackage[colorlinks=true, allcolors=blue]{hyperref}
\usepackage{multirow}
\usepackage{siunitx}
\DeclareSIUnit\parsec{pc}
\usepackage{comment}
\usepackage{float}
\usepackage{adjustbox}
\usepackage{graphicx}
\usepackage{txfonts}
\usepackage{natbib}
\usepackage[flushleft]{threeparttable}
\usepackage{booktabs}
\usepackage{etex}
\usepackage{amsmath}
\usepackage[english]{babel}
\usepackage{stfloats}
\usepackage{caption}
\usepackage[labelfont=bf]{caption}
\usepackage{subcaption}         
\usepackage{placeins}         
                        
\begin{document}
   \title{H II region filling factors in NGC~628: Luminosity–size relation and connection with polycyclic aromatic hydrocarbon emission} 
   \titlerunning{Filling Factors and Luminosity--Size Relation in NGC~628}
   \author{Lorena Aragüete Riesco\inst{1}
   \and
   José M. Vílchez\inst{2}
   \and
   Salvador Duarte Puertas\inst{1,3}
    \and 
    Laurie Rousseau-Nepton\inst{4,5}
    \and
    Carmelle Robert\inst{6,7}
    \and
    René Pierre Martin\inst{8}
    \and
    Philippe Amram\inst{9}
    \and
    Robert Kennicutt,~Jr.\inst{10,11,12} 
    \and 
    Christophe Morisset\inst{13,14}
    \and
    Ray Garner~III\inst{10,11} 
    }
   \institute{Universidad de Granada, Departamento de Física Teórica y del Cosmos, Avenida de Fuentenueva S/N,  18071, Granada, Spain\
    \email{lariesco@correo.ugr.es}
    \and
    Instituto de Astrofísica de Andalucía (IAA-CSIC), Glorieta de la Astronomía S/N, 18008 Granada, Spain\
    \and 
    Instituto Universitario Carlos I de Física Teórica y Computacional, Universidad de Granada, 18071 Granada, Spain\
    \and
    David A. Dunlap Department of Astronomy and Astrophysics, University of Toronto, 50 St-George Street, Toronto,M5S 3H4, Ontario, Canada\
    \and
    Dunlap Institute for Astronomy and Astrophysics, 50 St-George Street, Toronto, M5S 3H4, Ontario, Canada\
    \and
    Département de Physique, de Génie Physique et d’Optique, Université Laval, Québec, QC G1V 0A6, Canada\
    \and
    Centre de Recherche en Astrophysique du Québec (CRAQ), Québec, QC G1V 0A6, Canada\
    \and
    Department of Physics and Astronomy, University of Hawaii at Hilo, Hilo, HI 96720, USA\
    \and
    Laboratoire d’Astrophysique de Marseille, CNRS, CNES, Aix Marseille University, 38 Rue Frédéric Joliot Curie, F-13013 Marseille, France\
    \and
    Department of Physics and Astronomy, Texas A\&M University, 578 University Drive, College Station, TX 77843, USA\
    \and
    George P. and Cynthia W. Mitchell Institute for Fundamental Physics \& Astronomy, Texas A\&M University, 578 University Drive, College Station, TX 77843, USA\
    \and
    Department of Astronomy and Steward Observatory, University of Arizona, 933 N. Cherry Avenue, Tucson, AZ 85721, USA\
    \and
    Instituto de Astronomía, Universidad Nacional Autónoma de México, Unidad Académica en Ensenada, Km 103 Carr. Tijuana-Ensenada, Ensenada, B.C., C.P. 22860, México\
    \and
    Instituto de Ciencias Físicas, Universidad Nacional Autónoma de México, Av. Universidad s/n, 62210 Cuernavaca, Mor., México\
    }  
   \authorrunning{Aragüete Riesco et al.}
   \date{Received October, 14, 2025; Last revised July 31, 2026; Accepted August 3, 2026}
   \abstract
   {}
   {Understanding the internal structure of H\,II regions is fundamental for constraining star formation processes in galaxies. We investigated how the filling factor (FF) relates to luminosity, size, electron density, and H$\alpha$ equivalent width (EW(H$\alpha$)) in H\,II regions. We also explored the association between star-forming regions and the polycyclic aromatic hydrocarbon (PAH)-to-dust emission, to assess their possible connection to the evolutionary stages of H\,II regions.} 
   {We investigated the FF in 622~H\,II regions in NGC~628 by combining 475 regions from SIGNALS and 147 from PHANGS-MUSE using a dedicated analysis pipeline. We derived the luminosity, emission line fluxes, radii, electron densities, and FF. We used PHANGS-JWST survey images to study the spatial relation between the FF and PAH-to-dust emission, quantified by the PAH-to-dust ratio $R_{PAH}$ using James Webb Space Telescope/Mid-Infrared Instrument (JWST/MIRI) emission bands at 7.7, 11.3, and 21~$\mu$m.} 
   {Higher Log(FF) and high EW(H$\alpha$) values are found in luminous regions, whereas more extended regions with lower EW(H$\alpha$) display lower Log(FF) values. We show that the H\,II radius definition significantly affects the $\mathrm{L_{H\alpha}}-\mathrm{R}$ relation. Low-luminosity compact H\,II regions appear to reflect a transition from cluster-powered regions to nebulae that are ionized by single massive stars, with the transition occurring around $\mathrm{log(L_{H\alpha}) \sim 37 \ erg \ s^{-1}}$. The PAH-to-dust ratio $\mathrm{R_{PAH}}$ correlates with the volumetric H$\alpha$ luminosity density, $\mathrm{L_{H\alpha}/R^{3}}$, with a transitional range centered around $\mathrm{log(L_{H\alpha}/R^{3}) \sim \ 32~\ erg~s^{-1}}$, corresponding to $\mathrm{log(FF) \approx -4.4}$. Regions with lower Log(FF) values exhibit higher $\mathrm{R_{PAH}}$ values, suggesting less efficient PAH processing in more porous structures.}
   {These results can be consistent with an evolutionary scenario in which the FF decreases with stellar cluster age in giant H\,II regions as they evolve toward fainter and more spatially extended states. The volumetric H$\alpha$ luminosity density, $L_{\mathrm{H}\alpha}/R^3$, provides a useful representation that reduces the covariance between $L_{\mathrm{H}\alpha}$ and $R$ induced by different region definition methods, enabling more consistent cross-catalog comparisons in segmentation approaches.} 
   \maketitle
   \nolinenumbers

\section{Introduction} \label{sec:introduction}
Star-forming galaxies actively form stars over long periods of time. Studies of galaxy star formation histories have shown that galaxies have been forming stars over billions of years \citep{Madau2014}. Massive and dense molecular clouds are the birthplaces of regions hosting hot massive stars (O- and early B-type stars). Feedback from these young stars on the surrounding interstellar medium (ISM) occurs in the form of stellar winds, ionizing and non-ionizing radiation, and eventually, supernova explosions. These feedback mechanisms locally enrich the ISM and are thought to regulate and, in some cases, trigger subsequent star formation episodes \citep{Fichtner2024}.
Photoionization plays a key role in this process. Ionizing photons interact with the surrounding gas, giving rise to H\,II regions that emit characteristic nebular emission line spectra. By analyzing these emission lines, we can infer the physical and chemical properties of the gas. Furthermore, examination of emission line profiles provides insight into the kinematics of the gas, including expansion and turbulence along the line of sight.
\par
These H\,II regions produce characteristic optical emission, such as H$\alpha$, which is commonly used as a tracer of recent star formation on timescales shorter than 10~Myr \citep{Kennicutt2012}. Observations reveal that H\,II regions exhibit complex internal structures, with dense zones embedded within lower-density gas and spanning a wide range of densities and temperatures \citep[e.g.,][]{Castaneda1992}. Several models have been proposed to describe these inhomogeneities. The first studies to characterize internal density fluctuations were conducted by \cite{Osterbrock1959}. They proposed that as a first approximation, H\,II regions can be modeled as spheres filled with low-density gas containing small optically thin high-density clumps or condensations. These clumps form parsec-scale substructures that are coherent in position-velocity space \citep{Williams1999}, and electron density variations within individual regions can be traced using density-sensitive emission-line ratios such as [S\,II]. 
\par
In a spherical nebula with uniform physical conditions within the ionized region, the H$\alpha$ luminosity is obtained by integrating the volume emissivity, $j_{H\alpha}(\mathbf{r})$, over the emitting volume,
\begin{equation}\tag{1}
	L(H\alpha) = \int_V j_{H\alpha}(\mathbf{r}) \, dV = \int_V
	h\nu_{H\alpha} \, \alpha^{\mathrm{eff}}_{H\alpha}(T) \,
	n_e(\mathbf{r}) \, n_p(\mathbf{r}) \, dV, 
\end{equation}
\noindent where $n_e$ is the electron density, $n_p$ is the proton density, $\alpha^{eff}_{H\alpha}(T_e)$ is the $H\alpha$ effective recombination coefficient, and $h\nu_{H\alpha}$ is the photon energy. In a fully ionized hydrogen nebula, $n_e \simeq n_p$ and when we consider a fully spherical volume uniformly filled with ionized gas of constant electron density, that is, with a filling factor, FF, as FF = 1, and ionization-bounded conditions, the H$\alpha$ luminosity can be written as
\begin{equation}\tag{2}
	L(H\alpha) = j_{H\alpha}(T_e) \, n_{e, \mathrm{hom}}^2 \, V. 
\end{equation}
\par
When we define $n_{e, \mathrm{hom}}^2$ as the root-mean-square ($\langle n_e^2 \rangle$; rms) the electron density averaged over the entire geometrical volume, the electron density in the homogeneous case or rms can be derived as
\begin{equation}\tag{3}
	\langle n_e^2 \rangle =\frac{1}{V} \int_V n_{e, \mathrm{hom}}^2(\mathbf{r}) \, dV 
\end{equation} 
\begin{equation}\tag{4} \label{ec:nehom}
	L(\mathrm{H}\alpha) = \langle n_e^2 \rangle \, j_{H\alpha}(T_e) \, V \quad \Rightarrow \quad \langle n_e^2 \rangle = n_{e,\mathrm{hom}}^2 =  \frac{L(\mathrm{H}\alpha)} {j_{H\alpha}(T_e)\,V}.
\end{equation}
\par
In real H\,II regions the ionized gas does not fill the entire geometrical volume. Instead, the emission arises from dense clumps embedded in a more diffuse medium. In this case, an $FF < 1$ must be introduced to account for the fraction of the total volume actually occupied by gas with the local electron density, $n_{e,\mathrm{loc}}^2$. The H$\alpha$ luminosity then becomes
\begin{equation} \tag{5} \label{ec:neloc}
	L(\mathrm{H}\alpha) =	FF \, n_{e,\mathrm{loc}}^2 \,
	j_{H\alpha}(T_e) \, V  \quad \Rightarrow \quad 
	n_{e,\mathrm{loc}}^2 =	\frac{L(\mathrm{H}\alpha)}
	{FF \, j_{H\alpha}(T_e)\,V}.
\end{equation}
\par
The quantity $n_{e,\mathrm{loc}}$ represents the local electron density within the gas clumps. It is typically estimated from optical spectroscopy by density-sensitive collisionally excited line ratios such as the doublet [S\,II]$\lambda \lambda$ 6716 /6731. This is the density of the gas that produces the emission lines. By contrast, the density derived under the homogeneous assumption, $n_{e,\mathrm{hom}}$, corresponds to the mean luminosity-weighted or root-mean-square (rms), that is the electron density averaged over the entire geometrical volume, derived from H$\alpha$, H$\mathrm{\beta}$, or radio continuum emission. Then, from Eq.~\ref{ec:nehom}, and Eq.~\ref{ec:neloc} we obtain
\begin{equation} \tag{6} \label{ec:ff}
	n_{e,\mathrm{hom}}^2 = FF \, n_{e,\mathrm{loc}}^2 \quad \Rightarrow \quad \mathrm{FF} = \left[\frac{\langle n_{e}\rangle}{n_{e}}\right]^{2}.
\end{equation}
\par
Equation~\ref{ec:ff} is based on the assumption of a two-phase medium, as discussed above, in which the dense clumps are characterized by a uniform density, while the inter-clump medium contributes negligibly to the observed emission. Thus, the relation between both quantities is the FF, or the fraction of the total geometrical volume occupied by ionized gas with the local density \citep{Osterbrock1959}. 
\par
In the following, we denote the local electron density $n_{e,\mathrm{loc}}$ simply as $n_e$, while the rms electron density, $\langle n_e \rangle$, denotes the $n_{e,\mathrm{hom}}$. By rearranging the variables in Eq.~\ref{ec:neloc}, we express the FF in terms of the other independent variables,
\begin{equation} \tag{7} \label{ec:linealff}
	\mathrm{log(FF)} =  \mathrm{log(L_{H\alpha})} - \mathrm{log(R^3)} - 2\,\mathrm{log(n_{e})} - \mathrm{31.6}.
\end{equation}
\noindent where 31.6 arises from combining the H$\alpha$ emissivity at $T_{e} = 10^4$~K, the geometrical factor $4\pi/3$ for a spherical volume, and the conversion from parsecs to centimeters.
\par
The difference between these two measurements can reach nearly two orders of magnitude. It quantifies the degree of internal inhomogeneity, or porosity, of the ionized gas within each individual nebula \citep{Kennicutt1984, Pathak2025}. Subsequent models incorporating more realistic physical conditions were developed to better reproduce observations of nearby galactic H\,II regions, such as the Orion nebula \citep[e.g.][]{ Rubin1991, O’Dell2008, SimonDiaz2011}, and to address radiative transfer effects and the fraction of Lyman photons that escape from the nebula \citep[e.g.][]{Giammanco2004}. More recently, computational approaches based on probabilistic density distributions have been developed to model internal density fluctuations within H\,II regions, and explore potential correlations between density, temperature, and the FF \citep{Bergerud2019}.
\par
However, from the point of view of H\,II region models when applied to spatially resolved extragalactic H\,II regions, the FF should be interpreted as a global parameter that includes at least three physically distinct components. These are i) a clumping factor, representing small-scale density inhomogeneities within the ionized gas volume and analogous to the FF used in photoionization models, $ff$; ii) a covering factor, corresponding to the fraction of the ionizing photon field intercepted by the nebula ($\Omega/4\pi$) as seen from the ionizing source; and iii) a geometric or cavity factor, describing the morphology of the region through the ratio of the volume of the sphere defined by the outer radius and the volume of the shell actually occupied by the ionized gas. In the case of a filled sphere, this factor approaches unity, whereas it is much lower than unity for thin shell-like geometries,
\begin{equation} \tag{8} \label{ec:ffphoto}
	FF = ff \times \frac{\Omega}{4 \pi} \times \left(\frac{R_{out}-R_{cavity}}{R_{out}}\right)^3 
\end{equation}
\par
The FF is therefore not only linked to the relative distribution of dense clumps and the physical processes that enabled their survival, it is also tightly connected to the spatial scale over which electron density variations are measured, and thus, to the ions used as tracers. Therefore, it is crucial to understand the scale of electron density variations for unraveling the mechanisms driving star formation. These variations are important for interpreting the gas structure and the physical processes regulating star formation because they directly affect the inferred FF \citep{Mathis2005}.
\par
A comprehensive understanding of the FF remains challenging, largely due to the historical lack of observations at sufficiently high spatial and spectral resolution. Instrumental limitations have long hindered us to resolve fine-scale nebular structures and accurately characterize density variations. However, the advent of new telescopes and integral-field spectrographs has enabled studies to routinely achieve sensitivities and angular resolutions well below $\sim$ \SI{1}{\arcsec}. In NGC~628, the Physics at High Angular Resolution in Nearby Galaxies Survey -- Multi Unit Spectroscopic Explorer (PHANGS-MUSE; \cite{Leroy2021, Emsellem2022}) reaches a spatial resolution of $\sim$ 43.78 pc, while the Star formation, Ionized Gas, and Nebular Abundances Legacy Survey (SIGNALS; \cite{RousseauNepton2019}) achieves $\sim$ 35 pc, enabling detailed characterization of H\,II region properties and robust estimates of the FF in large samples.
\par
Only a limited number of studies have investigated the FF in extragalactic H\,II regions. \cite{Hunt2009} linked the FF values in the range of $\sim$ $10^{-3} - 10^{-1}$ to dust optical depth by examining size-density relations and incorporating dust extinction effects in H\,II regions. They found that a decreasing FF corresponds to an increasing optical path length (${\tau \varpropto \delta^{-1/6}}$; a change of 1~dex in the FF resulted in a tripling of the optical path). They concluded that as the FF increases, the volume occupied by clumps is diminished, while the space between them expands. \cite{Kennicutt1984} reported a somewhat narrow FF range of ${(10^{-2} - 10^{-1})}$ for Galactic and supergiant extragalactic H\,II regions. More recently, \cite{Cedres2013} analyzed 58 H\,II regions in NGC~6946 and extended the FF range down to $\sim 10^{-3}$ for regions smaller than $\sim$ 50 pc, highlighting a dependence on galactocentric radius. A conservative lower limit of $\log(\mathrm{FF}) \sim -7$ can be considered when  extrapolations toward more diffuse ionized structures, consistent with earlier extragalactic studies \citep[e.g.,][]{Cedres2013, Hunt2009}. At larger scales, \cite{Reddy2023a} showed that high FF values in star-forming galaxies correlate with elevated star formation rates and surface densities, and significantly affect the ionization parameter. 
More recently, \cite{Zurita2026} derived the rms electron density, local electron density, and FF for H\,II regions in NGC~2403 and NGC~628. These two galaxies follow a similar size–density relation, $\langle n_e \rangle_{\rm rms} \propto R^{-0.3}$ for $R \lesssim 50$~pc, with no clear dependence of $n_e$ or FF on galactocentric radius. The authors also showed that rms densities and the slope of the $L_{H\alpha}$–$R$ relation are sensitive to the cataloging method, highlighting the need for homogeneous approaches when comparing H\,II region properties in galaxies. Despite these efforts, the spatial distribution of FF within galaxies, its dependence on local environment, and its potential association with filamentary dust structures remain poorly constrained \citep{Smith2015}.
\par
The star formation activity is distributed across the entire galaxy \citep[e.g.,][]{Ragan2018, Sun2024, Querejeta2024}. It is intertwined with an extended network of filamentary dust and gas structures that span scales from tens of parsecs to several kiloparsecs \citep{Kumar2020, Pineda2023}. Studies examining the independent evolution of star-forming regions and dust filaments have not found a one-to-one correspondence at all spatial scales \citep{Louie2013}. Using James Webb Space Telescope/Mid-InfraRed Instrument (JWST/MIRI) observations combined with PHANGS-H$\alpha$ maps of NGC~628, \cite{Kim2023} showed spatial offsets between different star formation tracers and molecular gas on $\sim$100~pc scale. This is indicative of distinct evolutionary stages in the star formation cycle. Similarly, \cite{Thilker2023}, used mid-infrared (MIR) flux measurements along with H$\alpha$ luminosity values of H\,II regions from the catalog of \cite{Groves2023} to report that at least 75\% of star formation is associated with the filament structures on $\sim$ 25~pc scales. 
\par
At sub-kiloparsec scales, corresponding to the typical scale of classic H\,II regions, dust and polycyclic aromatic hydrocarbons (PAHs) coexist with ionized gas and emit strongly in mid-infrared  \citep[e.g.,][]{Tielens2013, Shipley2016, Oey2017, Herrero2022}. PAH emission primarily arises from the photoevaporation or disintegration of larger carbonaceous grains by ultraviolet radiation from massive young stars, and it represents an important component of interstellar dust. PAHs undergo several processes, such as UV photodissociation, ion bombardment, and accretion reactions involving H, C+, and O. Of these, UV photodissociation exhibits the highest reaction rate \citep{Omont2021}. De-excitation through vibrational modes gives rise to the Aromatic Infrared Bands (AIBs) at 3.3, 6.2, 7.7, 8.6, 11.3, and 12.7 $\mu$m, where the band at 7.7 $\mu$m traces predominantly ionized PAH, and the 11.3 $\mu$m band traces neutral PAHs \citep{Rapacioli2004}.
\par
Early observations at 8 $\mu$m with \textit{Spitzer} Space Telescope/IRAC showed that mid-infrared emission is spatially concentrated around H\,II regions \citep{Phillips2008}. The advent of JWST/MIRI now enables high-resolution infrared studies of PAH emission, opening new opportunities to investigate PAH evolution as a function of local H\,II region properties. Several studies indicate that PAHs become more ionized and smaller with increasing star formation surface density \citep[e.g.,][]{Thilker2007, Egorov2023a, Rigopoulou2024, Ujjwal2024, Baron2025}, likely due to enhanced destruction by hard radiation fields. The PAH abundance relative to the total dust population can be assessed by comparing PAH-dominated MIR bands (7.7 $\mu$m and 11.3 $\mu$m) with the mid-infrared emission from warm dust at 21$\mu$m. This approach is related to the $q_{PAH}$ parameter, defined as the fraction of total dust mass in PAHs containing fewer than $10^{3}$ carbon atoms \citep{Draine2007a}. The parameter $q_{PAH}$ modulates the spectral shape of the infrared emission, preferentially affecting PAH-dominated bands \citep{Draine2021}. In this framework, the JWST F2100W band provides a reliable proxy for the total infrared emission \citep{Chastenet2023a}, while the F770W and the F1130W bands are selectively sensitive to PAH emission. Accordingly, we used an empirical tracer of the $q_{PAH}$ fraction defined as $R_{PAH} = \frac{F_{770W}+F_{1130W}}{F_{2100W}}$.
\par
Recent observations show that the PAH fraction decreases significantly within H\,II regions, consistent with the efficient destruction of small grains in ionized environments \citep[e.g.,][]{Egorov2023b, Chastenet2023a, Sutter2024, Dale2025, Egorov2025}. Moreover, the PAH fraction is negatively correlated with the H$\alpha$ equivalent width of H\,II regions \citep{Egorov2023a}, suggesting a link with evolutionary stage.  
\par
We sought to obtain the most extensive and complete census of the FF to date within a single galaxy. We focused our analysis on the nearby spiral galaxy NGC~628 (M74), an almost face-on grand-design galaxy (class 9 in the classification of \cite{Elmegreen1987}). The global properties of NGC~628 are summarized in Table~\ref{tab:NGC628}. We investigated its relation with the physical and chemical properties of H\,II regions and assessed its connection with PAH-to-dust emission.
\par
This paper is structured as follows. Section~\ref{sec:data} presents the data used. Section~\ref{sec:selection} describes the analysis procedures and the H\,II region selection procedures. Section~\ref{sec:LR} examines the dependence of the H$\alpha$ luminosity-size relation on the adopted region definition. Section~\ref{sec:physicalparams} introduces the derived physical parameters. Section~\ref{sec:analysis} presents the main results on the FF and its relation with PAH emission. Section~\ref{sec:discussion} compares our findings with previous studies. Finally, Section~\ref{sec:conclusions} summarizes our conclusions and outlines prospects for future work.
\section{Data sample} \label{sec:data}
We used public MUSE (Multi Unit Spectroscopic Explorer at the Very Large Telescope, VLT; ESO Paranal Observatory, Chile; \citep{Bacon2010}) data from the PHANGS survey\footnote{\url{https://www.phangs.org}} \citep{Emsellem2022}, which observed spectroscopically 19 nearby galaxies to study the star formation history and the ISM with MUSE, complemented by observations from the SIGNALS survey\footnote{\url{https://signal-survey.org}} \citep{RousseauNepton2019}, a program designed to study star formation and ionized regions in $\sim$ 35 local galaxies. 
\par
We focused on NGC~628 which has also been observed by the Spectro-Imageur à Transformée de Fourier pour l'Étude en Long et en Large des raies d'Émission  (SITELLE) at the Canada-France-Hawaii telescope (CFHT) \citep{Drissen2019}. We selected our H\,II region candidates from the published SIGNALS survey and identified PHANGS-MUSE regions by using H$\alpha$ images from the PHANGS-MUSE survey. Table~\ref{tab:NGC628} outlines the main properties of this galaxy.
\begin{table}[h]
	\centering
	\caption{Global properties of NGC~628}
	\label{tab:NGC628}
	\resizebox{\columnwidth}{!}{%
		\begin{tabular}{l c c c}
			\hline
			Property & Units & Value & References \\
			\hline
			R.A. (J2000)  & hh:mm:ss & 01:36:41.747 & NED \\
			DEC. (J2000)  & dd:mm:ss & +15:47:01.18 & NED \\
			Distance      & \si{\mega\parsec} & 9.84 & \cite{Anand2021} \\
			Inclination$^{a}$   & \si{\degree} & 8.9 & \citet{Lang2020} \\
			$R_{25}$      & \si{\kilo\parsec} & 10.95 & \shortstack{\cite{Berg2013}\\ \citep{Makarov2014}} \\
			\hline
		\end{tabular}%
	}
	\tablefoot{$a$ Inclination adopted in SIGNALS is $21^\circ$ \citep{RousseauNepton2018}.}
\end{table}
\par
We further combined these data with PHANGS-JWST observations (PHANGS-JWST; \cite{Lee2023}) to examine the relation between H\,II regions and PAH-to-dust emission.
\subsection{SIGNALS survey} \label{subsec:sitelle}
The first subset of our H\,II region sample was drawn from the catalog compiled by \cite{RousseauNepton2018}. NGC~628 was observed using the imaging Fourier transform spectrometer SITELLE. This instrument operates in the optical regime with a field-of-view (FoV) of 11~$\arcmin$ $\times$ 11~$\arcmin$, corresponding to 121~$\arcmin²$ ($\sim$ 992~$kpc^{2}$ at the distance of NGC~628), providing full spatial coverage of the galaxy in a single pointing. The mean pixel scale is 0.321$\arcsec$$\times$ 0.321$\arcsec$, and the spatial resolution is $\sim$ 35~pc corresponding to a full width at half maximum (FWHM) of $\sim$\SI{0.8}{\arcsec} at the distance of 9.84~Mpc.  
The spectral coverage is provided by three filters:  SN1 (\SIrange{364}{385}{nm}), SN2 (\SIrange{484}{512}{nm}), and SN3 (\SIrange{648}{686}{nm}), with resolving powers R = 600, 600 and 1800, respectively. Further details on the observational setup can be found in \cite{RousseauNepton2018}.
\par
The resulting data cube was processed following several steps using SITELLE's dedicated reduction pipeline ORBS \citep{Martin2012}, the spectral extraction engine ORCS \citep{Martin2015} and an H\,II region detection algorithm described in \cite{RousseauNepton2018}. We summarize the main steps here i) subtraction of the stellar continuum using a reference spectrum representative of the old underlying stellar population, extracted from the peak of the galaxy continuum and scaled to the local continuum level of each region; ii) identification of emission peaks by comparing the intensity of adjacent pixels and applying a detection threshold; iii) definition of the zone of influence (ZoI) around each emission peak, up to a maximum radius of \SI{425}{\parsec}; iv) determination of the H\,II region boundaries in $H\alpha$ by identifying the radius at which the slope of the total flux in the flux-distance diagram decreases by less than \SI{2}{\percent}; and v) correction for the local diffuse ionized gas (DIG) background using the median intensity of an annulus surrounding the outer boundary of each region.
Dust extinction was corrected using the \cite{Cardelli1989} extinction law and the theoretical $F(H\alpha) / F(H\beta)$ = 2.87, assuming Case B recombination, an electron temperature of $10^{4}$ K and low-density conditions. The final published SIGNALS catalog contains 4285~H\,II region candidates and is publicly available.
\subsection{PHANGS-MUSE survey} \label{subsec:phangssurvey}
The PHANGS-MUSE mosaic covers the central star-forming disk of NGC~628, spanning a total area of $\sim$ 89~$kpc^{2}$ assembled from multiple IFU pointings \citep{Li2024}. The observations were obtained using the MUSE wide-field mode and combined into a mosaicked data cube. To ensure uniform image quality, all cubes were convolved to match the largest point spread function (PSF) measured across the observation, resulting in a PSF-homogenized dataset for NGC~628 with a final FWHM of \SI{0.92}{\arcsec}. These products, referred to as the “copt” cubes, were reduced using the tailored pipeline described in \cite{Emsellem2022}. 
The native MUSE pixel scale in wide-field mode is \SI{0.2}{\arcsec}. With a PSF FWHM of \SI{0.92}{\arcsec}, corresponding to a physical resolution of \SI{43.8}{pc}, the MUSE resolution is slightly coarser than, but comparable to, the SIGNALS resolution \citep{Santoro2022}. The spectral range covered by MUSE spans from \SIrange{4750}{9300}{\angstrom}.
\subsection{Additional data} \label{subsec:archival}
We compared our measurements with ancillary data from three published catalogs of H\,II regions in NGC~628 based on different segmentation methods: \cite{Groves2023, Congiu2023} using PHANGS-MUSE data, \cite{Barnes2026} with PHANGS-HST data, and \cite{Zurita2026} with ground-based narrow-band H$\alpha$ imaging at a lower spatial resolution than the PHANGS-MUSE datasets. 
\par
The catalog of \cite{Zurita2026} for NGC\,628 does not provide measurements of the [S\,II] $\lambda\lambda6716,6731$ emission line fluxes, making it impossible to determine electron densities for individual regions. Because estimates of $n_{\mathrm{e}}$ are required to compute the FF, this catalog cannot be included in the FF scaling analysis. It is therefore considered only in the analysis of the $L_{\mathrm{H}\alpha}$--$R$ relation presented in Sect.~\ref{sec:LR}, which requires only luminosity and radius measurements.
\par
These catalogs rely on distinct segmentation and boundary definition strategies, including variations in the treatment of the DIG and spatial resolution. As a result, systematic differences in the derived sizes and luminosities of the identified regions may affect scaling relations such as the $L$--$R$ relation we explored.
\par
Finally, to extend our analysis into the infrared regime and trace PAH and thermal dust emissions, we used publicly available MIRI observations of NGC~628 from the PHANGS-JWST survey. Details on the observing strategy and imaging post-processing are provided in \cite{Lee2022} and \cite{Williams2024}. These data allow us to probe different components of the star-forming environment and obtain a comprehensive view of the birthplaces of H\,II regions in this galaxy.
We employed the MIRI F770W, F1130W, and F2100W filters. The angular resolution achieved is \SI{0.238}{\arcsec} for F770W (corresponding to $\sim$ 11.3 pc), \SI{0.36}{\arcsec} for F1130W ($\sim$ \SI{17}{pc}), and \SI{0.651}{\arcsec} for F2100W ($\sim$ 30 pc). The images correspond to the v1p0p1 data release and were processed with the standard JWST Science Calibration Pipeline \citep{Bushouse2024, Williams2024}, yielding maps in units of \si{MJy/sr}. 
We evaluated the background level in the full FoV of each image and identified a small residual baseline in the F2100W band. However, this offset was not subtracted, as the signal-to-noise ratio of the residual relative to the background dispersion is below 3, indicating that it is not statistically significant.
All the infrared images were convolved to the SIGNALS PSF of \SI{0.8}{\arcsec} or the PHANGS-MUSE PSF of \SI{0.92}{\arcsec} using the convolution kernels and procedures described in \cite{Aniano2011}, ensuring a consistent spatial resolution across all datasets. The F770W filter primarily traces ionized PAH emission at \SI{7.7}{\micro\meter}, the F1130W filter is sensitive to neutral PAHs at \SI{11.3}{\micro\meter}, and the F2100W filter traces warm dust emission at \SI{21}{\micro\meter} \citep[e.g.,][]{Galliano2018, Rigopoulou2021}.
\section{H\,II region identification and catalog construction} \label{sec:selection}
\subsection{Definition of the PHANGS-MUSE nebular region}
\label{subsec:phangsidentification}
We developed a modular analysis pipeline to identify and characterize nebular regions in nearby galaxies, incorporating a novel treatment of local DIG and morphological complexity. Regions are initially segmented using HIIDENTIFY \citep{Easeman2023} on the H$\alpha$ emission maps. We applied a signal-to-noise (S/N) ratio threshold of S/N > 3 in order to suppress noise-dominated spaxels. The code applies an iterative growth around the brightest pixel flux and incorporates neighboring pixels whose flux exceeds a user-defined background threshold. In NGC~628, we disable this background lower limit so that the flux growth is driven solely by local emission. The growth is halted when the fraction of discarded pixels in the surrounding annulus reaches 80\%. 
To refine the region selection, we imposed two additional constraints: i) we required at least eight contiguous pixels, and ii) the spatial resolution was set as the minimum separation defined by the FWHM.
The algorithm does not impose constraints on region geometry, allowing for complex and irregular morphologies. Spaxels associated with DIG are not masked at this stage, as their treatment is deferred to subsequent analysis steps. A total of 4830 regions were initially identified. To consolidate all spaxels belonging to the same region and mitigate  isolated pixel-level artifacts in the H$\alpha$ map, we applied a topological cleaning procedure based on connected-component analysis. 
\par
Each region is subsequently analyzed in flux–distance space, where radial profiles are fit with a pseudo-Voigt function. The fitted parameters include the peak intensity, background offset, standard deviation ($sigma$), Gaussian-to-Lorentzian ratio ($alpha$), and the coefficient of determination, $R^{2}_{fit}$, adopted as a goodness-of-fit indicator. These profiles are used to define the cutoff radius and derive morphological properties. 
\par
The spatial extent of each region of interest (ROI) or potential H\,II region spectrum is defined using a hierarchical decision tree based on the statistical robustness of the radial profile fit, $R^{2}_{fit}$, and on the presence of extended or secondary emission components. Rather than adopting a fixed aperture, we combine analytical profile fitting, dispersion information, morphological constraints, and conservative flux thresholds to determine the final integration radius (typically in the order of $\sqrt{2}\sigma_{Pvoigt}$), for each region and avoid contamination from neighboring structures. 
\par
Additionally, the final radii are selected to exceed the instrumental resolution. This is achieved by adopting the FWHM of the PSF as the threshold. A full description of the method is presented in a forthcoming paper (Aragüete Riesco et al., in prep.). H\,II region geometrical parameters, including centroid, position, semi-major and semi-minor axis lengths, eccentricity, solidity, and orientation, are derived from minimum bounding ellipse fits.
\par
The galactic DIG subtraction is performed by characterizing the galactic diffuse emission within adaptive annuli surrounding each region and scaling it relative to the ROI extent. We then extracted the ROI-integrated spectra from the optimized PHANGS-MUSE NGC~628 'copt' data cube. 
\par
We removed the contribution of the underlying stellar population by fitting the spectra of each region with the Penalized Pixel-Fitting (pPXF) code \citep{Cappellari2023} over the wavelength range \SIrange{4800}{9250}{\angstrom}, which includes the [S\,III]$\lambda$ 9069 emission line. The pPXF fitting yields rectified spectra with the stellar continuum subtracted. We adopted a constant instrumental resolution of MUSE = \SI{2.65}{\angstrom}, despite its mild wavelength dependence \citep{Bacon2017}, and enforced a tolerance of five consecutive missing spectral values in the spectra. We employed Surface Brightness Fluctuations templates from the E-MILES database \citep{Vazdekis2012, Vazdekis2020}, assuming a \cite{Chabrier2003} initial mass function with slope of 1.30 and adopting Teramo$/$BaSTI stellar isochrones. Since the E-MILES templates have higher intrinsic spectral resolution than the MUSE data, they are convolved to match the observational resolution of our spectra. The templates library spans ages from 0.03 to 14 Gyr and metallicities from [Z/H] = $-$2.27 to +0.40. The fitting scheme relies solely on multiplicative Legendre polynomials to avoid modifying the continuum. 
\par
Emission line fluxes were fitted for each ROI using the Line Measuring library (LiMe) \citep{Fernandez2024}, assuming pseudo-Voigt profiles and a minimum S/N of 4 for line detection. Fluxes were corrected for dust attenuation using the \citet{ODonnell1994} extinction law with $R_{v}$=3.1. We adopted a theoretical F(H$\alpha$)/F(H$\beta$) ratio of 2.86 corresponding to Case~B recombination \citep{Osterbrock2006}. The deprojected galactocentric distance of each region was computed assuming a face-on orientation and an inclination of \SI{8.9}{\degree} (Table~\ref{tab:NGC628}). 
\par
Although our algorithm returns the number of pixels defining each ROI, this quantity is not used to estimate the radius. Instead, region volumes are inferred assuming asymmetry from their two-dimensional projection. A minimum bounding ellipse is fitted to estimate the area. The third spatial axis is approximated as the geometric mean of the semi-major and semi-minor axes. 
\subsection{Definition of the SIGNALS nebular region} \label{subsec:signalsboundaries}
Within the SIGNALS survey in NGC~628, \citep{RousseauNepton2018} integrated fluxes are corrected for DIG emission and subsequently fitted with pseudo-Voigt profiles within the ZoI. However, while H$\alpha$ fluxes are integrated out to the outer boundary of each region, emission line fluxes are limited to an aperture of radius $\sigma_{\mathrm{PVoigt}}$ to minimize DIG contamination and crowding effects. This can lead to an overestimation of H$\alpha$ luminosity relative to emission lines fluxes, which could affect their ratios in diagnostic diagrams. 
\par
In the SIGNALS public catalog, each nebular region is modeled as circular, with the radius estimated by convolving the pseudo-Voigt $\sigma$ with a Gaussian kernel to facilitate comparison with the classical spherical Strömgren model. To obtain radii consistent with the H$\alpha$ integration limit in Subsect.~\ref{subsec:phangsidentification}, we reconstructed the pseudo-Voigt $H\alpha$ intensity profile, assumed spherical symmetry, and adopted the radius of $\sqrt{2}\sigma_{Pvoigt}$, as a conservative size estimate. To take this into account in the derivation of the H$\alpha$ flux of the H\,II regions, we computed a flux correction factor quantifying the fraction of the integrated H$\alpha$ flux within the $\sqrt{2}\sigma_{Pvoigt}$ radius relative to the total within the outer boundary. The resulting H$\alpha$ correction factors derived have a mean value of $\sim$0.75 $\pm$ 0.05. Equivalently, all selected radii exceed the instrumental resolution, taking the SIGNALS survey FWHM as the effective resolution limit.
\par
Lastly, the galactocentric distances, $R_{g}$, are deprojected assuming a face-on orientation. While the public SIGNALS catalog adopts an inclination of \SI{21}{\degree}, we recalculated it for all regions using the same position angle and inclination as those adopted for the PHANGS-MUSE sample to ensure consistency. 
\subsection{BPT classification} \label{subsec:BPT}
We used the Baldwin–Phillips–Terlevich (BPT-type) diagnostic diagrams \citep{Baldwin1981}, together with the classification curves from \citet{Kewley2001} (Ke01) and \citet{Kauffmann2003} (Ka03), as an initial step to define our H\,II region subsample from the two surveys (Fig.~\ref{fig:BPT}). This approach discriminates photoionized H\,II regions from nebulae dominated by alternative ionization mechanisms. Three diagnostic diagrams are commonly used: $\mathrm{[N\,II]\lambda6583/H\alpha}$ versus $\mathrm{[O\,III]\lambda5007/H\beta}$ (BPT-NII);  $\mathrm{[S\,II]\lambda6716+\lambda6731/H\alpha}$ versus $\mathrm{[O\,III]\lambda5007/H\beta}$ (BPT-SII); and  $\mathrm{[O\,I]\lambda6300/H\alpha}$ versus $\mathrm{[O\,III]\lambda5007/H\beta}$ (BPT-OI). Since the SIGNALS survey does not include the [O\,I] $\lambda$6300 emission line, our analysis is restricted to the BPT-NII and BPT-SII diagrams. We applied conservative selection criteria following \cite{Kewley2006}, retaining only regions located below and to the left of the Ka03 classification line in the BPT-NII diagram, and below and to the left of the Ke01 line in the BPT-SII diagram.
\par
To minimize the impact of low signal-to-noise ratios, we required S/N > 4 for all relevant strong emission lines. The adopted criteria are summarized in Table~\ref{tab:HIIcriteria} of Appendix~\ref{app:selection_criteria}.
\par
In the SIGNALS survey, a small difference in the spatial integration of fluxes could remain after H$\alpha$ flux correction, since the forbidden lines are measured within $\sigma_{\mathrm{PVoigt}}$ \citep{RousseauNepton2018} while the corrected H$\alpha$ corresponds to $\sqrt{2}\,\sigma_{\mathrm{PVoigt}}$. Therefore, a slight underestimation of the [N\,II]/H$\alpha$ and [S\,II]/H$\alpha$ ratios may still be present. 
\par
The SIGNALS sample and the PHANGS sample we present (hereafter PHANGS) consist of H\,II regions with an $R^2_{fit}$ > 0.65 to ensure a consistent comparison between them. This threshold discards regions with poorly constrained radial profiles in SIGNALS and ensures that the retained H\,II regions in both catalogs have comparably well-constrained radii and H$\alpha$ luminosities.
\subsubsection{Identification of supernova remnants} \label{subsubsec:SNRs}
To ensure that all regions included in our sample are H\,II regions, we identified and removed potential supernova remnants (SNRs). 
\par
SNRs can contribute significantly to the H$\alpha$ emission through shock-driven ionization. In spatially integrated Integral Field Unit (IFU) spectra, the coexistence of photoionized and shock-ionized gas, according to recent work \citep{Points2019, DuartePuertas2024} can complicate the identification of SNRs using BPT-type diagnostics alone. We therefore adopted a set of complementary selection criteria, following \citet{Moumen2019}, to minimize contamination from shock-dominated sources.
\par
Additionally, we examined the velocity dispersion of the $\mathrm{H\alpha}$ and [S\,II]$\lambda$6716 emission lines, correcting for instrumental broadening as detailed in \cite{Li2024}. Shock-heated gas in SNRs produces broader emission line profiles than those typical of H\,II regions, which generally exhibit a velocity dispersion of \SI{20}{\kilo\metre\per\second} \citep{Winkler2017}. This additional kinematic criterion is applied only to the PHANGS-MUSE sample, as velocity dispersion measurements are not available for the SIGNALS survey. For H$\alpha$ and [S\,II]$\lambda$6716, the observed velocity dispersion was defined as an equivalent dispersion derived from the Gaussian FWHM profile ($\sigma = FWHM/(2\sqrt{2\ln2})$. Instrumental broadening is removed in quadrature using the wavelength-dependent line-spread function from \cite{Bacon2017}, yielding mean instrumental dispersions of $\sigma_{\text{inst}}$$(H\alpha)$ = \SI{48.9}{\kilo\metre\per\second} and $\sigma_{\text{inst}}$$([SII]6716)$ = \SI{48.7}{\kilo\metre\per\second}. 
\par
The adopted SNR identification criteria are listed below:
\begin{enumerate}
	\item Emission line ratio $\mathrm{[S\,II]\lambda6716+\lambda6731/H\alpha \geq 0.4}$ \citep{Mathewson1973}.
	\item Maximum physical size of $\sim$ 120 pc \citep{Cajko2009}.
	\item Classification in Sabbadin diagnostic diagrams \citep{Sabbadin1977}, following \cite{Leonidaki2012}
	\item Location outside the H\,II region domain in at least one BPT diagram.   
	\item $\mathrm{[O\,III]\lambda5007/H\beta < 6}$ \citep{Cox1985,Hartigan1987}.
	\item S/N([S\,II]$\lambda$6716) > 20 and $\sigma_{[S\,II]\lambda6716} > 50~\mathrm{km\,s^{-1}}$ \citep{Li2024}.
\end{enumerate}
\subsubsection{DIG regions and visual inspection} \label{subsubsec:DIGvisual}
Previous studies have shown that forbidden line ratios such as $\mathrm{[S\,II]\lambda6716/H\alpha}$ and $\mathrm{[N\,II]\lambda6584/H\alpha}$ reach significantly higher values in the DIG than in classical H\,II regions \citep[e.g.,][]{Madsen2006, Haffner2009,DellaBruna2020}. DIG emission may also exhibit line ratios similar to those of SNRs \citep{Elwert2005}. We therefore examined whether our regions display elevated ratios indicative of substantial DIG contamination, noting that low FF values are also characteristic of DIG structures, despite their typically low electron densities of $\sim$ 0.1 $cm^{-3}$ \citep{Ferriere2001}. 
We adopted conservative DIG-line thresholds of $Log([S\,II]\lambda6716/H\alpha$) > $-$0.55 \citep{ Belfiore2022} and $Log([N\,II]\lambda6583/H\beta$) > $-$0.35 \citep{FloresFajardo2009}, corresponding to values 0.5 $\mathrm{dex}$ below the canonical DIG limits. 
\par
We visually inspected all H\,II regions included in the PHANGS sample to remove cases in which the analytically defined region extended significantly into the DIG, despite being classified as H\,II regions based on their BPT line ratios. As a result, approximately 7\% of the initial PHANGS-H\,II region sample was discarded, primarily due to very irregular $H\alpha$ morphologies (e.g., crossing or filamentary structures) or excessively extended radii dominated by diffuse emission.
\section{Luminosity–size coupling induced by the definition of the H\,II region} \label{sec:LR}
\begin{figure*}[!t]
	\centering
	\includegraphics[width=\linewidth]{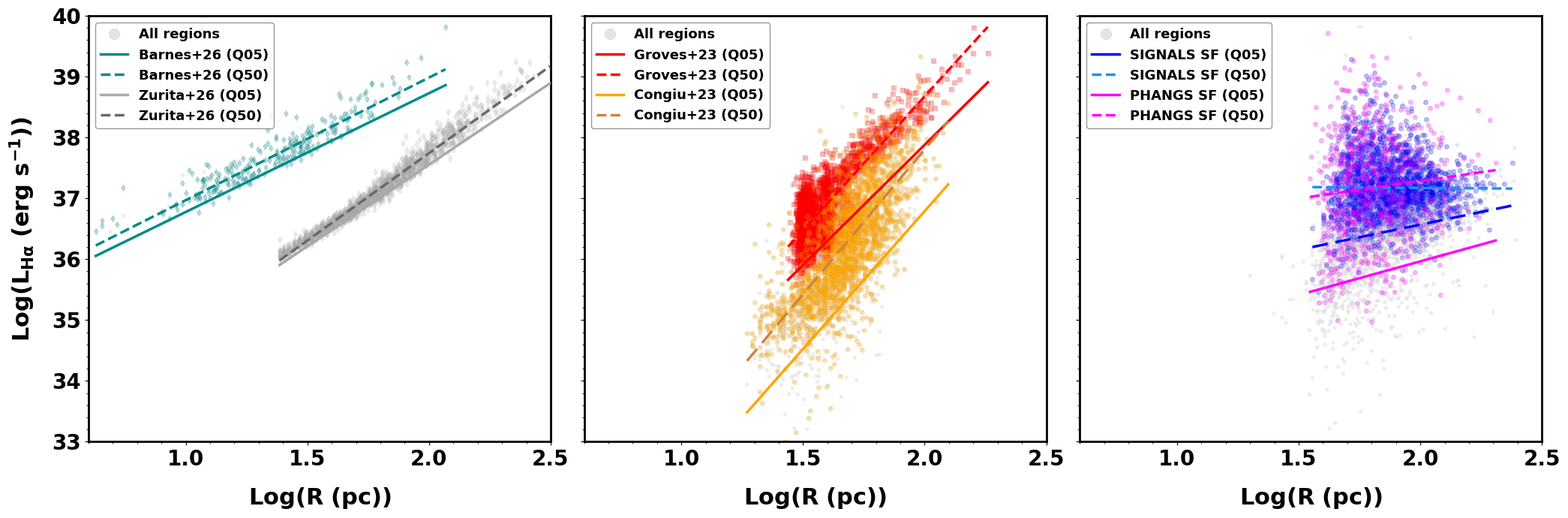}
	\caption{H$\alpha$ luminosity, $L_{H\alpha}$, against the nebula radius in NGC~628 for six published H\,II region catalogs. A homogeneous comparison via lower $5\%$ (Q05, solid lines) and median (Q50, dashed lines) quantile regressions of $\log(L_{H\alpha})$ on $\log(R)$ is shown. 
		Left panel (a): \cite{Barnes2026} (darkcyan diamonds, $\kappa_{05, 50} \sim 1.95, 2.01$) and \cite{Zurita2026} (dark gray diamonds, uncorrected fluxes, $\kappa_{05, 50} \sim 2.69, 2.87$), both with constrained absolute-threshold radius definitions; the \cite{Zurita2026} catalog already consists exclusively of H\,II regions and therefore has no gray background population. Middle panel (b): \cite{Groves2023} (red squares, not DIG-corrected, $\kappa_{05, 50} \sim 3.95, 4.40$) and \cite{Congiu2023} (orange circles, DIG-corrected, $\kappa_{05, 50} \sim 4.53, 4.76$), both with absolute-threshold radius definitions. Right panel (c): \cite{RousseauNepton2018} (SIGNALS, blue dots, $\kappa_{05, 50} \sim 0.83, -0.03$) and the PHANGS sample (magenta circles, $\kappa_{05, 50} \sim 1.66, 0.96$), both with profile-truncated radius definitions ($R^2_{\mathrm{fit}} > 0.65$ for SIGNALS). In each panel, objects beyond the BPT H\,II region zone of the parent catalog are shown as light gray points (see Sect.~\ref{sec:LR} and Sect.~\ref{app:LR_derivation} for details).}	
	\label{fig:PHANGSSIGNALS_ChristmasTree}
\end{figure*}
\par
The H$\alpha$ luminosity–size relation is commonly used to characterize the physical scaling of H\,II regions. Here, we compared the resulting luminosity--size relation for all considered catalogs of nebular regions, restricting each sample to star-forming (H\,II) regions identified using standard BPT-type diagnostics in order to minimize contamination from non-photoionized sources. The H$\alpha$ luminosities used in this comparison have not been corrected for the observed dependence on galactocentric radius. The effect of this correction on the luminosity--size relation is examined separately in Appendix~\ref{app:galactocentric_correction}.
Fig.~\ref{fig:PHANGSSIGNALS_ChristmasTree} presents the $\mathrm{log(L_{H\alpha})}$-$\mathrm{log(R)}$ relation for H\,II regions in NGC~628 in the six catalogs we considered, arranged in three panels. For each catalog, we performed the corresponding lower $5\%$ (Q05) and median (Q50) quantile-regressions. We denote the corresponding slopes as $\kappa_{05}$ and $\kappa_{50}$\footnote{$\kappa_{50}$ is shown for comparison with standard fits.}, respectively. The fitted slopes and intercepts, with 1000 bootstrap uncertainties, are listed in Table~\ref{tab:lr_quantile_fits}. For comparisons among catalogs, we adopted $\kappa_{05}$ as the reference estimator because it traces the lower envelope of the coherent H\,II region population at a given radius, while being less sensitive to the vertical scatter introduced by variations in emissivity among individual regions (Aragüete Riesco et al., in prep.).
\par
The three panels show different behavior: the power-law slope $\kappa$ of the $\mathrm{log(L_{H\alpha})}$-$\mathrm{R}$ relation does not appear to be a universal physical quantity, but rather a methodological observable that encodes the degree of coupling between the adopted radius definition and the $\mathrm{log(L_{H\alpha})}$ of each region. The catalogs span nearly two orders of magnitude in their range of H\,II region radius, from the $\sim$10~pc resolution limit of \citet{Barnes2026} to the $\sim$170~pc scale of the MUSE-based catalogs. The most striking differences are found in the scatter of the $\mathrm{log(L_{H\alpha})}$-$\mathrm{R}$ relation itself. In \cite{RousseauNepton2018} and the PHANGS sample (panel c), the slopes are shallow and even compatible with zero considering $\kappa_{50}$, with $\kappa_{05} \sim 0.83$--$1.66$ at Q05 flattening to $\kappa_{50} \sim -0.03$--$0.96$, accompanied by a pronounced vertical scatter at fixed $R$ and a triangular "Christmas-tree" distribution. On the other hand, \cite{Groves2023} and \cite{Congiu2023} (panel b) produce steep slopes of $\kappa_{05} \sim 3.95$--$4.53$ with a tighter, collapsed distribution. \cite{Barnes2026} and \cite{Zurita2026} (panel a) occupy intermediate positions with $\kappa_{05} \sim 1.95$--$2.69$.
\par
This diversity does not seem to reflect intrinsic differences in the physical properties of nebular regions in the catalogs. This is illustrated by the fact that \cite{Congiu2023} and the present work construct their nebular catalogs from the same H$\alpha$ frame but employ different pipelines for the adopted radius definition. These differences arise because each segmentation algorithm defines a region radius using criteria that introduce different levels of coupling between the adopted radius and the underlying H$\alpha$ surface brightness \citep{Wisnioski2012, McClain2026}, which can systematically tune the resulting luminosity–radius relation. 
\par
None of the methods discussed here should be intrinsically superior to the others from the standpoint of physical fidelity. Each approach makes well-defined, if different, operational choices about what constitutes the radius of a nebula, and each is internally self-consistent. As demonstrated analytically in Appendices~\ref{app:gamma_derivation} and \ref{app:kappa_L_derivation}, the slope $\kappa$ propagates the methodological contribution to the observed slopes of the $\mathrm{log(FF)}$-$\mathrm{log(R)}$ and $\mathrm{log(FF)}$-$\mathrm{log(L_{H\alpha})}$. Any departure of an observed slope from its predicted methodological $\mathrm{log(L_{H\alpha})}$-$\mathrm{R}$ coupling value therefore constitutes a physical signal, independent of the catalog construction (Eq.~\ref{ec:gamma_R_phys} in Appendix~\ref{app:LR_derivation}, and \ref{ec:eta_R_phys} in Appendix~\ref{app:kappa_L_derivation}). 
\par
We provide the basis for a direct, quantitative separation of physical and methodological effects in catalogs with fundamentally different radius definitions, predicting how the adopted radius definition propagates into derived relations such as $\mathrm{log(FF)}$–$\mathrm{log(R)}$ and $\mathrm{log(FF)}$–$\mathrm{log(L_{H\alpha})}$. We now describe the specific segmentation approach adopted by each catalog.
\par
\cite{RousseauNepton2018} and the PHANGS sample pipeline determine the integration radius based on the local flux–density diagram or radial profile structure, independent of the absolute flux level. In \cite{RousseauNepton2018}, the threshold is universal, set to $R_{\mathrm{cut}} = \sqrt{2}\sigma_\mathrm{Pvoigt}$ (Sect.~\ref{subsec:signalsboundaries}), whereas in the PHANGS sample we adopted an adaptive, hierarchical approach combining profile fitting, morphology, and flux thresholds (Sect.~\ref{subsec:phangsidentification}), yielding $R_{\mathrm{cut}}$ values comparable to the previous method because it is also based on geometrical refinements applied to initial radii derived from $\sigma_\mathrm{Pvoigt}$. Accordingly, in the two profile-truncated methods, the region radius is only partially coupled to the total H$\alpha$ luminosity, since the truncation is set relative to the local profile structure rather than to an absolute surface brightness threshold.
\par
In contrast, \cite{Groves2023} identify nebular regions in PHANGS-MUSE data using a HIIphot-based pipeline \citep{Thilker2000}, which grows regions iteratively from local maxima until the surface brightness falls below an absolute emission measure threshold of $EM_{min}$ = 0.5 $pc \ cm^{-6}$. The estimated radius is defined as the circularized equivalent of the segmented area, $R_{\mathrm{cir}} = \sqrt{\frac{A}{\pi}}$. \cite{Congiu2023} apply CLUMPFIND \citep{Williams1994} to the same PHANGS-MUSE data, identifying regions through intensity contours down to a threshold of background plus noise dispersion, and then applying a probabilistic, machine-learning algorithm to refine the classification of detected nebulae and distinguish H\,II regions from other ionized sources. In both cases the radius of nebular regions is defined as the circularized equivalent of the segmented area, A, $R_{\mathrm{cir}} = \sqrt{A/\pi}$, and $R_{\mathrm{cir}}$ grows proportionally with the absolute flux level because the radius of the nebulae is determined by an absolute threshold applied to the same flux map that is integrated to obtain the H$\alpha$ luminosity. This introduces a strong coupling between $R$ and $L_{H\alpha}$.
\begin{table}[H]
	\centering
	\caption{Quantile regression fits to the $\log L_{\mathrm{H}\alpha}$--$\log R$ relation for star-forming regions in each catalog.}
	\label{tab:lr_quantile_fits}
	\begin{tabular}{l c c c c}
		\hline\hline
		Catalog & Quantile & Slope ($\kappa$)  & Intercept & $N$ \\
		\hline
		Barnes+26 & $\kappa_{05}$ & $1.95 \pm 0.09$ & $34.82 \pm 0.14$ & 245 \\
		Barnes+26 & $\kappa_{50}$ & $2.01 \pm 0.07$ & $34.96 \pm 0.09$ & 245 \\
		Zurita+26 & $\kappa_{05}$ & $2.69 \pm 0.03$ & $32.18 \pm 0.06$ & 1458 \\
		Zurita+26 & $\kappa_{50}$ & $2.87 \pm 0.02$ & $32.00 \pm 0.03$ & 1458 \\
		Groves+23 & $\kappa_{05}$ & $3.95 \pm 0.35$ & $29.98 \pm 0.55$ & 2459 \\
		Groves+23 & $\kappa_{50}$ & $4.40 \pm 0.06$ & $29.87 \pm 0.10$ & 2459 \\
		Congiu+23 & $\kappa_{05}$ & $4.53 \pm 0.18$ & $27.73 \pm 0.29$ & 2342 \\
		Congiu+23 & $\kappa_{50}$ & $4.76 \pm 0.09$ & $28.29 \pm 0.14$ & 2342 \\
		SIGNALS & $\kappa_{05}$ & $0.83 \pm 0.21$ & $34.91 \pm 0.39$ & 1780 \\
		SIGNALS & $\kappa_{50}$ & $-0.03 \pm 0.09$ & $37.23 \pm 0.18$ & 1780 \\
		PHANGS & $\kappa_{05}$ & $1.66 \pm 0.35$ & $32.72 \pm 0.63$ & 761 \\
		PHANGS & $\kappa_{50}$ & $0.96 \pm 0.29$ & $35.41 \pm 0.53$ & 761 \\
		\hline
	\end{tabular}
\end{table}
\cite{Barnes2026} construct a catalog of H\,II regions in PHANGS galaxies by applying a top-down segmentation to the PHANGS-MUSE nebula catalog \cite{Groves2023} to HST H$\alpha$ imaging homogenized to a physical resolution of 10~pc and a spatially constant noise level $\sigma_{\mathrm{final,10pc}}$. Then, source identification follows a two-threshold procedure: compact structures detected above a high threshold of $5\sigma_{\mathrm{final}}$ are grown down a lower threshold of $2\sigma_{\mathrm{final}}$. $\sigma_{\mathrm{final}}$ is spatially uniform by selection, then the effective detection threshold is absolute rather than local. However, the growth of each region is topologically confined within the parent MUSE nebula mask, which prevents the radii from expanding as in the pure and unconstrained absolute threshold methods of \cite{Barnes2026} and \cite{Congiu2023}. This topological constraint partially decouples $R_{\mathrm{cir}}$ from the absolute flux level, yielding an intermediate $L_{H\alpha}$-$R$ slope of $\kappa_{05} \sim 1.95$. Here, the radius is defined identically as $R_{\mathrm{cir}}$ from the segmented area. In the \cite{Barnes2026} catalog only 216 regions were ultimately retained and positively identified as H\,II regions. However, the dominant factor distinguishing this catalog from the others is its higher spatial resolution. At $\sim10$ pc, HST H$\alpha$ imaging resolves H\,II region substructure approximately four times more finely than the $\sim35$--44 pc PSF-limited SIGNALS and PHANGS-MUSE data, yielding intrinsically more precise size measurements and, consequently, a tighter and systematically offset $L_{\mathrm{H}\alpha}$--$R$ relation (see also Sect.~\ref{subsec:fillingfactor}). The additional selection criteria likely contribute to the higher luminosities of the retained regions, whereas the improved spatial resolution primarily explains the tighter relation and its offset with respect to the other catalogs.
\par
Finally, \citet{Zurita2026} constructed H\,II region catalogs for NGC~628 using image segmentation based on the spatial gradient of the H$\alpha$ surface brightness. A background map tracing the diffuse ionized gas (DIG) emission is created by masking pixels above a gradient cut-off. H\,II regions are identified as structures with emission exceeding $3\sigma_{\rm rms-noise}$ in the background-subtracted image, where $\sigma_{\rm rms-noise}$ is approximately uniform. The method belongs to the class of photometrically constrained absolute threshold techniques, partially decoupling $R_{\mathrm{cir}}$ from the absolute flux and limiting the expansion of the radii compared to pure absolute threshold methods. Radii are circularized, defined as $R_{\mathrm{eq}} = \sqrt{\mathrm{A}/\pi}$. This approach yields an intermediate H$\alpha$ luminosity–radius slope of $\kappa_{05} \approx$ 2.69 for NGC~628.
\par
Based on the type of coupling between the adopted radius definition and the H$\alpha$ luminosity, the six catalogs compared in this section fall into three distinct methodological categories, which we analyzed in Appendix~\ref{app:LR_derivation}:
\begin{enumerate}
	\item Profile-truncated methods (\cite{RousseauNepton2018} and the PHANGS sample): the integration radius is defined by the local structure of the radial brightness profile, fixing $R_{\mathrm{cut}} \approx k\sigma$ independently of the absolute surface brightness $\Sigma_0$. The $L_{H\alpha}$-$R$ slope reflects the intrinsic variation of the peak surface brightness with region size, yielding $\kappa_{05} \sim 0.83$-$1.66$.	
	\item Absolute-threshold methods (\cite{Groves2023} and \citealt{Congiu2023}): the radius grows until the surface brightness falls below a fixed, universal threshold $\Sigma_{\min}$, introducing a strong coupling between $R_{\mathrm{cir}}$ and $L_{H\alpha}$ that depends on the ratio $\Sigma_0/\Sigma_{\min}$. This produces steep $L_{H\alpha}$-$R$ slopes of $\kappa_{05} \sim 3.95$-$4.53$.
	\item Hybrid methods (\cite{Barnes2026} and \cite{Zurita2026}): the coupling between $R$ and $\mathrm{L_{H\alpha}}$ is partially suppressed by a secondary constraint, yielding intermediate slopes $\kappa_{05} \sim 1.95$-$2.69$. In \cite{Barnes2026} this suppression is topological: the region growth is confined within a parent MUSE mask, limiting the maximum extent independently of the flux level. In \cite{Zurita2026} the suppression is photometric: the detection threshold adapts to the background-subtracted image. Despite having similar $\kappa_{05}$, the two mechanisms are different and are expected to produce distinct behaviors.
\end{enumerate}
\noindent The three families of methods yield systematically different power-law $\mathrm{log(L_{H\alpha})}$-$\mathrm{R}$ slopes, which reflect the extent to which the adopted radius definition couples the measured radius to the intrinsic flux distribution of each region. This is derived analytically in Appendix~\ref{app:LR_derivation}. Any deviation of an observed slope from its predicted methodological value therefore constitutes a physical measurement, independent of the radius definition. This interpretation should, however, be considered in the context of additional methodological differences among catalogs beyond the adopted radius definition, such as the application of DIG corrections (not applied in the case of the \cite{Groves2023} catalog) and differences in the post-processing pipelines used to derive integrated fluxes.
\par
We next examine the impact of different radius definitions on the
$\log(L_{\mathrm{H}\alpha})$--$\log(R)$ distribution. The vertical scatter in the $\mathrm{log(L_{H\alpha})}$–$\mathrm{log(R)}$ plane at fixed radius (the “Christmas-tree” distribution) visible in the left panel of Fig.~\ref{fig:PHANGSSIGNALS_ChristmasTree} illustrates that H\,II regions of a given H$\alpha$ luminosity can span a broad range of radii, and conversely, regions of similar size can exhibit a wide range of H$\alpha$ luminosities.
\par
Eq.~\ref{ec:linealff} shows that at fixed R, the luminosity varies because $\mathrm{FF}$ and $n_e$ both differ in the population. This diversity would be preserved only when the radius definition does not couple $R$ to $L_{H\alpha}$. Profile-truncated methods (\cite{RousseauNepton2018} and this work) determine the integration radius from the local radial brightness profile. Consequently, $R$ does not systematically increase with $L_{H\alpha}$, allowing the full range of FF and $n_e$ to appear as vertical scatter. The triangular shape arises because the lowest luminosities measured are limited by the detection limit (PSF). The observed upper-right limit implies that any region above this limit would be a very large, highly luminous region and would require an H$\alpha$ luminosity so extreme that either no coherent region can exist or stellar feedback disperses the gas before it can be observed. 
\par
By contrast, absolute surface brightness thresholds (\cite{Groves2023} and \cite{Congiu2023}) assign radii that grow with peak flux. Brighter regions have larger $R_{\rm cir}$, stretching the scatter. Here, faint compact and luminous extended regions cannot coexist at the same radius, collapsing the "Christmas-tree" into a narrow, nearly one-dimensional distribution.
\par
An intermediate case is found in \citet{Zurita2026} ($\kappa \approx 2.69$). When $\kappa $ approaches 3, it implies that the assigned radius for each region scales approximately as $R \propto L_{H\alpha}^{1/3}$. In practice, faint regions are assigned small radii while luminous regions receive large radii. The radius definition thus imposes a tight functional link between $\mathrm{L_{H\alpha}}$ and R, leaving little room for vertical scatter at fixed $R$. The Christmas-tree shape collapses because the method prevents compact regions from being highly luminous and extended regions from being faint, by assigning radii that closely track the intrinsic surface brightness of each region, compressing scatter perpendicular to the $L_{H\alpha}$–$R$ relation.
\par
The analysis from this point onwards is restricted to five catalogs, as the [S\,II] emission line fluxes in \cite{Zurita2026} are not provided for NGC\,628.
\section{Physical parameters and sample definition} \label{sec:physicalparams}
In this section, we describe how we determined the parameters involved in the FF calculation, together with the physical properties of the H\,II regions. We first estimate the local electron density, which is an important property of H\,II regions and provides key information for characterizing star-forming regions and studying the spatial variation of the interstellar medium (ISM) within the host galaxy.
We derived the mean luminosity-weighted electron density $\langle n_{e} \rangle$ and selection criteria required for the FF analysis, including the definition of the characteristic radius and the final sample selection. Finally, we then compare the local electron density, luminosity and size distributions of H\,II regions across the five catalogs.
\subsection{Local electron density} \label{subsec:neparams}
To compare H\,II regions based on their intrinsic physical characteristics, we first estimate their local, or $in situ$, electron density using the ratio of the collisionally excited [S\,II] emission lines at optical wavelengths: $\mathrm{[S\,II]}\,\lambda6716$ and $\mathrm{[S\,II]}\,\lambda6731$. 
\begin{figure}[H]
	\centering  
	\includegraphics[width=0.9\columnwidth]{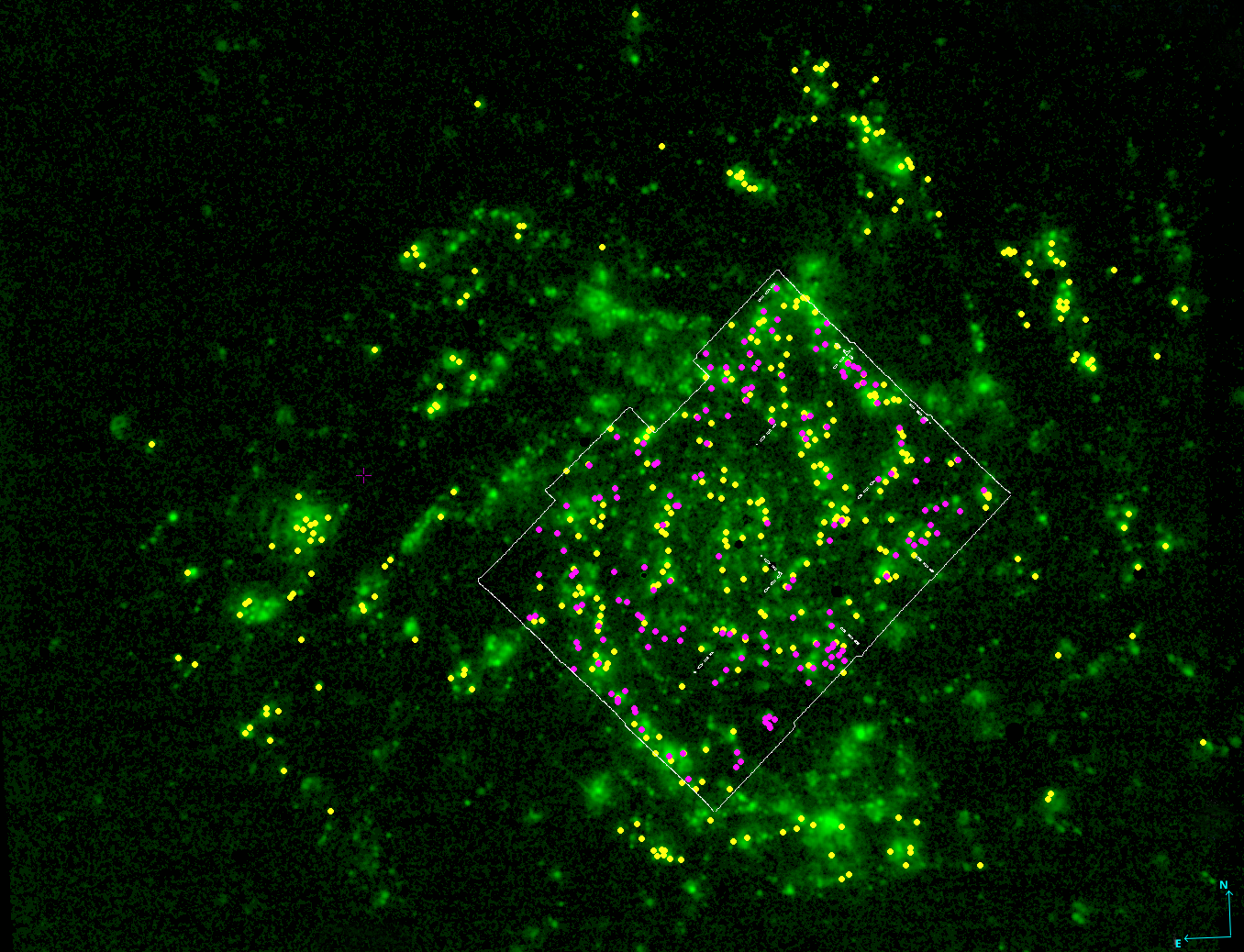}
	\caption{SIGNALS H$\alpha$ image of NGC\protect~628 showing the final selection of H\,II regions from SIGNALS-H\,II (yellow dots) and from PHANGS-H\,II (magenta dots). The white contour delineates the PHANGS-MUSE mosaicked field of view. North is up, and east is to the left.}
	\label{fig:2D}
\end{figure}
For the SIGNALS and PHANGS samples we estimated the local electron density ($n_e$) from the intensity ratio of the [S\,II] emission lines $\lambda\lambda6716,6731$, using the analytical solution for a three-level atom \citep{McCall1984}. This ratio is related to the electron density through
\begin{equation} \tag{11} \label{R_sii}
	\mathrm{R_{[S_{II}]} \sim 1.49\left[\frac{1 + 3.77x}{1 + 12.8x}\right]}
\end{equation}
where \quad $\mathrm{x=10^{-4}n_{e}t^{-1/2}}$ with $\mathrm{t=T_{e}/10000}$.
Assuming a typical gas temperature in the nebula of $\mathrm{T_{e} \sim 10^{4}~K}$ for star-forming regions (thus t = 1), the expression simplifies to
\begin{equation} \tag{12} \label{ne_sii}
	\mathrm{n_{e}=\left[\frac{1.49 - R_{[S_{II}]}}{12.8 R_{[S_{II}]}-5.6173}\right]\times10^4}.
\end{equation}
This method is valid over the density range $\mathrm{10 \  - \ 10^{4}~\ cm^{-3}}$ \citep{Castaneda1992, Zaritsky1994}. 
\subsection{rms electron density}\label{subsec:rms}
We calculated the mean luminosity-weighted $rms$ electron density, $\langle n_{e} \rangle$, for each region from the $H\alpha$ emission line flux. In PHANGS-H\,II, the radius $R$ was calculated as the geometric mean of the semi-major and semi-minor axes of the minimum bounding ellipse for each region, which corresponds to the radius of a circle with the same area. We assumed a constant electron temperature of $10^{4}~K$ for the H\,II regions \citep{Osterbrock1989, Gutierrez2010}, 
\begin{equation} \tag{13}
	\label{ec:rms}
	\langle n_{e} \rangle = 1.5\times10^{-16}\left[\frac{L_{H\alpha}}{R^3}
	\right]^{1/2} \quad \mathrm{cm^{-3}}.
\end{equation}
The $\langle n_{e} \rangle$ measurements show no significant variation with galactocentric radius, suggesting that the luminosity-weighted electron density is largely insensitive to the large-scale galactic environment before correcting for the radial luminosity gradient.
\par
Direct cross-catalog comparisons of the $\langle n_e \rangle_{\rm rms}$–$R$ slope are not straightforward because the observed gradient depends on the $L_{H\alpha}$–$R$ coupling parameter $\kappa$, which varies with radius definition. 
\subsection{Final H\,II region samples}\label{subsec:sample_selection}
We defined the final H\,II region samples of SIGNALS and PHANGS by applying a set of quality and physical criteria. We adopted a conservative lower limit of $n_e \sim 80\ \mathrm{cm^{-3}}$ to ensure reliable electron density estimates. In addition, we required a minimum signal-to-noise ratio in the emission lines used for the derivation of physical parameters, as summarized in Table~\ref{tab:HIIcriteria}. 
\par
For the FF calculation, the characteristic radius $R$ is defined according to the catalog method. In SIGNALS, $R$ corresponds to the profile-truncated radius derived from the pseudo-Voigt fit, while in PHANGS it is defined as the geometric mean of the semi-major and semi-minor axes of the minimum bounding ellipse, yielding an area-equivalent circular radius. The filling factor is then computed following Eq.~\ref{ec:ff}. 
We further excluded regions with $\log(\mathrm{FF}) < -7$, which are likely dominated by observational uncertainties. The resulting samples consist of 475~H\,II regions in SIGNALS-H\,II and 147 in PHANGS-H\,II.  
\par
To assess the consistency between electron density measurements in the two surveys, we constructed a new subsample (hereafter SIGNALS-MUSE) containing 227 regions from SIGNALS-H\,II sample. We created this subsample by limiting it to regions fully enclosed within the PHANGS-MUSE FoV, maintaining a minimum distance of 1~PSF from the mosaicked boundaries. Inspection of the right histogram in Figure~\ref{fig:ratioSII} (see Appendix) shows that both subsamples show a concentration of values near the adopted lower density limit $80~\mathrm{cm^{-3}}$. Fig.~\ref{fig:2D} shows the H$\alpha$ image of NGC~628 with the final selection of H\,II regions from SIGNALS-MUSE and PHANGS-H\,II.
To enable a direct comparison of the electron densities in the catalogs, we derived $n_e$ for the samples of \citet{Groves2023}, \citet{Congiu2023}, and \citet{Barnes2026}, as described in Subsubsect.~\ref{subsubsec:localdensity}. We applied a common set of quality criteria following the same procedure used for the SIGNALS-H\,II and PHANGS-H\,II catalogs (i) classification as H\,II according to BPT-type diagnostics; (ii) positive fluxes in the two [S\,II] $\lambda\lambda 6716, 6731$ lines; (iii) no edge or stellar-contamination flags when available; (iv) signal-to-noise ratios greater than four in [S\,II] $\lambda6716$, [S\,II] $\lambda6730$, [N\,II] $\lambda6583$, [O\,III] $\lambda5007$, H$\alpha$, and H$\beta$; and (v) H\,II region radii larger than the PSF FWHM. Furthermore, for the catalogs of \citet{Congiu2023}, \citet{Groves2023}, and \citet{Barnes2026}, the FF calculation assumes that $R$ corresponds to the circularized radius derived from the segmented area.
\begin{figure}[H]
	\includegraphics[width=0.495\columnwidth]{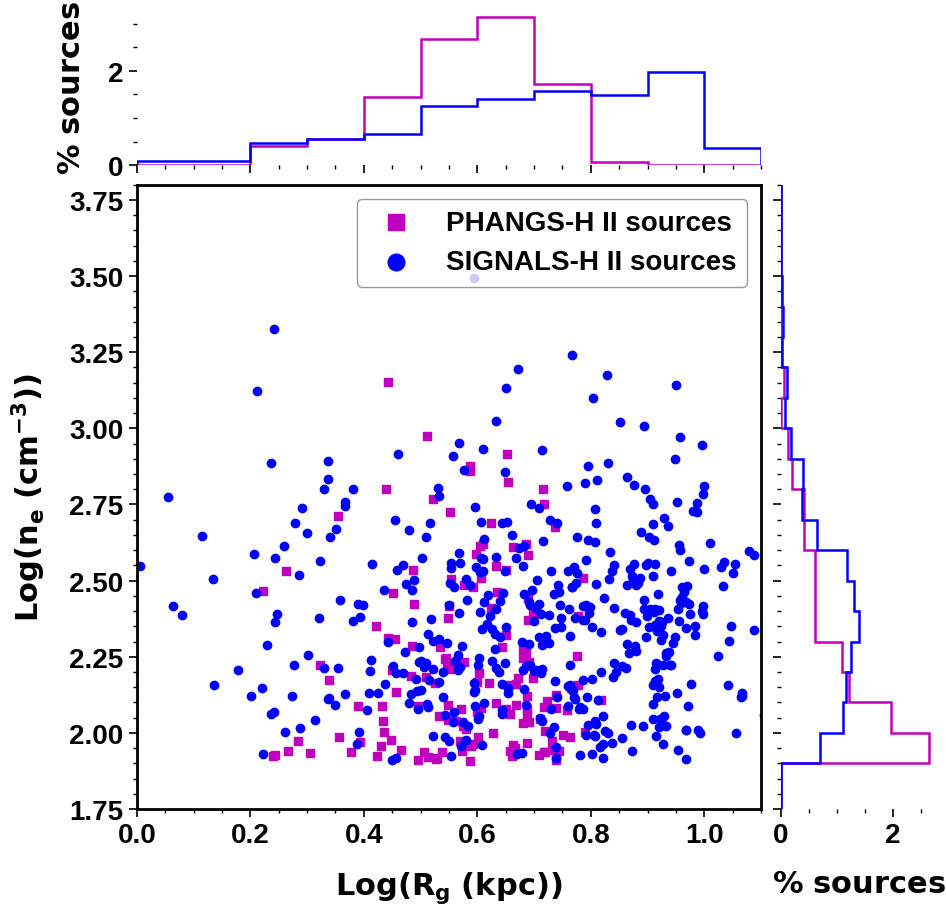}
	\includegraphics[width=0.495\columnwidth]{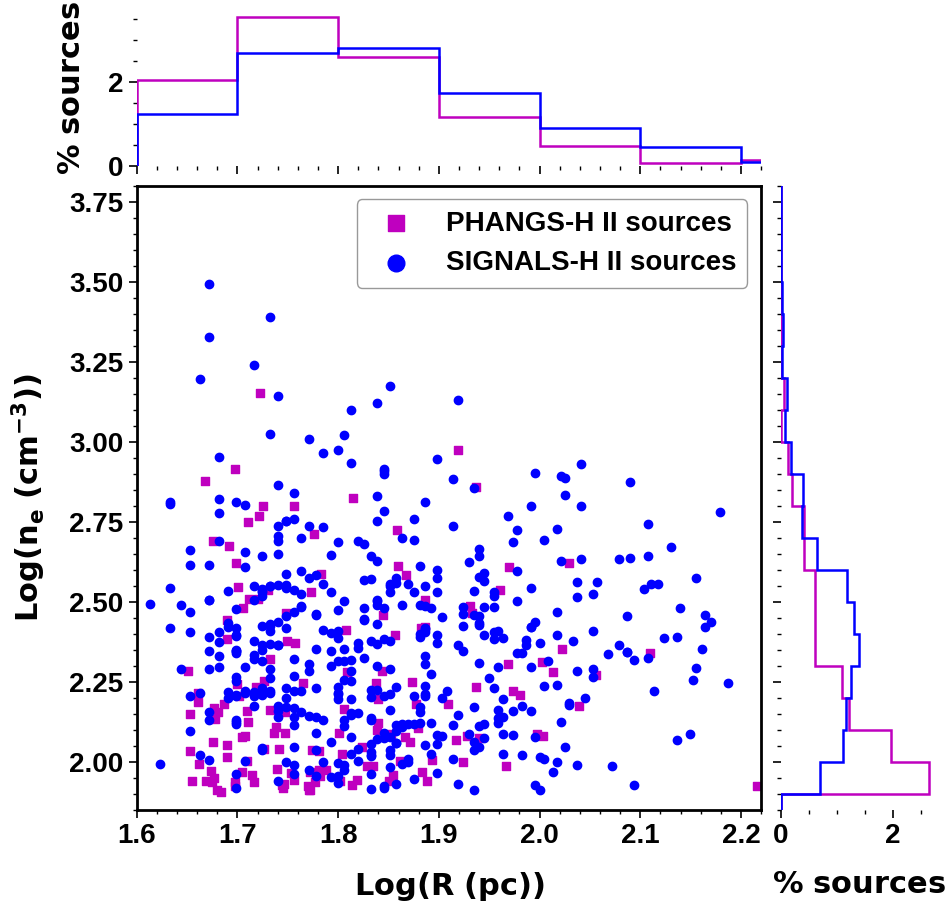}
	\caption {In situ electron density, $\mathrm{n_{e}}$, against the galactocentric radius, $\mathrm{R_{g}}$ of NGC~628 (left panel) and against the size of the region estimated through the profile-truncated radii, R, (right panel) for \cite{RousseauNepton2018} (SIGNALS, blue dots) and this work (magenta squares). The radii are defined by a criterion relative to the radial profile structure in the flux-distance diagram. In \cite{RousseauNepton2018}, R corresponds to a universal $\sqrt{2}\sigma_{Pvoigt}$ for all the segmented regions, and in this work, R emerges from the pseudo-Voigt adjustment and the hierarchical selection of the integration radius based on geometric and dispersion criteria. Regions below the low-density limit ($\mathrm{n_{e} \ < \ 80 \ cm^{-3}}$) are not shown.}
	\label{fig:neRgyR}
\end{figure}
These criteria reduced the parent samples to 391 regions out of 2855 in \citet{Groves2023}, 704 out of 3502 in \citet{Congiu2023}, and 216 out of 258 in \citet{Barnes2026}. The FF analysis was further restricted to regions with electron densities based on the \citet{McCall1984}, $n_{\mathrm e}\geq 80~\mathrm{cm^{-3}}$, yielding final samples of 27, 192, and 29 regions for \citet{Groves2023}, \citet{Congiu2023},
and \citet{Barnes2026}, respectively.
\subsection{Comparison of the H\,II region catalogs} \label{subsec:comparparams}
\subsubsection{Local electron density} \label{subsubsec:localdensity}
The left panel of Fig.~\ref{fig:neRgyR} shows that the local electron density exhibits no significant dependence on galactocentric radius. In the right panel of the same figure, the PHANGS-H\,II regions appear somewhat more populated at H\,II smaller radii. This trend likely reflects improved detection techniques within the mosaicked MUSE field, which enables the detection of a larger number of compact regions at small galactocentric radii. 
\par
\begin{figure}[H]
	\centering
	\includegraphics[width=\columnwidth]{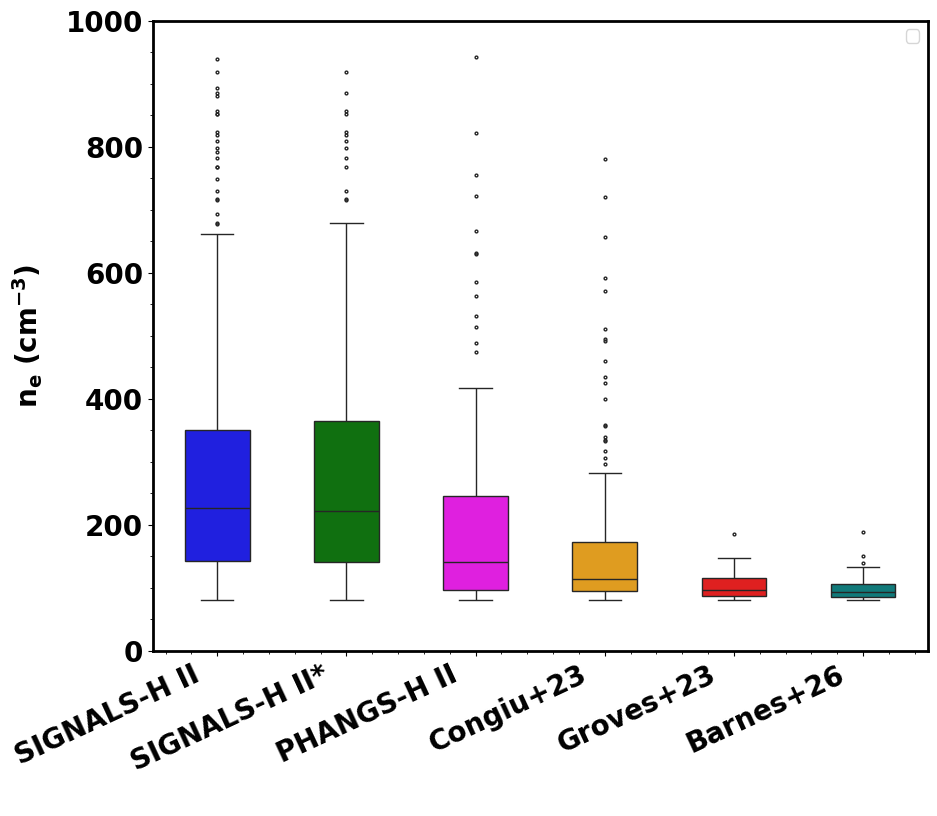}
	\caption{$n_e$ boxplot distributions for H\,II regions in NGC\,628 in five catalogs: SIGNALS-H\,II, SIGNALS-H\,II* (where the asterisk indicates the subsample limited to sources located within the MUSE-mosaicked FoV), PHANGS-H\,II, \cite{Groves2023, Congiu2023, Barnes2026}. The electron density is derived from the [S\,II]\,$\lambda\lambda$\,6716/6731 doublet ratio following the analytical solution of \citet{McCall1984}, assuming $T_e = 10^4$\,K in all catalogs.}
	\label{fig:comparne} 
\end{figure}

Fig.~\ref{fig:comparne} shows the $n_e$ distributions for the five catalogs. \cite{Groves2023} and \cite{Barnes2026} show the lowest median densities (96.8 $\pm$ 11.3 and 93.8 $\pm$ 10.6 $cm^{-3}$ respectively, with uncertainties given by the median absolute deviation), \cite{Congiu2023} and PHANGS-H\,II show intermediate median comparable densities (113.8 $\pm$ 27.2 and 140.84 $\pm$ 50.1 $cm^{-3}$ respectively), consistent with samples spanning a wider range of physical conditions. The SIGNALS-H\,II catalog shows the highest median densities (231.1 $\pm$ 97.5 $cm^{-3}$). When the comparison is restricted to the spatial MUSE FoV, the median yields (211.9 $\pm$ 88.4 $cm^{-3}$; see also Fig.~\ref{fig:ratioSII}). 
\par
The observed offset should not be interpreted solely as evidence of intrinsic differences between the H\,II regions sampled. Instead, it likely results from a combination of observational and methodological effects, including differences in spatial resolution, sensitivity, DIG treatment, flux integration, and region segmentation. At the same time, the PHANGS and SIGNALS catalogs are not fully overlapping samples owing to their different FoV and selection functions (Sect.~\ref{sec:data}).
\par
This interpretation is consistent with the electron density distributions shown in Fig.~\ref{fig:ratioSII}.
\subsubsection{Luminosity and size distributions in the catalogs}\label{subsubsec:LRdistcatalogs}
\begin{figure}[H]
	\includegraphics[width=0.495\columnwidth]{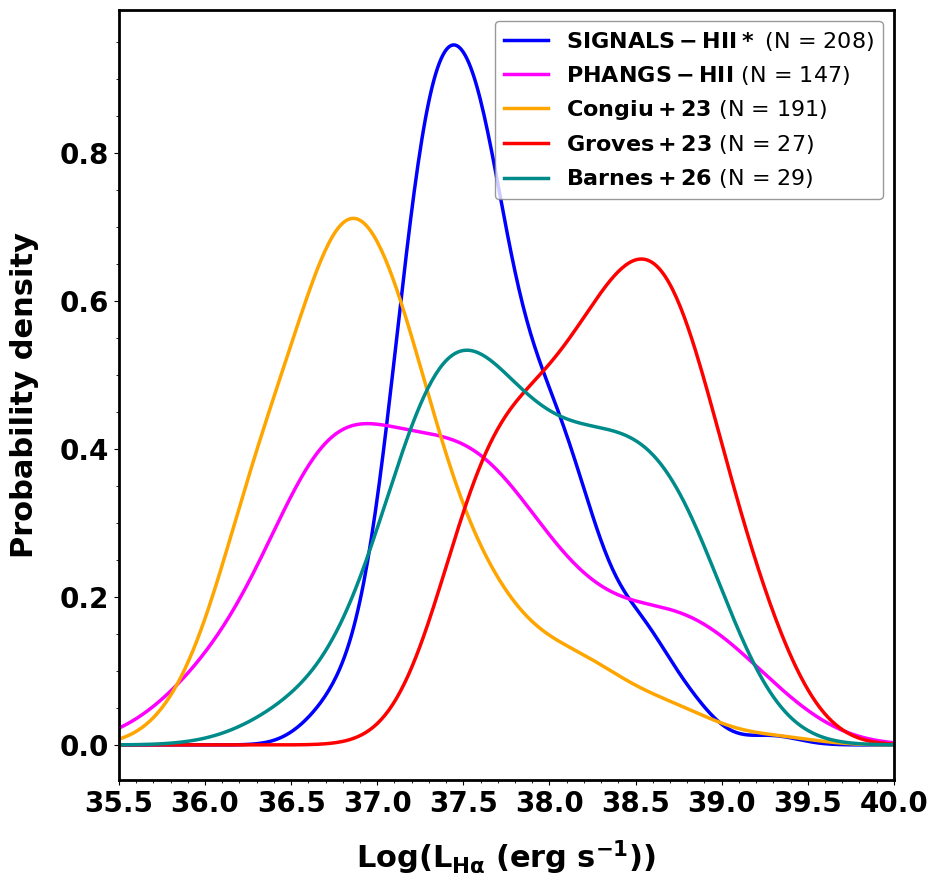}
	\includegraphics[width=0.495\columnwidth]{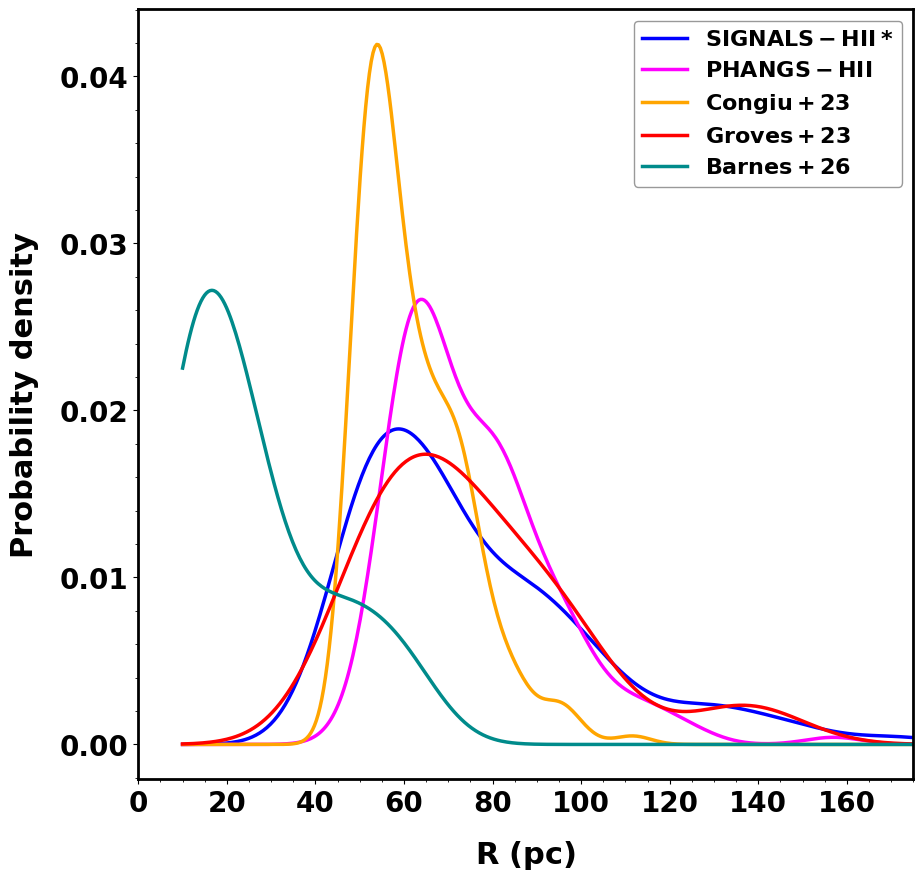}
	\caption{Left panel: KDE of the H$\alpha$ luminosity distribution, $\mathrm{log(L_{H\alpha})}$, for H\,II regions in NGC\,628 in all five catalogs: SIGNALS-H\,II* (where the asterisk denotes the subsample restricted to sources within the MUSE mosaic footprint), PHANGS-H\,II, and the catalogs of \cite{Congiu2023, Groves2023, Barnes2026}. Right panel: Distributions of the H\,II region radii in the same catalogs. The radius $R$ corresponds to the profile-truncated radius for SIGNALS-H\,II. For PHANGS-H\,II, $R$ is computed as the geometric mean of the semi-major and semi-minor axes of the minimum bounding ellipse fitted to each region, providing an equivalent circular radius. For the catalogs of \cite{Congiu2023}, \cite{Groves2023}, and \cite{Barnes2026}, the radii correspond to the equivalent circular radius derived from the number of pixels within the segmented area. In \cite{Barnes2026}, below 10~pc, the radius definition is dominated by resolution effects.} 
	\label{fig:comparLR}
\end{figure}

We compared the kernel density estimate (KDE) H$\alpha$ luminosity distribution in all catalogs. Fig.~\ref{fig:comparLR} illustrates the differences in luminosity and size distributions in all catalogs. In the left panel, the median H$\alpha$ luminosities are $\sim$ 37.6, 38.3, and 37.9 $\mathrm{erg \ s^{-1}}$ for SIGNALS-H\,II restricted to the MUSE mosaicked FoV, \cite{Groves2023, Barnes2026}, respectively, while \cite{Congiu2023} and PHANGS-H\,II reach fainter populations with medians of 36.9 and 37.4 $\mathrm{erg \ s^{-1}}$. 
Considering the luminosity histograms in Fig.~\ref{fig:signalsphangsLHa} for SIGNALS-H\,II and PHANGS-H\,II, the SIGNALS-H\,II sources populate a similar luminosity range than PHANGS-H\,II but the latter recovers a larger number of intermediate- to low-luminosity regions.
\par
Conversely, the DIG-uncorrected catalog of \cite{Groves2023} integrates fluxes out to nearly the full radial extent of the regions, which expand for brighter regions and increases the luminosity measurement, whereas PHANGS-H\,II truncates the integration at a profile-defined radius, which enables the detection of compact faint regions. The size distributions in the right panel of Fig.~\ref{fig:comparLR} show similar behavior. The radii histogram for SIGNALS-H\,II and PHANGS-H\,II catalogs in  Fig.~\ref{fig:signalsphangsSize} shows in the left panel that nearly 60\% of SIGNALS-H\,II sources have radii smaller than 80~pc, with a long tail toward larger sizes. The PHANGS-H\,II subsample reveals that nearly 80\% of regions have radii below 80~pc, with a notably higher fraction of small resolved regions at 50~pc. 
\cite{Barnes2026} yields the smallest mean radius (26 pc), consistent with the HST-resolved substructures at $\sim$10 pc resolution. SIGNALS-H\,II and \cite{Groves2023} reach the largest mean radii (77.9 and 76.1 pc), reflecting boundaries that extend to large fractions of the Gaussian profile in bright regions. PHANGS-H\,II and \cite{Congiu2023} show intermediate values (60.3 and 62.1 pc), with \cite{Congiu2023} displaying the smallest dispersion (11.7 pc) among the MUSE-based catalogs.
\section{Analysis and results} \label{sec:analysis}
In this section, we present the FF values of H\,II regions in NGC~628 and investigate their relation with physical quantities. We analyze how FF varies with the radius of H\,II regions, local electron density, and H$\alpha$ luminosity. We further compare its distribution in the different H\,II region catalogs and assess the range of physical conditions covered by the classical FF formulation. Finally, we investigate possible correlations with other structural properties, including a possible connection with PAH-to-dust emission.
\subsection{Filling factor in H\,II regions} \label{subsec:fillingfactor}
\begin{figure}[H]
	\centering  
	\includegraphics[width=\columnwidth]{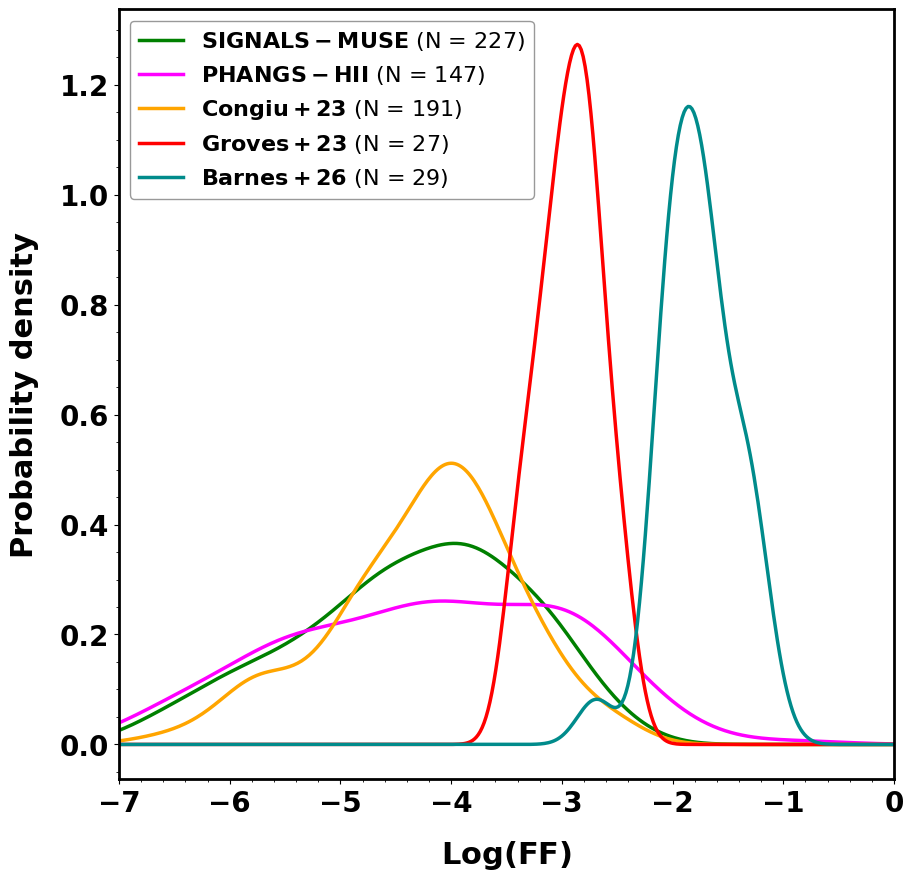}
	\caption{KDE of the $\mathrm{log(FF)}$ distribution of H\,II regions in NGC~628 among all the different catalogs: SIGNALS-H\,II, PHANGS-H\,II, \cite{Groves2023,Congiu2023, Barnes2026}. The overall range of FF values is comparable in the PHANGS-MUSE and SIGNALS catalogs, although the median values shift systematically depending on the adopted boundary definition and spatial resolution.}
	\label{fig:comparFF}
\end{figure}

The range of FF values in Fig.~\ref{fig:comparFF} exhibits consistency among the SIGNALS-H\,II,  PHANGS-H\,II and \cite{Congiu2023} catalogs (The specific histogram comparing the FF distributions of SIGNALS-H\,II and PHANGS-H\,II is presented in Fig.~\ref{fig:signalsphangsff}). However, the  median values shift systematically depending on the adopted boundary definition and spatial resolution in SIGNALS-H\,II, PHANGS-H\,II, and the catalogs of \cite{ Congiu2023, Groves2023, Barnes2026}. \cite{Barnes2026} shows the highest median value, $\mathrm{log(FF)}$ = -1.8, followed by \cite{Groves2023} with $-2.9$, while SIGNALS-H\,II, PHANGS-H\,II and \cite{Congiu2023} go down to a lower median value of $\mathrm{log(FF)} \sim$ -4.4, -4.2 and -4.1 respectively, with PHANGS-H\,II having the largest dispersion ($\sigma = 1.4$ dex). This ordering follows the differences in the characteristic radii discussed above (see Sect.~\ref{sec:LR}). \cite{Barnes2026}, which reports the smallest median radius, yields the highest FF values, whereas catalogs adopting larger boundary radii shift the distribution toward lower FF. There is substantial overlap between the SIGNALS-H\,II, PHANGS-H\,II, and \cite{Congiu2023} distributions.
\par
The \citet{Groves2023} catalog does not include an explicit DIG subtraction, which may lead to systematically higher FF estimates, since diffuse emission within the aperture increases the measured H$\alpha$ luminosity. In addition, DIG contamination may affect the [S\,II] emission line fluxes, which can bias the inferred electron densities toward lower values. These combined effects may therefore contribute to higher inferred FF values relative to the other catalogs.
\par
To assess whether the systematically higher FF values reported by \citet{Barnes2026} reflect an extension of the same H\,II regions identified in PHANGS-MUSE catalogs toward smaller physical scales or are partly driven by the higher angular resolution of HST relative to MUSE/SITELLE, we estimate the impact of beam smearing on the inferred radii and derived FF values. Under the assumption of Gaussian profiles, the observed size of a region with intrinsic (HST-resolved) radius $R_{\rm true}$ can be approximated as the quadratic combination of the intrinsic size and the PSF full width at half maximum, $\Phi$:
\begin{equation} \tag{14}
	R_{\rm obs} \approx \frac{1}{2}\sqrt{(2R_{\rm true})^2 + \Phi^2},
	\label{eq:beamsmear}
\end{equation}
Adopting $\Phi \approx 35$~pc for SIGNALS/SITELLE and $\Phi \approx 43.78$~pc for PHANGS--MUSE (Sect.~\ref{sec:data}), and considering a region in \cite{Barnes2026} with a mean radius in NGC~628 of $\sim26$~pc, Eq.~(\ref{eq:beamsmear}) yields $R_{\rm obs} \approx 34$~pc, corresponding to $\log(R_{\rm obs}/R_{\rm true}) \approx 0.12$~dex. Since ${\rm FF} \propto R^{-3}$ at fixed $L_{\rm H\alpha}$ and $n_e$, this implies $\Delta\log({\rm FF}) \approx -3 \log(R_{\rm obs}/R_{\rm true}) \approx -0.35~{\rm dex}$. The observed offset between the median $\log({\rm FF})$ measured by \citet{Barnes2026} and those derived for SIGNALS-H\,II, PHANGS-H\,II, and \citet{Congiu2023}, amounting to $\sim2.1$--2.3~dex, can therefore be partially attributed to beam smearing. However, this effect accounts for a somewhat small fraction of the offset for sources near the population mean, increasing for the most compact regions close to the $\sim10$~pc HST resolution limit. The remaining discrepancy likely reflects the inclusion of high-surface-brightness H\,II regions that are resolved by HST but only partially resolved, or unresolved, in MUSE/SITELLE observations. We therefore interpret the elevated FF values reported by \citet{Barnes2026} as arising from a combination of resolution effects and a shift in the sampled FF distribution toward smaller and denser H\,II regions, without requiring a fundamentally distinct FF regime.
\par
Beam smearing affects all seeing-limited datasets we considered, namely the SIGNALS and PHANGS-MUSE catalogs, although its impact is most pronounced when comparing these samples to the higher-resolution HST-based catalog of \citet{Barnes2026}.
\subsection{Filling factor relations}\label{subsec:FFrelations}
\begin{figure}[htbp]
	\begin{center}
		\includegraphics[width=\columnwidth, keepaspectratio]{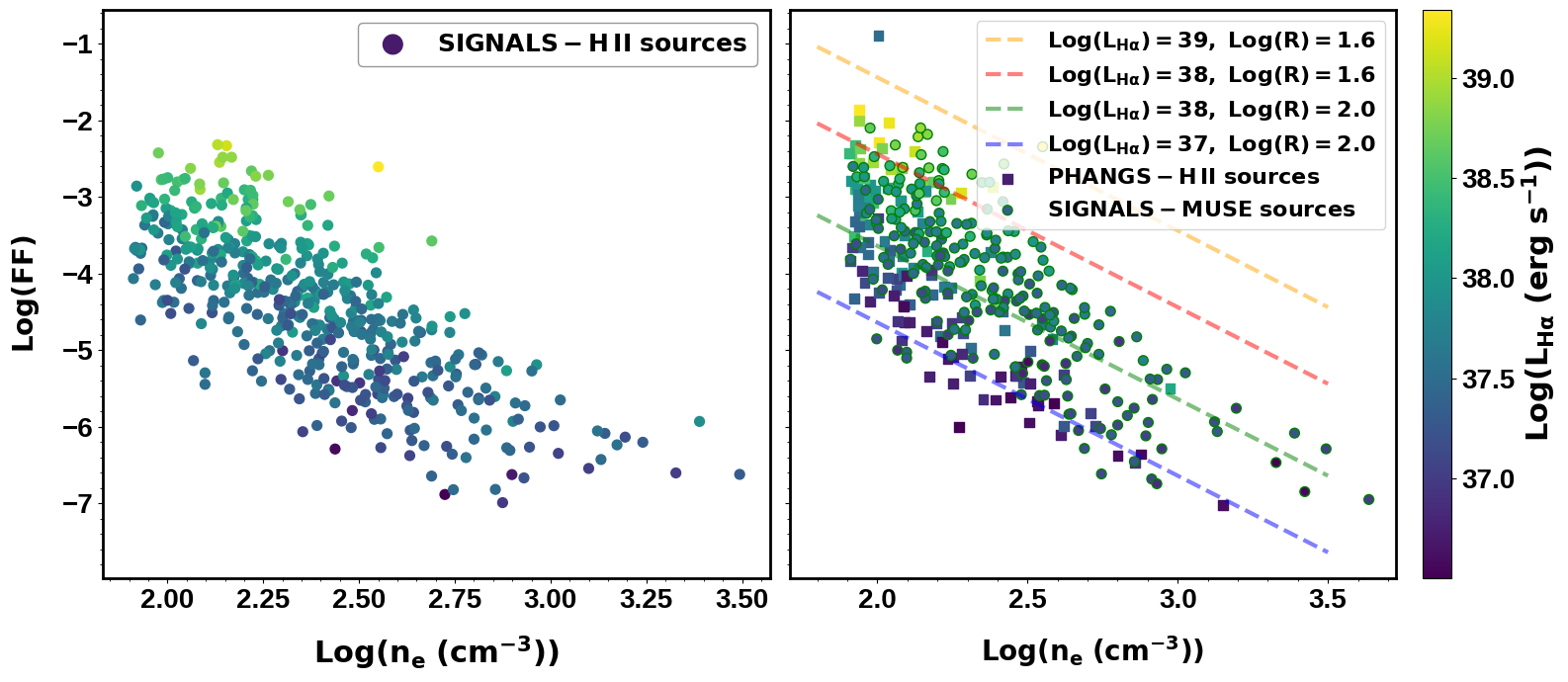}	
		\caption{$\mathrm{Log(FF)}$ against the local electron density, color-coded according to the H\,II region’s luminosity. In the left panel, we show the SIGNALS-MUSE regions, and in the right panel the subsample of SIGNALS-MUSE H\,II regions along with PHANGS-H\,II regions. The dashed lines show the theoretical $\mathrm{log(FF)}$–$\mathrm{log(n_e)}$ for selected $\mathrm{log(L_{H\alpha})}$ and $\mathrm{log(R)}$ values assuming the Strömgren sphere model.}
		\label{fig:FFne}
	\end{center}
\end{figure}

The definitional scalings presented in Eq.~\ref{ec:linealff} show that, by construction, FF is not independent of $\mathrm{L_{H\alpha}}$, R, and $n_e$. However, the observed trends may still carry physical information beyond this definitional scaling through the scatter, and we can use these relations as diagnostics of the regime in which the classical FF formulation provides a physically meaningful description. It is also necessary to account for the artificial trend introduced by the $Log(L_{H\alpha})$–$Log(R)$ coupling, which is addressed in Sect.~\ref{sec:LR}. 
Within this framework, we performed linear regression analyses to investigate correlations with L, R, and $n_e$. The fitted slopes and intercepts for the $\mathrm{log(FF)}$–$\mathrm{log(R^{3})}$, $\mathrm{log(FF)}$ – $\mathrm{log(n_e)}$, and $\mathrm{log(FF)}$–$\mathrm{log(L_{H\alpha})}$ relations are summarized in Table~\ref{tab:FF_fit} for the SIGNALS-H\,II and PHANGS-H\,II samples. The associated plots are Figs.~\ref{fig:FFvol}, \ref{fig:FFne}, \ref{fig:FFLHa}.
\begin{table*}[ht bp]
	\centering
	\caption{Linear regression models for the FF parameters}
	\label{tab:FF_fit}
	\begin{adjustbox}{max width=\textwidth}
		\begin{tabular}{@{} l c c S[table-format=-1.4] S[table-format=1.4] S[table-format=-1.3] S[table-format=1.4] l @{}}
			\toprule
			\multicolumn{1}{c}{Catalog} & \multicolumn{1}{c}{Model} & \multicolumn{1}{c}{Parameter} & {Values} & {Std. Error} & {$t$-value} & {$p$-value} & \multicolumn{1}{c}{Model Statistics} \\
			\midrule
			\multirow{6}{*}{SIGNALS-H\,II} & \multirow{2}{*}{$\mathrm{log(FF)} = \alpha \, \mathrm{log(n_e)} + \beta$} & $\beta$ & 1.84 & 0.25 & 7.38 & \text{< 0.001} & \multirow{2}{*}{$R^2_{fit}=0.58$, F-stat.=646.9} \\
			& & $\alpha$ & -2.65 & 0.10 & -27.51 & \text{< 0.001}  & \\
			\addlinespace
			& \multirow{2}{*}{$\mathrm{log(FF)} = \gamma \, \mathrm{log(R^3)} + \delta$} & $\delta$ & 1.86 & 0.56 & 3.34 &  0.001 & \multirow{2}{*}{$R^2_{fit}$= 0.21, F-stat.= 130.2} \\
			& & $\gamma$ & -1.13 & 0.10 & -11.41 & \text{< 0.001} & \\
			\addlinespace
			& \multirow{2}{*}{$\mathrm{log(FF)} = \eta \, \mathrm{log(L_{H\alpha})} + \zeta$} & $\zeta$ & -63.35 & 2.27 & -28.41 & \text{< 0.001} & \multirow{2}{*}{$R^2_{fit}=0.60$, F-stat.=699.1} \\
			& & $\eta$ & 1.60 & 0.06 & 26.44 & \text{< 0.001} & \\
			\midrule
			\multirow{6}{*}{PHANGS-H\,II} & \multirow{2}{*}{$\mathrm{log(FF)} = \alpha \, \mathrm{log(n_e)} + \beta$} & $\beta$ & 2.48 & 0.84 & 2.97 & 0.004 & \multirow{2}{*}{$R^2_{fit}=0.42$, F-stat.=46.64} \\
			& & $\alpha$ & -2.72 & 0.40 & -6.83 & \text{< 0.001} & \\
			\addlinespace
			& \multirow{2}{*}{$\mathrm{log(FF)} = \gamma \, \mathrm{log(R^3)} + \delta$} & $\delta$ & 2.24 & 1.96 & 1.15 & 0.25 & \multirow{2}{*}{$R^2_{fit}$= 0.07, F-stat.= 10.36} \\
			& & $\gamma$ & -1.18 & 0.37 & -3.22 & \text{< 0.001} & \\
			\addlinespace
			& \multirow{2}{*}{$\mathrm{log(FF)} = \eta \, \mathrm{log(L_{H\alpha})} + \zeta$} & $\zeta$ & -49.89 & 3.07 & -16.29 & \text{< 0.001} & \multirow{2}{*}{$R^2_{fit}=0.63$, F-stat.=224.0} \\
			& & $\eta$ & 1.22 & 0.08 & 14.97 & \text{< 0.001} & \\
			\bottomrule
		\end{tabular}
	\end{adjustbox}
	\tablefoot{
		\tablefoottext{a}{Luminosities are $H\alpha$ in $\mathrm{erg\,s^{-1}}$, electron densities in $\mathrm{cm^{-3}}$, volumes in $\mathrm{pc^3}$; only PHANGS-H\,II regions with $Log(L_{\mathrm{H}\alpha}) > 36.5~\mathrm{erg\,s^{-1}}$ are included in the fits. }
		\tablefoottext{b}{The Coefficient (column 4) corresponds to the best-fit regression coefficient, with Std. Error (column 5) indicating its 1$\sigma$ uncertainty. The t-value (column 6) is the ratio between the coefficient and its standard error, used to test whether the parameter is significantly different from zero, and the associated p-value (column 7) gives the corresponding significance level. The coefficient of determination ($R^{2}_{fit}$; column 8) quantifies the fraction of variance explained by the model, while the F-statistic (column 9) assesses the overall significance of the linear regression.} 
		\tablefoottext{c}{$R$ denotes the profile-truncated radius for SIGNALS-H\,II, and the geometric mean of the semi-major and semi-minor axes of the minimum bounding ellipse fitted to each region for PHANGS-H\,II. } }
\end{table*}
\subsubsection{Relation with the volume}\label{subsubsec:FFR}
\begin{figure}[htbp]
	\begin{center}
		\includegraphics[width=\columnwidth, keepaspectratio]{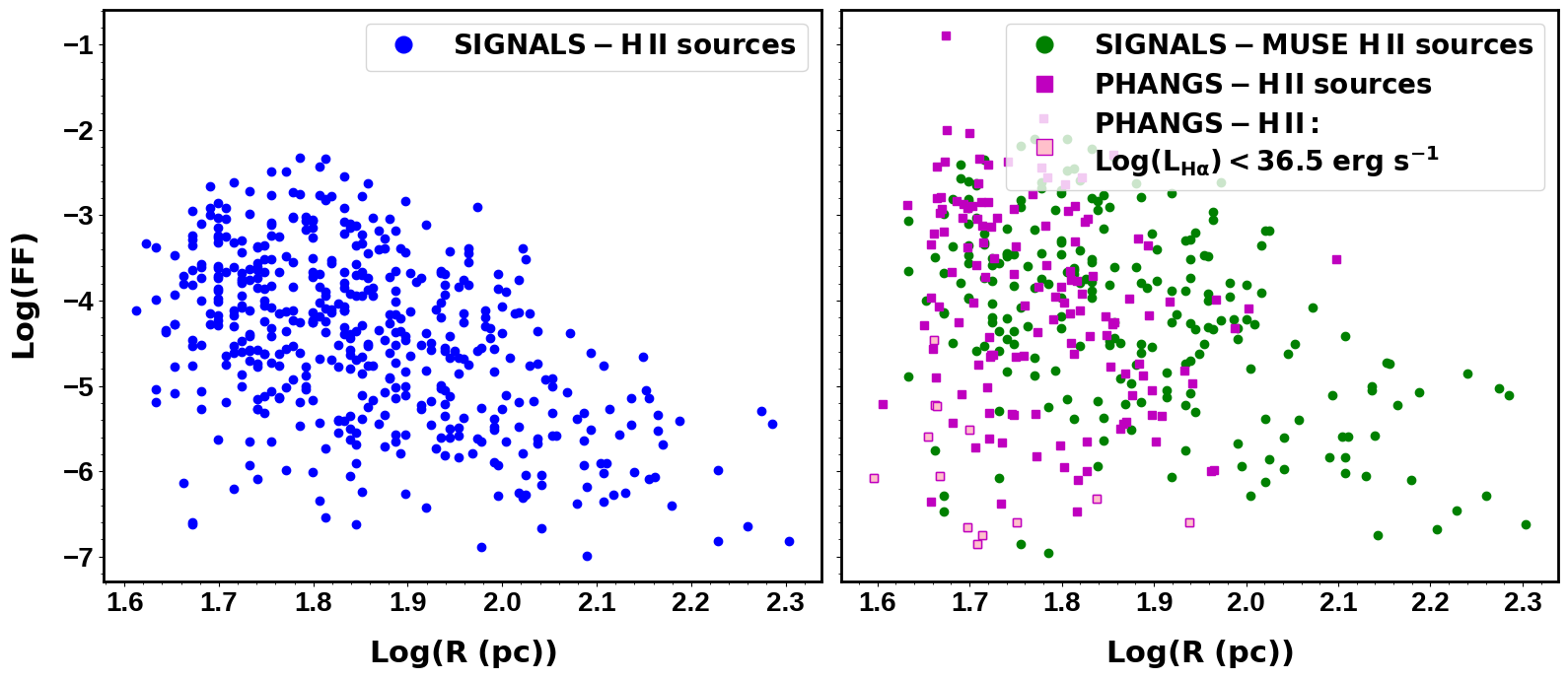}
		\caption{FF against the radii of H\,II regions in log-log scale. We show the SIGNALS-H\,II regions in the left panel, and PHANGS-H\,II regions (magenta squares) along with SIGNALS-MUSE H\,II regions (green dots) in the right panel. The light magenta squares represent PHANGS-H\,II fainter regions with $\mathrm{log(L_H\alpha) < 36.5~\ erg \ s^{-1}}$. These fainter regions exhibit the greatest deviation from the trend.}
		\label{fig:FFvol}
	\end{center}
\end{figure}
Fig.~\ref{fig:FFvol} presents the correlation between FF and radii of H\,II regions in logarithmic scale. The low coefficients of determination obtained for the FF-volume relation ($R^2_{fit} \sim 0.07-0.21$; Table~\ref{tab:FF_fit}) indicate that this variable explains only a small fraction of the variance in FF. Although the definitional scaling is recovered in the slope values, the volume alone does not capture the intrinsic scatter of FF within the sample, which reflects variations in the $\mathrm{L_{H\alpha}/n_e^2}$ and can provide insight into secondary trends among the other parameters. This behavior is consistent with Fig.~\ref{fig:comparLR}, where for a given radius (or equivalently volume) the regions display a vertical dispersion reflecting variations in $FF \, n_{e}^2 \, j_{H\alpha}(T_e)$ (Eq.~\ref{ec:neloc}). 
\begin{figure}[htbp]
	\begin{center}
		\includegraphics[width=0.495\columnwidth, keepaspectratio]{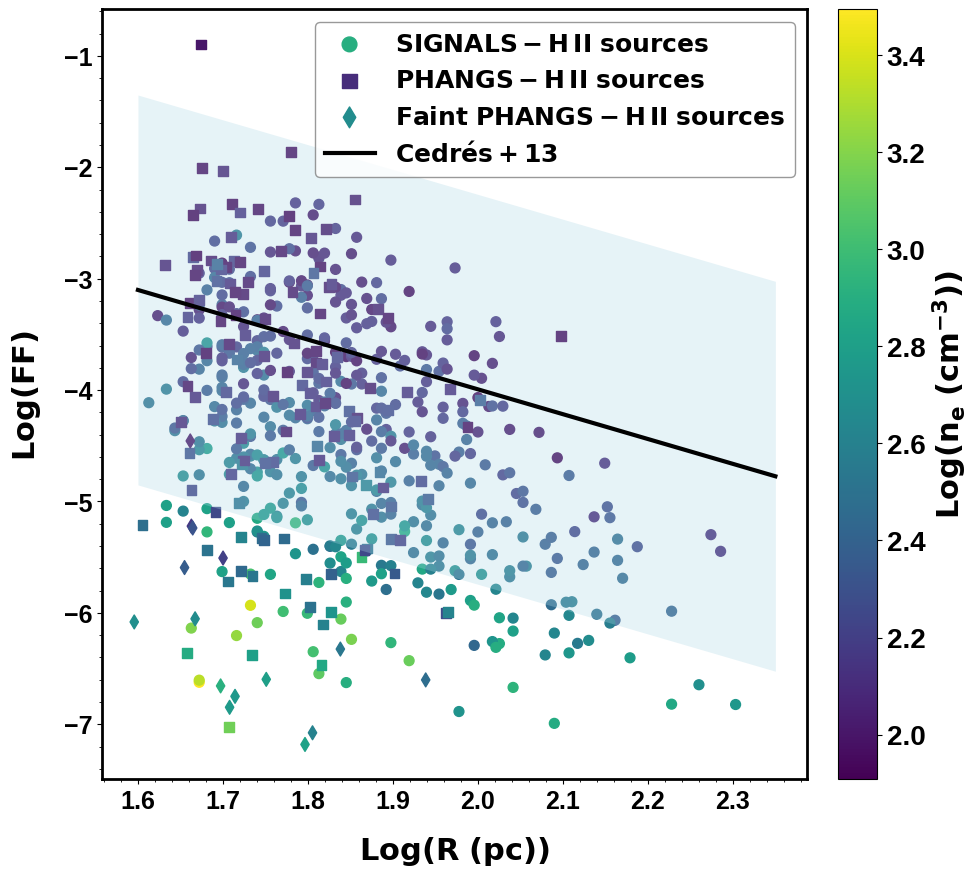}
		\includegraphics[width=0.495\columnwidth, keepaspectratio]{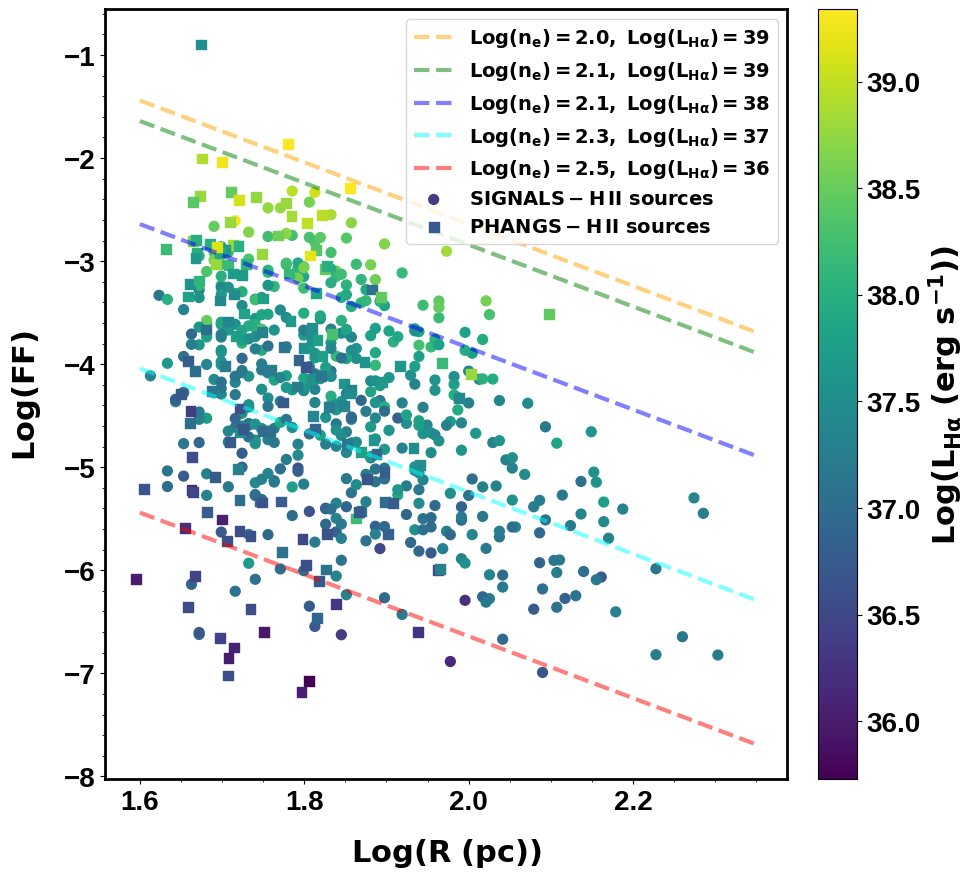}
		\caption{Log(FF) as a function of the circularized equivalent radius, R, for SIGNALS-H\,II and PHANGS-H\,II regions in NGC\,628, color-coded by the local electron density $\mathrm{log(n_e)}$ (left panel) and by the $\mathrm{log(L_{H\alpha})}$ luminosity (right panel). The trend reported by \citet{Cedres2013} for H\,II regions in NGC\,6946 is overplotted in the left panel with its $1\sigma$ dispersion for comparison at similar physical scales. Faint PHANGS-H\,II regions with $\mathrm{log(L_{H\alpha})} < 36.5~\mathrm{\ erg\,s^{-1}}$ are highlighted with diamonds and exhibit the largest deviations from the trend, indicating a H\,II regime that is preferentially powered by single massive stars. (Sect.~\ref{subsubsec:FFR}). The dashed lines in the right panel show the theoretical $\mathrm{log(FF)}-\mathrm{log(R)}$ for selected $\mathrm{log(L_{H\alpha})}$ and $\mathrm{log(n_e)}$ values, assuming the Strömgren sphere model.}
		\label{fig:FFRne}
	\end{center}
\end{figure}
The right panel of Fig.~\ref{fig:FFRne} shows the relation between $\mathrm{log(FF)}$ and radius, color-coded by $n_e$ (left panel) and $\mathrm{log(L_{H\alpha})}$ (right panel). The color gradients confirm that the vertical scatter at fixed radius is physically meaningful: at a given radius, regions with higher $n_e$ tend to exhibit lower FF, whereas more luminous H\,II regions cluster toward higher FF values. This suggests that regions with higher H$\alpha$ luminosity tend to occupy the upper envelope of the FF distribution, but this relation is modulated by variations in electron density and region size. At the same time, lower-luminosity regions display a larger scatter around the classical FF relation. 
\par
To quantify the physical contribution to the observed $\mathrm{log(FF)}$--$\mathrm{log(R^{3})}$ relation beyond the methodological effect introduced by the $\mathrm{L_{H\alpha}}$-R coupling, we consider the observed slope as $\gamma^{\mathrm{obs}} = \gamma^{\mathrm{pred}} + \gamma^{\mathrm{phys}}$, where $\gamma^{\mathrm{pred}} = (\kappa_{05} - 3)/3$ represents the true physical slope expected (Appendix~\ref{app:gamma_derivation}) and $\kappa_{05}$ is the slope of the $\mathrm{L_{H\alpha}}$-R relation obtained from the 5th-quantile regression. 
For SIGNALS-H\,II, we obtain $\gamma^{\mathrm{obs}} =$ -1.13 and $\gamma^{\mathrm{pred}} =$ -0.72, which yield a physical residual of $\gamma^{\mathrm{phys}} =$ -0.41. For PHANGS-H\,II, the observed value is $\gamma^{\mathrm{obs}} = -1.18$, with $\gamma^{\mathrm{pred}} =$ -0.45, and $\gamma^{\mathrm{phys}} =$ -0.73. The sign agreement between the two surveys is remarkable given their different instruments and spatial resolutions. This indicates that larger H\,II regions intrinsically exhibit lower FF, beyond the effect imposed by the radius definition. This behavior is consistent with an evolutionary scenario in which more extended regions correspond to later stages of expansion and dispersal of the ionized gas (Sect.~\ref{sec:discussion}).
\par
The trend reported by \citet{Cedres2013} for NGC~6946 is also shown in the left panel of Fig.~\ref{fig:FFRne} and the value reported is consistent with the more luminous H\,II regions. Low-luminosity PHANGS-H\,II regions ($\mathrm{log(L_{H\alpha})} < 36.5~\mathrm{\ erg\,s^{-1}}$, diamonds) systematically deviate below the \cite{Cedres2013} relation and the theoretical Strömgren scaling. We calculated the average luminosity of these fainter regions in solar luminosity units\footnote{$L_\odot = 3.826\times10^{33}\ \mathrm{erg\ s^{-1}}$.}, and inferred that the ionizing sources were consistent with single late O-type or early B-type stars (B0.5 or earlier, \cite{Vacca1996}). This behavior was further examined in the $\mathrm{log(FF)}$–$\mathrm{log(n_e)}$ and $\mathrm{log(FF)}$–$\mathrm{log(L_{H\alpha})}$ relations below, and departures in the three scaling relations were analyzed in the context of the classical FF framework in Sect.~\ref{sec:discussion}.
\par
The apparent deviation of low-luminosity H\,II regions from the expected scaling does not necessarily imply a breakdown of the classical FF framework. Instead, it likely reflects observational sampling effects, including the limited spatial coverage of the MUSE observations for this galaxy. In other systems observed with the same pipeline, such as NGC~5068, a significantly larger population of faint and compact H\,II regions is recovered, showing that the classical FF formulation remains valid in this luminosity regime (Aragüete Riesco et al., in prep). 
\par
More generally, the observed trend may indicate a transition between different ionizing-source regimes. Depending on the distance of the galaxy, the detected H\,II regions preferentially sample nebulae powered by clusters or stellar associations rather than those ionized by single massive stars. 
\begin{figure}[H]
	\centering
	\includegraphics[width=\columnwidth]{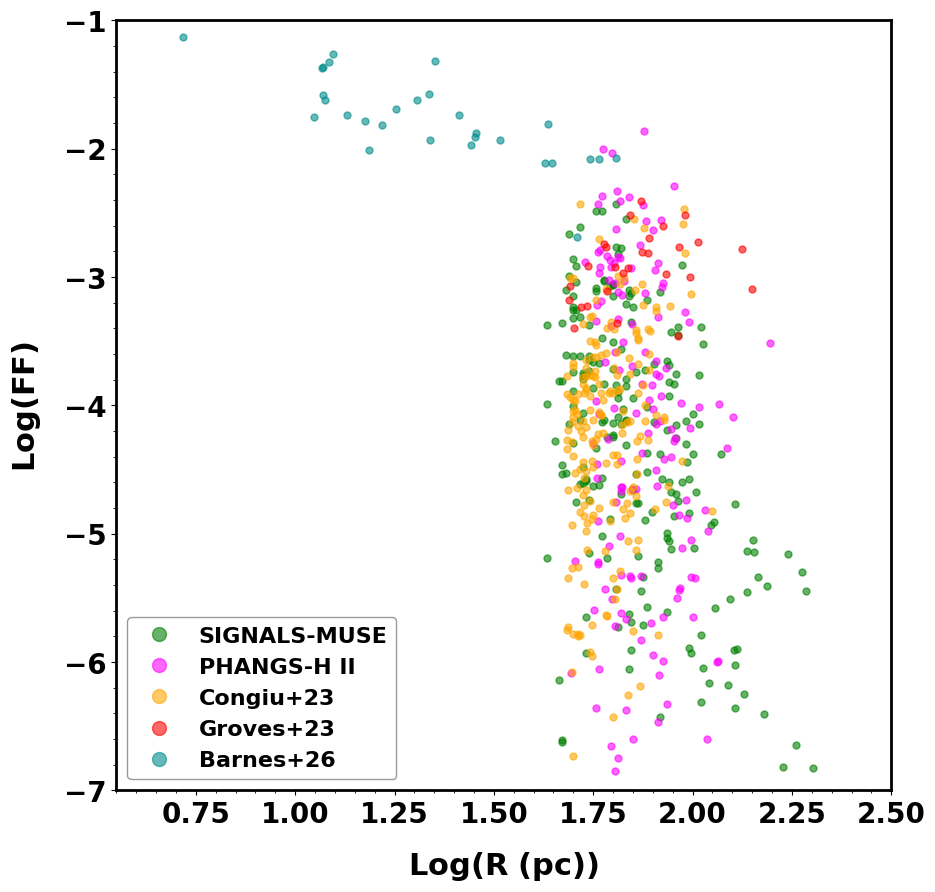}
	\caption{Distributions of $\mathrm{log(FF)}$–$\mathrm{log(R)}$ in NGC~628 among five catalogs: SIGNALS-MUSE, PHANGS-H\,II, \cite{Congiu2023, Groves2023} and \cite{Barnes2026}.}
	\label{fig:comparALLFFR}
\end{figure}
Fig.~\ref{fig:comparALLFFR} compares the $\log({\rm FF})$ distribution as a function of H\,II region radius for the five catalogs analyzed in NGC~628. The SIGNALS-H\,II, PHANGS-H\,II, and the catalogs of \citet{Congiu2023} and \citet{Groves2023} span a similar range in radius, from approximately $\log(R) \sim 1.6$ to $2.1$, and show a large dispersion in FF values. In contrast, the catalog of \citet{Barnes2026} probes smaller regions and shows comparatively higher FF values, likely reflecting resolution-driven effects. This suggests that differences among the seeing-limited catalogs are primarily methodological, while the HST-based catalog extends the analysis toward more compact and better-resolved H\,II regions.
\par
Figure~\ref{fig:FF_4plots} (see Appendix~\ref{app:kappa_L_derivation}) shows the $\log({\rm FF})$--$\log(R^{3})$ relation for the four catalogs at comparable spatial resolution. Catalogs using truncated-profile radii (SIGNALS-H\,II and PHANGS-H\,II) show a consistent negative trend with substantial scatter, whereas absolute-threshold catalogs (\citealt{Congiu2023}, \citealt{Groves2023}) show flatter relations, with small scatter in \citet{Groves2023} and no clear trend in \citet{Congiu2023}.
\par
Analytically (see Appendix~\ref{app:gamma_derivation}), the expected slope of the $\log({\rm FF})$--$\log(R^{3})$ relation is given by $\gamma = (\kappa_{05} - 3)/3$. After correcting for this methodological contribution, the residual slopes are $\gamma^{\rm phys} = -0.41$ for SIGNALS-H\,II and $\gamma^{\rm phys} = -0.73$ for PHANGS-H\,II, both indicating a decreasing trend of FF with region size. The difference between the two profile-truncated catalogs reflects the larger fraction of compact low-luminosity regions in PHANGS-H\,II, indicating that larger H\,II regions in NGC~628 tend to have intrinsically lower FF beyond what is imposed by the radius definition in all catalogs.
\par
To investigate the apparent discrepancy in \citet{Congiu2023} ($\gamma^{\rm phys} = +0.26$), we repeat the analysis in bins of electron density ($n_e < 100$, $100 \le n_e < 140$, and $n_e \ge 140$ cm$^{-3}$). In all bins, \citet{Congiu2023} yields $\gamma^{\rm phys} < 0$ ($-0.33$, $-0.39$, $0.05$), consistent with SIGNALS-H\,II in the lower-density regimes. The positive value in the highest-density bin is driven by the limited dynamic range at large $n_e$ (Fig.~\ref{fig:comparne}).
PHANGS-H\,II shows larger bin-to-bin variations in $\gamma^{\rm phys}$ ($-0.13$, $-0.78$, $+0.40$), whereas SIGNALS-H\,II remains more stable ($-0.59$, $-0.57$, $-0.75$). This likely reflects differences in the luminosity range probed by each survey, with PHANGS extending toward lower-luminosity H\,II regions, while SIGNALS predominantly samples brighter regions. These trends are consistent with the behaviour discussed in Sect.~\ref{subsubsec:FFR} and may indicate a gradual change in the dominant ionizing sources, from single massive stars to those powered by stellar clusters or associations, across the luminosity range sampled in each catalog, rather than a single universal scaling.
\par
For \citet{Groves2023}, comparatively narrow range of local electron densities effectively restricts the sample to a single $n_e$ regime, resulting in a $\gamma^{\rm phys} = -0.09$, representative of the luminous regime probed by this catalog. This behavior likely reflects the specific characteristics of the dataset, including the lack of an explicit DIG subtraction and the resulting impact on the flux distribution.
\subsubsection{Relation with the local electron density}\label{subsubsec:FFne}
We observe a statistically significant negative trend between FF and $n_e$ in SIGNALS-H\,II and PHANGS-H\,II catalogs, reflecting the combined influence of density, luminosity, and volume variations on the scatter. Deviations from the exact theoretical exponents are consistent with the impact of density inhomogeneities and the coupling between H$\alpha$ and the volume. 
\par
The fitted slope of the $\mathrm{log(FF)}$–$\mathrm{log(n_e)}$ relation differs between the two surveys. The SIGNALS-H\,II sample yields a slope of -2.65 $\pm 0.08$. In contrast, the PHANGS-H\,II sample exhibits a steeper slope of $-3.61 \pm 0.23$. However, when the PHANGS-H\,II regions are restricted to higher luminosities ($\mathrm{log(L_{H\alpha}) > 37.5 \ erg~s^{-1}}$), the slope decreases to $-2.7$, approaching the value measured for the SIGNALS-H\,II sample. The steeper slope obtained therefore still reflects the influence of faint, compact regions at the low-luminosity end.
The moderate $R^{2}_{fit}$ values of $\sim$ 0.60, indicate that the $n_e$ is the parameter most closely associated with the variance in FF. Within the high-luminosity regime, the classical FF formulation provides a consistent description of the data.
\par
Fig.~\ref{fig:FFne} shows $\mathrm{log(FF)}$ versus $\mathrm{log(n_e)}$ for the two surveys. The figure illustrates the interplay between H$\alpha$ luminosity, local electron density, and region size, and overplots theoretical relations from the Strömgren sphere model for different combinations of the independent variables.
\begin{figure}[H]
	\centering
	\includegraphics[width=\columnwidth]{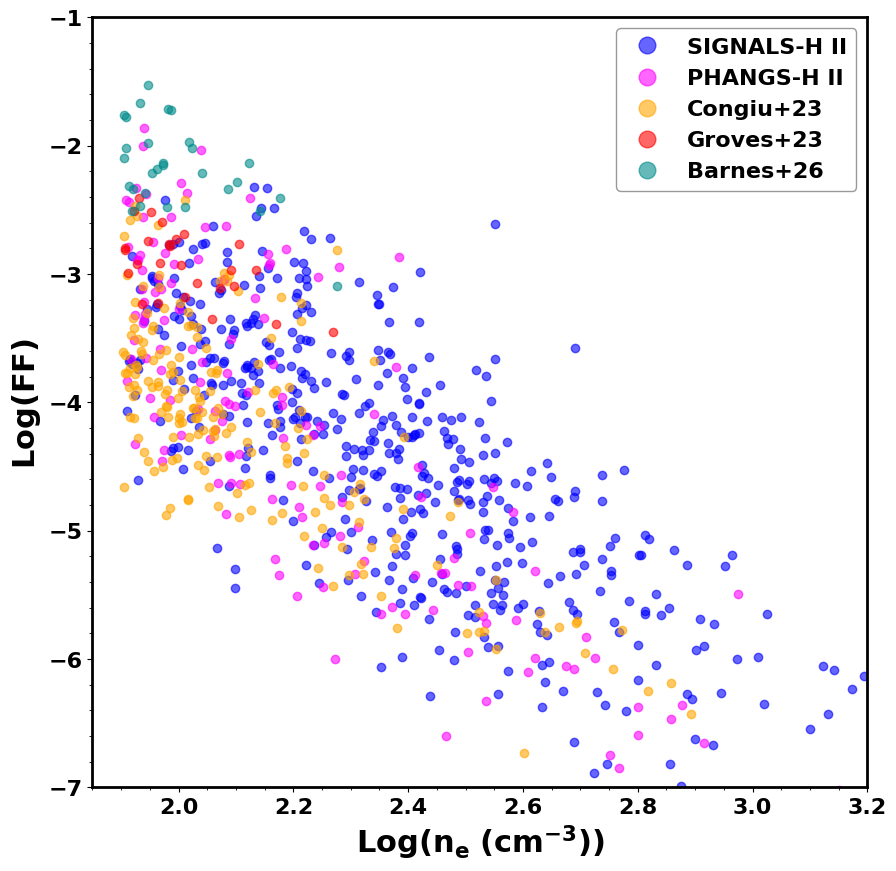}
	\caption{Distributions of $\mathrm{log(FF)}$ vs. $\mathrm{log(n_e)}$ in NGC~628 for five catalogs: SIGNALS-H\,II, PHANGS-H\,II, and the catalogs of \cite{Congiu2023, Groves2023} and \cite{Barnes2026}.}
	\label{fig:comparALLFFne}
\end{figure}
Fig.~\ref{fig:comparALLFFne} shows the $\mathrm{log(FF)}$ distribution as a function of the electron density for all the catalogs analyzed in NGC~628. All catalogs display the previously described anti-correlation between $\mathrm{log(FF)}$ and $\mathrm{log(n_e)}$. The SIGNALS-H\,II catalog spans the widest range of $n_e$ and FF values. The PHANGS-H\,II and \cite{Congiu2023} samples follow a similar trend but over a slightly narrower FF parameter range. The catalog from \cite{Groves2023} mainly populates the high-FF, low-density regime due to its limited range of local electron densities, while \cite{Barnes2026}, with local electron densities derived by \cite{Groves2023}, occupies a similar low-density regime but at systematically higher FF values. Overall, the $\mathrm{log(FF)}$–$\mathrm{log(n_e)}$ relation remains consistent among the different catalogs.
\subsubsection{Relation with the luminosity}\label{subsubsec:FFLHa}
\begin{figure}[htbp]
	\begin{center}
		\includegraphics[width=0.515\columnwidth, keepaspectratio]{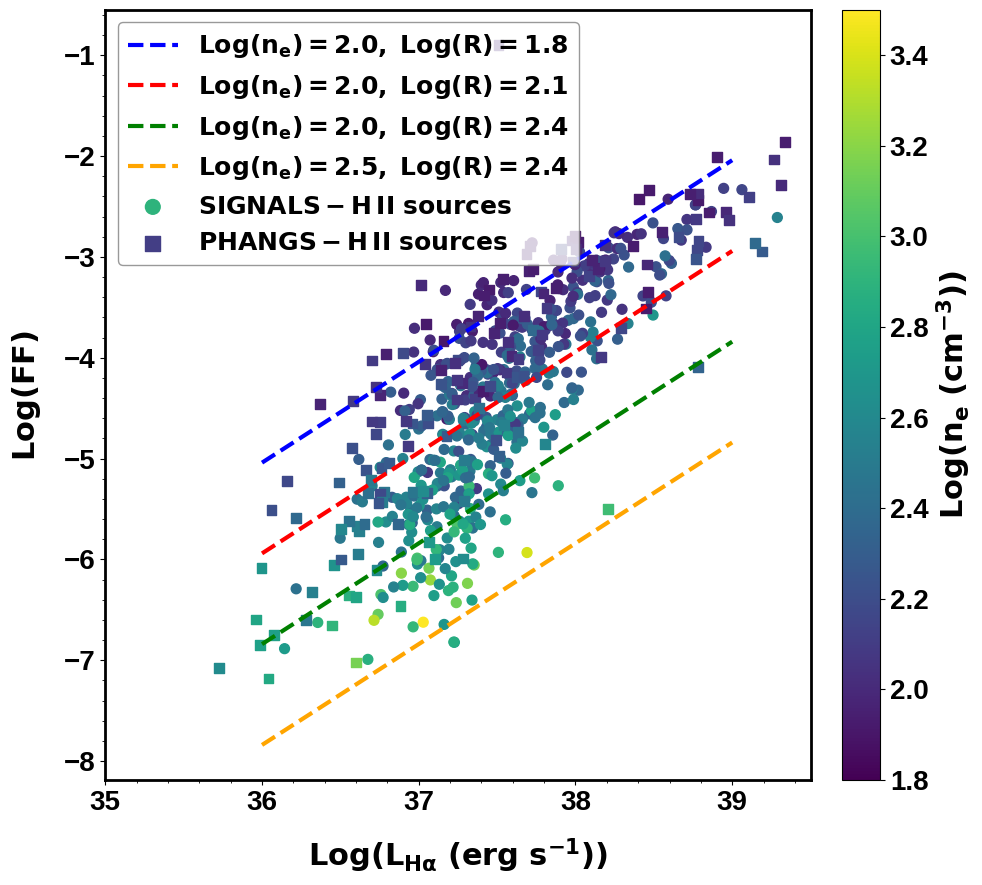}\hfill
		\includegraphics[width=0.485\columnwidth, keepaspectratio]{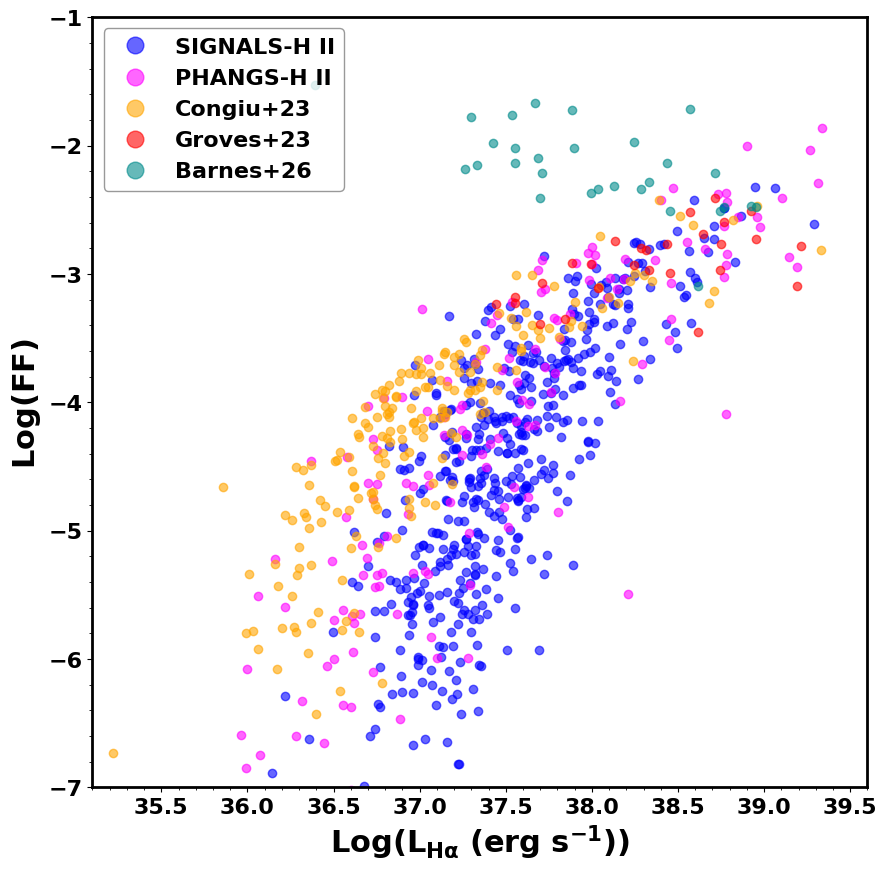}
		\caption{$\mathrm{Log(FF)}$ against the H\,II region H$\alpha$ luminosity, $\mathrm{log(L_{H\alpha})}$, color-coded by the local electron density (left panel). We differentiate the regions by survey: SIGNALS-H\,II (dots), PHANGS-H\,II (squares). The dashed lines show the theoretical $\mathrm{log(FF)}$-$\mathrm{log(L_{H\alpha})}$ relation for selected Log $n_e$ and $\mathrm{log(R)}$ values, assuming the Str\"omgren sphere model. The distributions of $\mathrm{log(FF)}$ vs. $\mathrm{log(L_{H\alpha})}$ in NGC~628 for five catalogs: SIGNALS-H\,II, PHANGS-H\,II, and the catalogs of \cite{Congiu2023, Groves2023} and \cite{Barnes2026} are shown in the right panel.}
		\label{fig:FFLHa}
	\end{center}
\end{figure}

H$\alpha$ luminosity presents a strong and statistically significant positive correlation with the FF (left panel of Fig.~\ref{fig:FFLHa}). Here, we use the FF-H$\alpha$ luminosity relation as a diagnostic tool to assess the regime in which the classical FF formulation provides a physically meaningful description of H\,II regions. The linear fits yield moderate-to-high $R^2_{fit}$ values (Table~\ref{tab:FF_fit}), yet still leaves $\sim 40\%$ where the combination of physical conditions contributes to the remaining scatter in the data. The color-coded representation for SIGNALS-H\,II and PHANGS-H\,II samples illustrates that similar luminosities span a wide range of local electron densities, highlighting the coupled role of $n_e$, size, and H$\alpha$ luminosity in shaping the observed FF values.
In the high-luminosity regime, for a given electron density, luminous giant and supergiant H\,II regions, which are ionized by massive clusters and generally more massive and limited in volume, tend to have higher FF. At lower luminosities, approaching the regime typical of single massive stars, both surveys show increased scatter and systematic departures from the dominant trend. In SIGNALS-H\,II, enhanced scatter below $\mathrm{log(L_{H\alpha}) < 37.5 \ erg~s^{-1}}$ may be influenced by aspects of the data reduction adopted in this survey. In PHANGS-H\,II, the regions with $\mathrm{log(L_{H\alpha}) < 36.5 \ erg~s^{-1}}$ reflect the larger diversity of physical conditions, as discussed in Sect.~\ref{subsubsec:FFR} and Sect.~\ref{subsubsec:FFne}, this defines a coherent low-luminosity regime that is analyzed in Sect.~\ref{sec:discussion}.
\par
The right panel of Fig.~\ref{fig:FFLHa} extends this analysis to all five catalogs. Despite differences in the boundary definition and spatial resolution, the positive FF-$\mathrm{L_{H\alpha}}$ trend is consistent in catalogs within the scatter, with \cite{Barnes2026} occupying the high-$\mathrm{L_{H\alpha}}$, high-FF end as expected from its high HST-resolution, which resolves compact, high-surface-brightness regions. The systematic offset in median FF between catalogs (Fig.~\ref{fig:comparFF}) is substantially reduced when comparing regions at fixed luminosity above $\mathrm{log(L_{H\alpha}) > 37.5 \ erg~s^{-1}}$, confirming that the FF-$L$ relation is more robust against methodological differences than the FF distributions considered alone. After subtracting the methodological contribution, $\eta^{\rm pred} = (\kappa_{05}-3)/\kappa_{05}$ (Sect.~\ref{app:kappa_L_derivation}, the positive correlation between FF and $\mathrm{L_{H\alpha}}$ is largely physical in all catalogs with available $\eta^{\rm obs}$ (Table~\ref{tab:slopes_allcatalogs}). The physical residuals remain large in SIGNALS-H\,II and PHANGS-H\,II ($\eta^{\rm phys} =$ +4.21 and $+2.06$), indicating that more luminous H\,II regions intrinsically have higher FF. $\eta^{\rm phys}$ is lower in catalogs with restricted or electron density ranges \citep{Groves2023, Congiu2023}. 
\subsubsection{The FF-L/R³ relation: a diagnostic of internal gas structure}\label{subsubsec:FFLR3}
\begin{figure}[H]
	\begin{center}
		\includegraphics[width=\columnwidth, keepaspectratio]{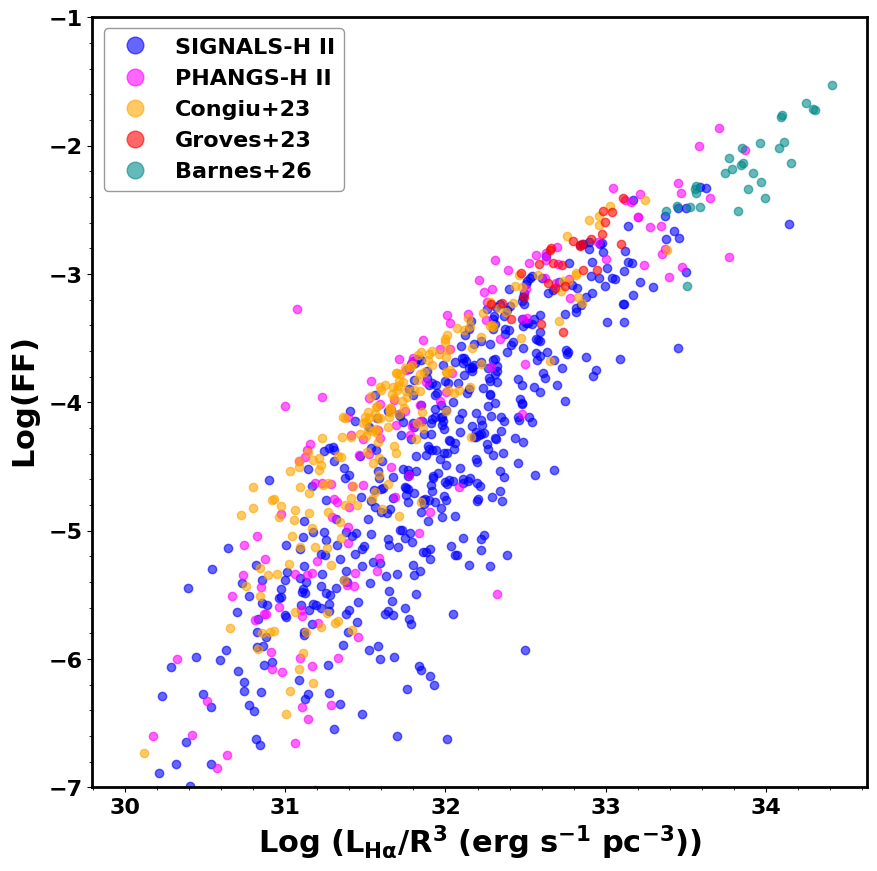}	
		\caption{$\mathrm{Log(FF)}$ as a function of the volumetric H$\alpha$ luminosity density, $\mathrm{log(L_{H\alpha}/R^{3})}$, for all five catalogs in NGC~628. The scatter around the relation reflects the variation in $n_e^2$ in the population of H\,II regions.}
		\label{fig:ALLcatalogsFFLR3}
	\end{center}
\end{figure}
Fig.~\ref{fig:ALLcatalogsFFLR3} shows $\log(\mathrm{FF})$ as a function of the volumetric luminosity density $\log(L_{\mathrm{H}\alpha}/R^{3})$ for all five catalogs. Despite differences in radius definitions and spatial resolution, spanning nearly two orders of magnitude in characteristic size between \citet{Barnes2026} and the MUSE- and SIGNALS-based catalogs, all datasets follow a common linear relation with similar slope.
This behavior follows directly from the definition of the FF, which yields a unit slope in the $\log({\rm FF})$--$\log(L_{\mathrm{H}\alpha}/R^{3})$ plane under the assumption of fixed $n_e$, that is, neglecting additional coupling terms such as $B$ (see Appendix~\ref{app:LR3_derivation}). In this plot, the departure from the definitional slope is a physical result. 
\par
This behavior demonstrates that $\mathrm{L_{H\alpha}/R^3}$, the volumetric luminosity density of the ionized gas, minimizes the methodological variance introduced by different radius definition criteria, and thus the FF is physically driven by the internal gas structure rather than by the segmentation and definition of boundary method. This makes the volumetric luminosity density a useful quantity to explore how the internal density structure of H\,II regions varies in different galactic environments. 
\par
The primary source of scatter in this representation is the intrinsic variation in electron density in the H\,II region population. As shown in Eq.~\ref{ec:D.4}, the dispersion in $\log(\mathrm{FF})$ at fixed $L_{\mathrm{H}\alpha}/R^{3}$ directly traces variations in $n_e$. It is a direct measure of local electron density heterogeneity. Secondary contributions may include variations in metallicity, electron temperature or nonuniform distribution of clumps, which affect the H$\alpha$ emissivity coefficient, and departures from the assumption of a uniform clump distribution underlying the classical FF formulation.
\par
Having established the internal consistency and the limitations of the classical FF formulation, we next explore how the FF relates to additional physical tracers. In the following sections, we examine the connection between FF and the $H\alpha$ equivalent width, and the dust properties of H\,II regions, to assess whether FF encodes information about evolutionary stage and interaction with the surrounding ISM.
\subsection{Relation with EW(H$\alpha$)}\label{subsec:FFEWHa}
\begin{figure}[H]
	\begin{center}
		\includegraphics[width=\columnwidth]{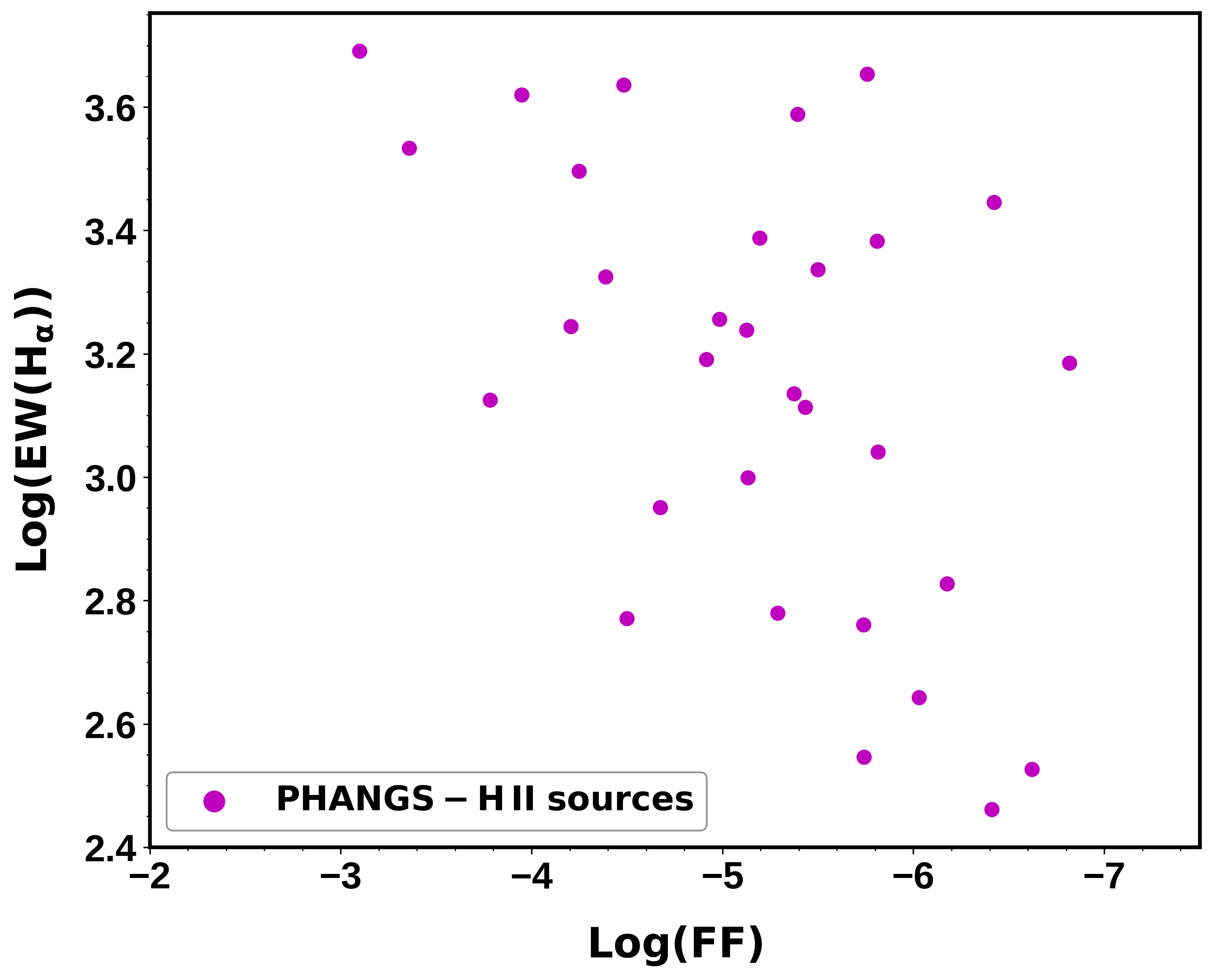}
		\caption{$\mathrm{Log(FF)}$ of H\,II regions vs. $\mathrm{EW(H\alpha)}$ equivalent width. The age estimated for the PHANGS-H\,II star forming regions ranges from $< 3$ to  6.3 Myr compared with the Starbust99 model \citep{Leitherer1999}, as described in \cite{Ramya2007} assuming instantaneous-burst with a Salpeter IMF in the mass range 1 - 10 $M\textsubscript{\(\odot\)}$. The SIGNALS catalog does not include the $\mathrm{EW(H\alpha)}$ parameter.}
		\label{fig:FFEWHa}
	\end{center}
\end{figure}
Determining the age of H\,II regions plays a key role in understanding their evolutionary stage. H$\alpha$ emission traces recent star formation, such that high H$\alpha$ flux relative to the stellar continuum indicates a recent episode. The equivalent width of H$\alpha$, or $\mathrm{EW(H\alpha)}$, represents the ratio of H$\alpha$ luminosity to the underlying stellar continuum of the young stellar cluster, therefore providing a widely used observational proxy for the age of the ionizing stellar population. 
\par
For the PHANGS-H\,II sample, $\mathrm{EW(H\alpha)}$ values were measured and corrected for the contribution of the underlying galactic stellar population associated with each region. Regions for which the measured stellar continuum was consistent with zero within the uncertainties were excluded from the analysis. This requirement reduces the available sample to 18~H\,II regions. This result should therefore be regarded as exploratory, given the small number of regions available. Future studies will extend this analysis to other nearby galaxies.
\par
Fig.~\ref{fig:FFEWHa} presents the distribution of $\mathrm{log(FF)}$ as a function of $\mathrm{log(EW(H\alpha))}$ for the PHANGS-H\,II regions. When compared with Starburst99 instantaneous-burst models \citep{Leitherer1999}, the inferred ages for the sample span from $< 3$ to 6.3 Myr. We find a positive correlation between the $\log(\mathrm{FF})$--$\log(\mathrm{EW}(\mathrm{H}\alpha))$ relation in PHANGS H\,II regions ($\rho = 0.56$, p-value $< 0.001$), suggesting that high-FF regions tend to show high $\mathrm{EW}(\mathrm{H}\alpha)$, characteristic of younger populations.
\par
However, we note that the two quantities are not fully independent, as they share an implicit dependence on the H$\alpha$ luminosity, and therefore part of the observed correlation may be driven by this underlying covariance. Nonetheless, other independent tracers of the H\,II region evolutionary state have been proposed in the literature (e.g., excitation, kinematics) that could be used in future work.
\par
We identified three broad regimes in this parameter space: (i) high $\mathrm{EW(H\alpha)}$ and high FF ($\mathrm{log(EW(H\alpha)) \gtrsim 3; \ Log(FF) \gtrsim -3}$), corresponding to very young, compact H\,II regions with high filling factors; (ii) intermediate $\mathrm{log(EW(H\alpha))}$ and moderate FF ($\mathrm{log(EW(H\alpha)) \lesssim 3 \ to \ 2; \ Log(FF) \sim -3 \ to \ -5}$), associated with more evolved regions where the ionizing flux remains significant but the nebular gas becomes more spatially extended; and (iii) low $\mathrm{log(EW(H\alpha))}$ and low FF ($\mathrm{log(EW(H\alpha)) \lesssim 1.5; \ Log(FF) \lesssim -5}$), characteristic of older or more evolved H\,II regions, in which nebular emission decreases relative to the stellar continuum and the ionized gas occupies a smaller fraction of the volume. We note that this regime is distinct from the low-luminosity, compact regions discussed in Sect.~\ref{subsubsec:FFR}, which exhibit low FF values due to the nature of their ionizing sources. Extending this analysis into the single massive star regime would require larger samples of faint, compact H\,II regions observed at higher spatial resolution.
\par
However, the evolutionary sequence of giant and supergiant H\,II regions does not necessarily follow a simple linear sequence, as their ionization could be driven by stellar associations or clusters comprising multiple stars at different evolutionary stages, which can introduce additional scatter in the FF-$\mathrm{EW(H\alpha)}$ relation. Further work is underway to study this sequence for a larger sample of galaxies.
\subsection{Relation with the PAH-to-dust ratio}\label{subsec:PAH}
Using the enhanced angular resolution provided by JWST/MIRI, reaching scales of a few tens of parsecs, we are able to investigate the presence of small-scale correlations between the FF and the local dusty environment of H\,II regions.
\par
We performed circular aperture photometry for H\,II regions of SIGNALS and elliptical aperture for H\,II regions of PHANGS on each MIRI image of NGC~628 (F770W, F1130W, and F2100W), using the Photutils Python package, adopting the calibration coefficients listed in Table~4 of \cite{Lee2023} and the conversion factors PHOTMJSR and PHOTUJA2 provided in the FITS files. Photometric measurements were carried out for all locations corresponding to the sample of PHANGS-H\,II and SIGNALS-H\,II regions within the MIRI FoV. For each region, the flux was computed by summing the contribution of all pixels (in $\mu$$Jy/arcsec^{2}$) enclosed within the projected area of each H\,II region. 
\par
We did not regrid the infrared images to the astrometric grid of the PHANGS-MUSE NGC~628 data in order to avoid introducing potential additional errors from interpolation effects. Instead, we rescaled and reprojected the apertures to match the geometrical properties (size, orientation, and position) of each H\,II region as defined in the corresponding catalogs. As a consequence, we adopted a variable aperture radius, matched to the equivalent radius of each H\,II region, rescaling and reprojecting it individually to match the resolution and orientation of each image. In all cases, the adopted aperture size exceeds the instrumental PSF.
\par
We focused on a subset of 158 H\,II regions from the SIGNALS-H\,II sample (approximately 33\% of the sample) that are covered by the JWST footprint and exhibit nonzero PAH emission. This subsample is hereafter referred to as SIGNALS-JWST. For the PHANGS-H\,II catalog, 105 regions (about 71\% of the 147 selected regions) fall within the JWST coverage and are hereafter labeled PHANGS-JWST. The distribution of radii, $H\alpha$ luminosities, and FF of the regions within the JWST mask closely mirrors those obtained for the MUSE samples, as discussed in Section \ref{subsec:FFrelations} and we illustrate in Fig.~\ref{fig:LHaRFFJWST}.
\par
Following \cite{Chastenet2023a}, we quantified the relative contribution of PAHs using the combined emission in the 7.7 $\mu$m and 11.3 $\mu$m bands, normalized by the thermal dust emission traced by the 21 $\mu$m band:
\begin{equation} \tag{15} \label{ec:Rpah}
	R_{\mathrm{PAH}} = \frac{F_{F770W} + F_{F1130W}}{F_{F2100W}}
\end{equation}
We analyzed each catalog independently to investigate the correlation between $R_{PAH}$ and $\mathrm{log(L_{H\alpha}/R^{3})}$. This will minimize potential biases from differences in the H\,II region detection techniques and/or emission line flux measurement methods.
\par
\begin{figure}
	\begin{center}
		\centering    
		\includegraphics[width=\columnwidth]{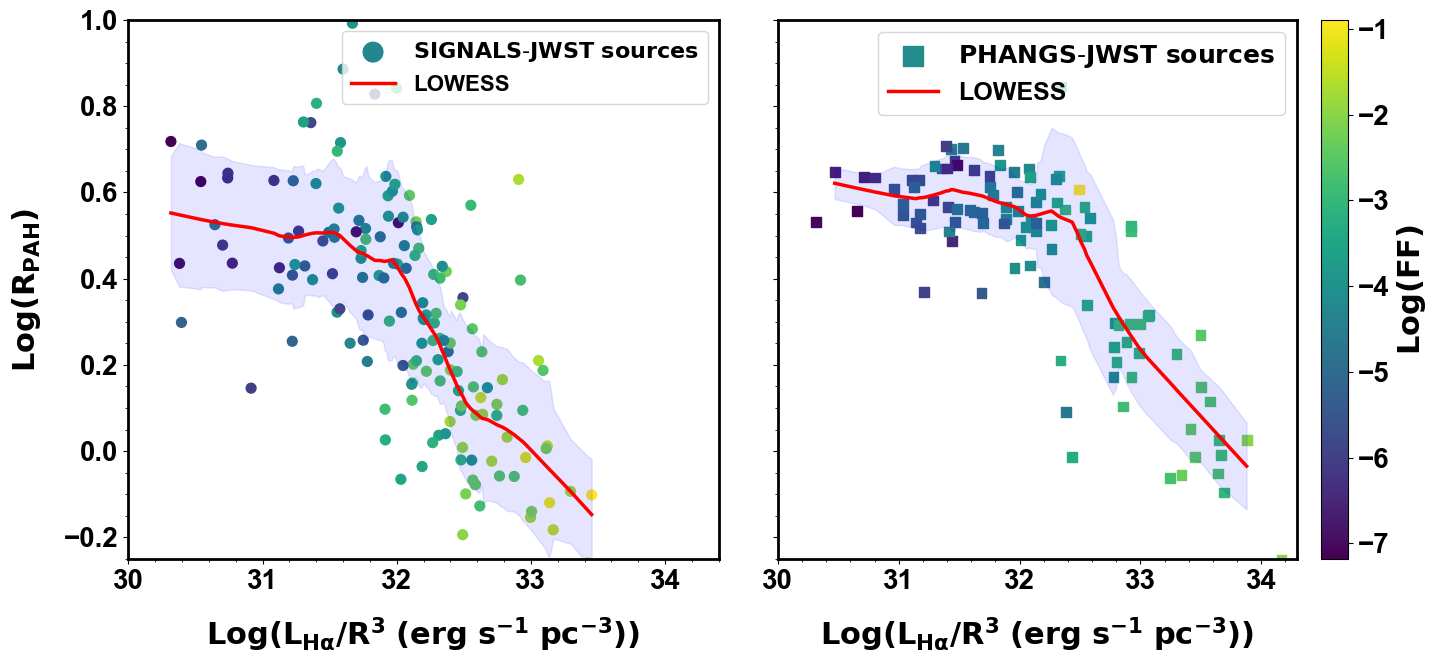}
		
		\caption{Correlation of the PAH abundance traced with $R_{PAH}$ with $\mathrm{log(L_{H\alpha}/R^{3})}$ in SIGNALS-JWST (left panel) and PHANGS-JWST H\,II regions (right panel). The red curve shows a LOWESS smoothing of the data, highlighting the global trend. }
		\label{fig:RpahLR3}
	\end{center}
\end{figure}

In Fig.~\ref{fig:RpahLR3}, the PHANGS-JWST and SIGNALS-JWST samples suggest the presence of two distinct regimes in the relation between the $\mathrm{Log(R_{PAH})}$--$\mathrm{log(L_{H\alpha}/R^{3})}$, with a noticeable change in the slope. To characterize the global behavior of the relation between $\mathrm{log(R_{PAH})}$ and the $\mathrm{log(L_{H\alpha}/R^{3})}$, we applied a locally weighted scatterplot smoothing (LOWESS) regression to the data ($f=0.25$). LOWESS is a nonparametric method that estimates the underlying trend by performing local weighted linear fits over neighboring subsets of the data, without assuming a specific functional form for the relation. 
\par
The resulting LOWESS-smoothed relation reveals a clear change in slope around $\mathrm{log(L_{H\alpha}/R^{3})}$ $\sim 32 \ erg \ s^{-1} \ pc^{-3} $ in both samples. To quantify the range of this transition, we performed bootstrap resampling of the datasets with 1000 iterations. We obtained distributions of transition values with mean and standard deviation of $\mathrm{log(L_{H\alpha}/R^{3})} =  $31.64 $\pm$ 0.16 for the SIGNALS–JWST sample and $\mathrm{log(L_{H\alpha}/R^{3})} = 32.03 \pm 0.20$ for the PHANGS–JWST sample. The global shape of the two samples appears to be consistent within the uncertainties. The comparatively large dispersions indicate that the transition is not a sharp breakpoint but rather a gradual change over an extended range in FF. Given the small offset and comparable dispersions, we defined a common transition regime as 31.5 $ \ \leq \ \mathrm{log(L_{H\alpha}/R^{3})} \ \leq 32.2$, encompassing the bulk of both distributions. This choice enables a consistent comparison while acknowledging that the transition does not occur at a single universal value.
The H\,II regions within the transition emissivity range span hydrogen-ionizing photon rates of $\mathrm{Q_H \sim 10^{49}--10^{50} \ s^{-1}}$, with median values of $\mathrm{Q_H \sim 2.58 \times 10^{49} \ s^{-1}}$ (SIGNALS-JWST) and $\mathrm{Q_H \sim 1.56 \times 10^{49} \ s^{-1}}$ (PHANGS-JWST), calculated using Case~B relation of \cite{Osterbrock2006} ($\mathrm{Q_H \ = \ 7.35 \times 10^{11} LH\alpha)}$. These values are consistent with the threshold reported by \citet{Egorov2023b}, above which $\mathrm{Q_H \sim 10^{50} \ s^{-1}}$ PAH destruction becomes efficient. The transition identified in the $\mathrm{log(R_{PAH})}$-$\mathrm{log(L_{H\alpha}/R^{3})}$ relation thus corresponds to the regime where the ionizing output of H\,II regions approaches this critical rate, marking the onset of efficient PAH photodissociation rather than its fully established phase. This interpretation is further supported by the upper envelopes of both samples, which reach $\mathrm{Q_H \sim 10^{50} \ s^{-1}}$ at the high-volumetric H$\alpha$ luminosity density end of the transition range. Future studies applying this analysis to additional nearby galaxies will provide a more precise determination of the characteristic volumetric H$\alpha$ luminosity density at which PAH depletion begins.
\par
The FF values of H\,II regions within the transition emissivity range are consistent between the two catalogs. For SIGNALS-JWST, the defined transition interval contains 51 regions with a median FF of $\mathrm{log(FF)}$ = -4.58 and an interquartile range of [-5.01, -4.09]. 
For PHANGS-JWST, the transition interval contains 27 regions with a median of $\mathrm{log(FF)}$ = -4.28$~[-4.80, -4.02]$, with a median offset of  0.22~dex, well within the standard deviation of $\sim$ 0.59 and 0.63~dex respectively present in both samples. Despite the difference in the transitional range in the two catalogs, the FF distributions are statistically consistent, indicating that the physical characterization of the transition is robust.
\par
For simplicity, we adopt a first-order approximation of the transition point at $\mathrm{log(FF)} \sim -4.4$. We used Spearman’s correlation coefficient to test whether $R_{PAH}$ and $\mathrm{log(FF)}$ are monotonically associated, regardless of the functional form of the relation. For PHANGS-JWST, 42 regions with $\mathrm{log(FF)}$ $<$ -4.4 show no significant correlation ($\rho$ = 0.12, p-value = 0.77), with a characteristic value of $\mathrm{log(R_{PAH})}$ = 0.60 $\pm$ 0.05. In contrast, 63 regions with $\mathrm{log(FF)}$ $>$ -4.4 exhibit a strong statistically significant negative correlation ($\rho$ = -0.69, p-value = $\text{< 0.001}$), indicating that the relation strengthens above this threshold. 
A similar behavior is found for the SIGNALS-JWST sample. 70 regions with $\mathrm{log(FF)}$ $<$ -4.4 do not display a significant correlation ($\rho$ = -0.18, p-value = 0.13) with $\mathrm{log(R_{PAH})}$ = 0.47 $\pm$ 0.23, whereas 88 regions with $\mathrm{log(FF)}\ > -4.4$ show a significant negative correlation ($\rho$ = -0.54, p-value = $\text{< 0.001}$).
\par
Overall, these results indicate that in both samples, the strength of the $\mathrm{log(R_{PAH})}$-$\mathrm{log(FF)}$ correlation increases for regions above $\mathrm{log(FF)}$ value of -4.4, revealing a clear transitional change in behavior around this threshold. The subset of high-luminosity H\,II regions with larger FF tends to exhibit lower PAHs ratios, whereas fainter regions show higher PAH abundances. This behavior could be understood as a consequence of the weaker ionizing radiation field, which is insufficient to efficiently produce photodissociation and destruction of PAHs. However, we should bear in mind that part of this trend may also be driven by observational effects in marginally resolved regions.
\par
This result is fully consistent with the values reported by \cite{Egorov2023b}. H\,II regions with $f_{ion}$ < 0.95 are associated with lower $\mathrm{log(FF)}$ and show no evidence of PAH destruction.
\begin{figure}
	\begin{center}
		\centering
		\includegraphics[width=\columnwidth]{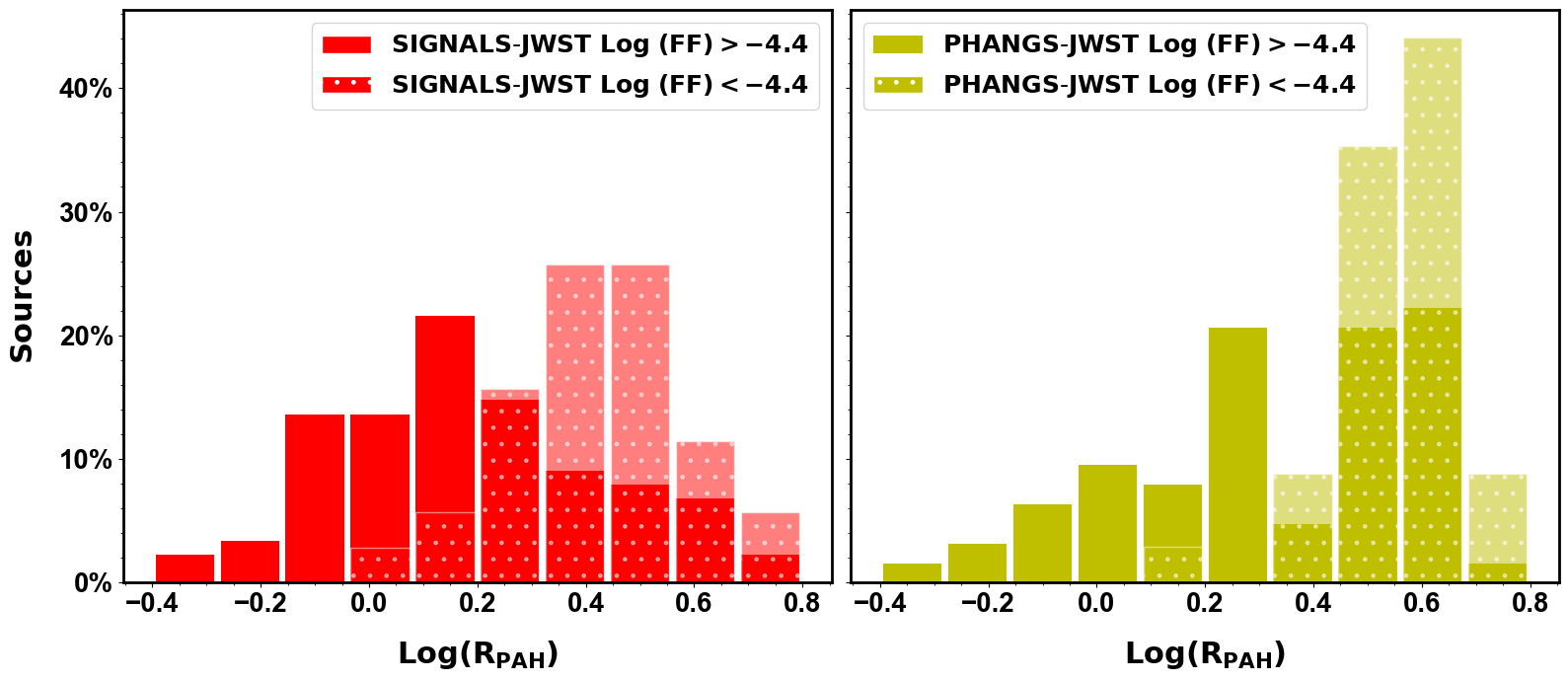} 	
		\caption{Histogram showing the number of regions with the PAH abundance traced with $\mathrm{R_{PAH}}$ as a function of the FF $\lessgtr$ $-$ 4.4 in SIGNALS-JWST (left panel) and PHANGS-JWST (right panel) H\,II regions contained in the FoV of the JWST. The vertical axis represents the total number of regions in percentage.}  
		\label{fig:histRpahLR3}
	\end{center}
\end{figure}
For the PHANGS-JWST and SIGNALS-JWST samples $R_{PAH}$ distributions of Fig.~\ref{fig:histRpahLR3}, we test the equality of variances using Levene’s test and rejected the null hypothesis of equal variances in both samples, (for PHANGS-JWST, p-value = 0.001 and for SIGNALS-JWST, p-value 0.71). However, a Shapiro–Wilk test indicated at least one subsamples deviates significantly from normality. Therefore, we used the nonparametric Mann–Whitney U test to assess differences between the $\mathrm{R_{PAH}}$ distributions in the two FF regimes in the two subsamples.
\par
In the PHANGS–JWST sample, a Mann–Whitney U test reveals a statistically significant difference between the two FF regimes (U = 510, p-value $< 0.001$). The effect size is large (rank-biserial correlation $r_{rb} = -0.52$), with a "Common Language Effect Size" (CLES) $\sim$ 0.76, indicating that values in the low-FF sample exceed those in the high-FF sample in approximately 76\% of pairwise comparisons. A consistent result is obtained for the SIGNALS–JWST sample, where the Mann–Whitney U test also indicates a highly significant difference between the two populations (U = 1052, p-value $ < 0.001$). The effect size remains large ($r_{rb} \approx$ - 0.59), with an effect size of CLES of $\sim$ 0.79, confirming that regions in the low-FF regime systematically exhibit higher PAH ratios and that the effect does not depend on the specific catalog.
\par
\begin{figure*}[htbp]
	\begin{center}
		\includegraphics[width=\linewidth]{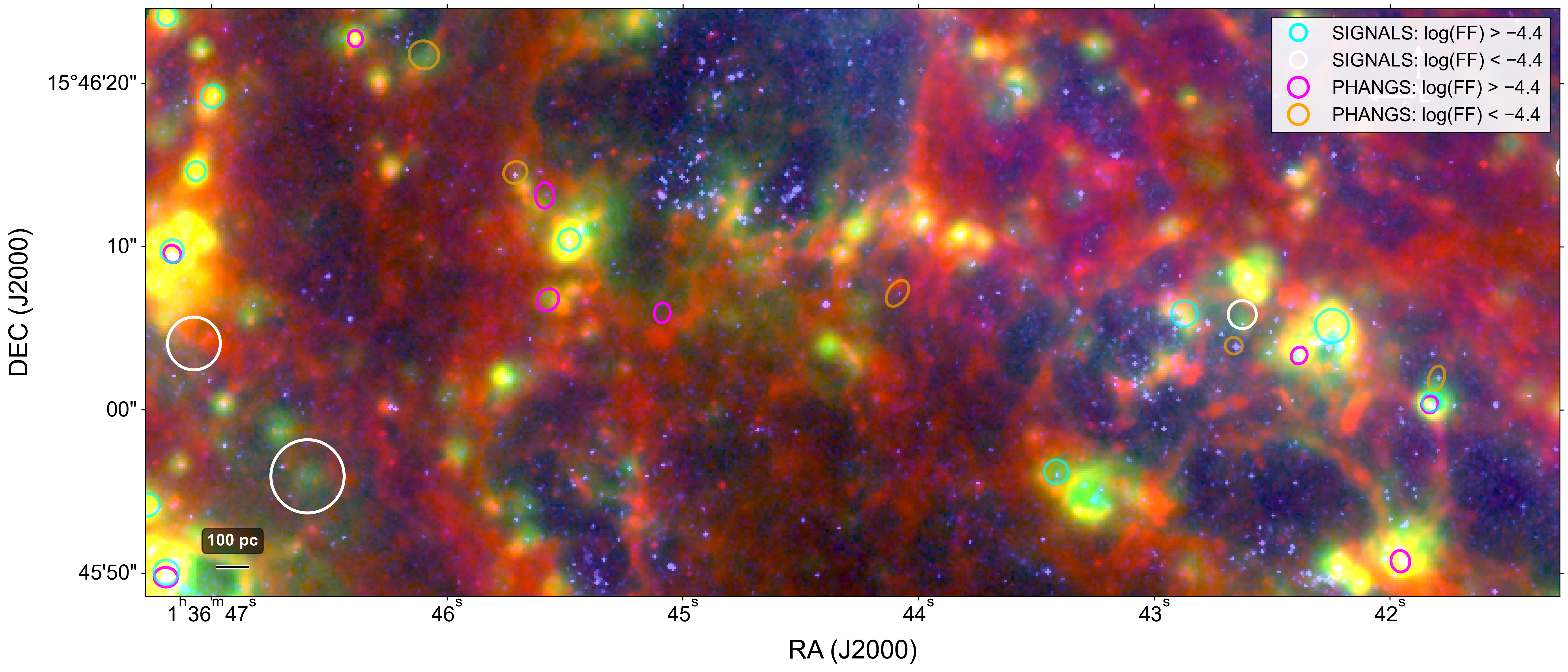}
		\includegraphics[width=\linewidth]{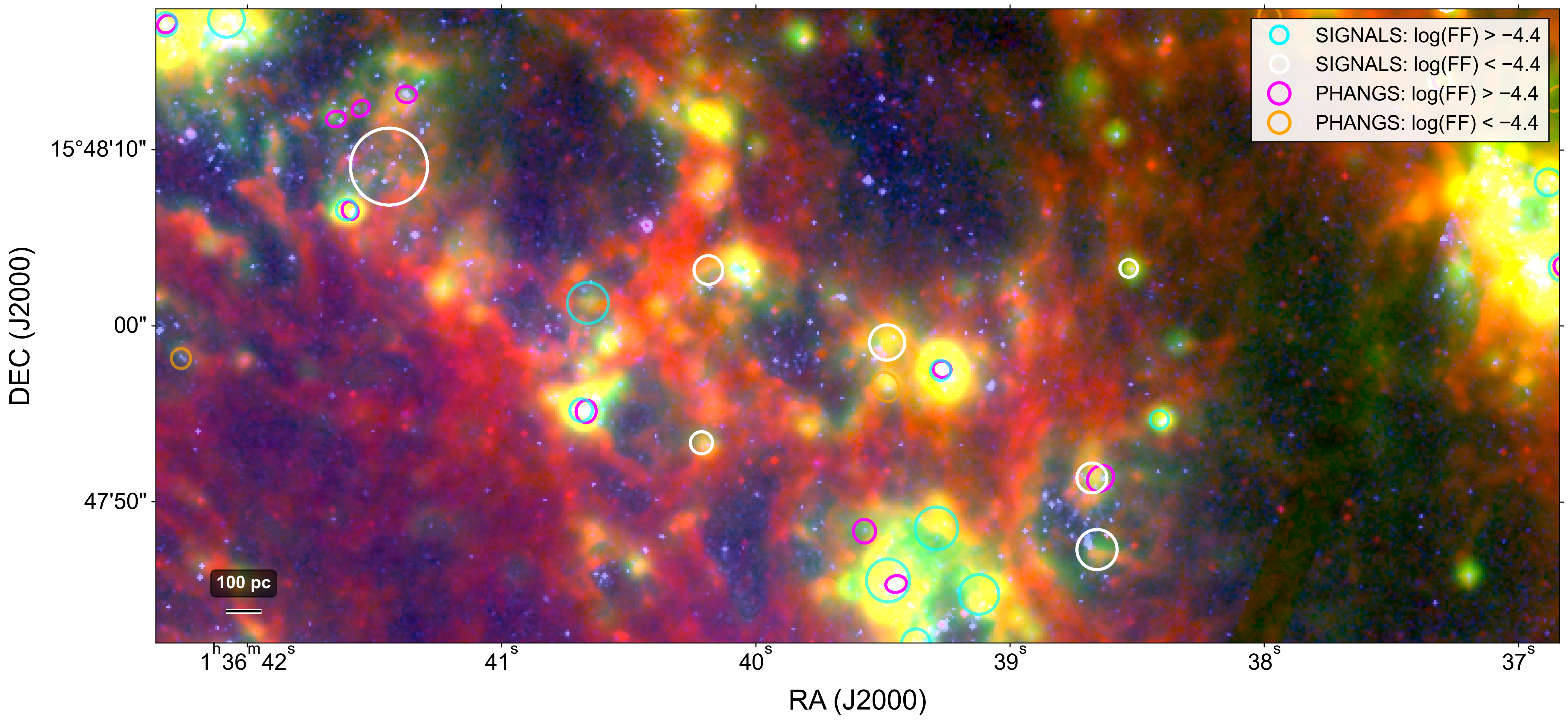}  
		\caption{Close-up of the central southwest area of the galaxy NGC~628 showing $\mathrm{log(FF)}$ of each H\,II region in a local RGB composite image. The circles measured are plotted in light blue for SIGNALS-JWST regions with $\mathrm{log(FF)}$ > $-$4.4 and as white circles for regions with $\mathrm{log(FF)}$ < $-$4.4. In PHANGS-JWST, magenta ellipses illustrate regions with $\mathrm{log(FF)}$ > $-$4 and orange circles those with $\mathrm{log(FF)}$ < $-$4.4. Red corresponds to the MIRI F770W image, green to the PHANGS-MUSE H$\alpha$ image and blue to the HST F435W image of NGC~628. Low-$\log(\mathrm{FF})$ regions are preferentially associated with red MIRI F770W emission, whereas high-$\log(\mathrm{FF})$ regions appear associated within the brightest PHANGS-MUSE H$\alpha$ structures.}
		\label{fig:2DF770_FF_closeup}
	\end{center}
\end{figure*}

Finally, a qualitative visualization of this result is suggested in Fig.~\ref{fig:2DF770_FF_closeup} by the likely association between regions with the lowest FF and areas of enhanced PAH ionized emission. We combined PHANGS-H$\alpha$ with PHANGS-JWST/MIRI F770W image to produce a false-color composite image at their native resolutions. We overlay the $\mathrm{log(FF)}$ values of individual H\,II region to highlight those with $\mathrm{log(FF)}$ < $-$4.4 which are expected to most clearly exhibit the observed trends. 
\par
Lower $\log(\mathrm{FF})$ values (white circles for SIGNALS-JWST and orange ellipses for PHANGS-JWST) appear preferentially located within or near the red regions in the MIRI F770W image, which exhibit systematically higher $\mathrm{R_{PAH}}$ ratios (Fig.~\ref{fig:RpahLR3}). This is consistent with less efficient PAH processing in regions characterized by more porous ionized-gas structures.
\par
Assuming that regions with very high FF are, on average, powered by younger stellar populations, the intense radiation fields of central massive stars may initially produce small PAHs and subsequently destroy them. At later evolutionary stages, stellar winds and mechanical feedback can disperse the ionized gas, while reducing the efficiency of PAH destruction. This evolutionary scenario is not expected to be uniform, as it ultimately depends on the particular physical conditions of the ISM in each region because the mechanisms governing PAH destruction, survival, and migration remain poorly constrained \citep[e.g.,][]{GarciaBernete2022, Hrodmarsson2023}. 
\section{Discussion} \label{sec:discussion}
The main goal of this study is to establish a foundation for future investigations of H\,II regions at similar spatial resolution, exploring how FF reflects the internal structure and connects to the physical and chemical properties of extragalactic star-forming regions in NGC~628. We consider the FF as a global parameter, and acknowledge that it encompasses several distinct physical components (Sect.~\ref{sec:introduction}), while spherical ionization-bounded approximations together with uniform density and spherical symmetry, represent a first-order simplification.
\par
The typical spatial resolutions of the SIGNALS and the PHANGS-MUSE surveys in NGC~628 do not allow us to resolve substructures such as rings, shells, or cavities within the analyzed H\,II regions. We selected a total of 622 H\,II regions from the SIGNALS and the PHANGS-MUSE H\,II regions survey. After applying a consistent homogenization of the physical parameters and density measurements in the two surveys, we found that the derived FF values showed very good agreement. The statistical consistency between the distributions confirms the equivalence of the SIGNALS and PHANGS-MUSE datasets for this analysis and provides a robust basis for our subsequent results. 
We also compared our results with published H\,II region catalogs of the same galaxy based on PHANGS-MUSE and PHANGS-MUSE HST data. These catalogs use different segmentation pipelines and boundary definitions. We demonstrated that the apparent slope of the observed H$\alpha$ luminosity–R relation is not universal, but rather a methodological effect reflecting the position of the radius definition threshold relative to the intrinsic profile scale of each region (Sect.~\ref{sec:LR}). The cross-catalog analysis demonstrated in Fig.~\ref{fig:PHANGSSIGNALS_ChristmasTree} confirms that this scatter is not an effect of the boundary definition but reflects physical variation in $n_e^2$ in the population.
\par
Luminous giant and supergiant H\,II regions show the expected decrease in FF with increasing volume \citep{Gutierrez2010, Cedres2013}, consistent with the definitional scaling. This behavior indicates that larger regions are characterized by more widely spaced clumps and, consequently, by a larger fraction of low-density gas occupying the available volume \citep{Hunt2009}. However, the variance explained by volume alone is low ($R^{2}_{fit} \sim$ 0.16-0.21), indicating that volume alone cannot explain the intrinsic scatter of FF within this group of regions. 
\par
The local electron density exhibits a strong anti-correlation with FF, consistent with the $n_{e}^{-2}$ slope in the FF definition. SIGNALS-H\,II regions show excellent agreement with the expected slope, while PHANGS-H\,II regions display more scatter, particularly in low-luminosity and compact regions. Most of the brightest giant H\,II regions are found in the low-density regime, with lower FF (longer optical paths) than fainter giant H\,II regions. This suggests that in fainter regions, the ionized gas is distributed in a more porous and inhomogeneous medium with greater spacing between condensations. The increased separation between inhomogeneities could result in a substantial low-density volume, allowing ionizing photons to escape more efficiently from central stars through leaky paths, as suggested by \citet{Belfiore2022, Teh2023}. This effect supports a scenario in which the internal geometry and possibly the photon escape fraction play key roles in regulating the ionized gas structure \citep{Giammanco2004}. This process could, in turn, contribute to enhanced DIG emission in the surrounding environment. 
\par
H$\alpha$ luminosity correlates positively with FF, consistent with the analytical relation $FF \propto L/(n_{e}^{2}R^{3})$. Luminous H\,II regions, typically ionized by stellar associations or clusters containing multiple OB stars  \citep{Santoro2022, Scheuermann2023}, exhibit higher FF values, consistent with ionized gas confined to dense clumps embedded within a low-density medium. At lower luminosities, however, systematic departures from the dominant trends emerge. Regions with $\log(L_{\rm H\alpha}) \lesssim 37$--37.5~$\mathrm{\ erg\,s^{-1}}$ show increased scatter in the FF--$L_{\rm H\alpha}$ relation, fall below the FF--$R$ relation reported by \cite{Cedres2013}, and drive the steeper FF--$n_e$ slope measured for PHANGS-H\,II ($-3.61 \pm 0.23$); excluding them recovers a slope of $\sim -2.7$, consistent with the SIGNALS-H\,II sample. In the FF–$L_{H\alpha}$ relation, increased scatter and departures below the dominant trend appear at the faint end in both surveys. These objects comprise compact regions ($R<50$~pc) together with a population of somewhat larger nebulae, and their behavior appears largely independent of local ISM conditions owing to the DIG-corrected measurements. Their inferred ionizing luminosities are consistent with excitation by single late O-type or early B-type stars, or by low-mass clusters \citep{Lee2011}, making them analogous to classical Galactic H\,II regions despite being located in external galaxies \citep{Kennicutt1984}. The transition does not correspond to a sharp luminosity threshold, but rather to the luminosity range over which deviations from the classical FF scaling become apparent, in agreement with the regime change identified by \cite{Lee2011} around $\log(L_{\rm H\alpha}) \lesssim 37.1~\mathrm{erg\,s^{-1}}$. Overall, the classical FF framework remains valid throughout the full luminosity range we probed. In the luminous regime, regions with higher FF values, indicative of reduced interclump spacing and a larger fraction of ionized gas filling the nebular volume, tend to exhibit lower local electron densities than more extended, lower-FF regions. We adopted $\mathrm{log(L_{H\alpha})} \lesssim 37-37.5~\mathrm{\ erg \ s^{-1}}$ as an approximate luminosity range over which the transition between regimes becomes apparent, in line with the threshold reported by \citep{Lee2011} around $\mathrm{log(L_{H\alpha})} \lesssim 37.1~\mathrm{erg \ s^{-1}}$. This range does not represent a sharp threshold, but rather reflects that the departure becomes apparent at different luminosities depending on the scaling relation considered. 
\par
Strong radiation pressure in luminous H\,II regions drives gas expansion, concentrating material into dense shells where the [S\,II] lines probe the local electron density. In this regime, the rms-based FF provides a meaningful description of the nebular geometry and closely approximates the idealized Strömgren-sphere picture. At lower luminosities, however, ionized structures become increasingly affected by sampling and resolution effects, with filaments and condensations contributing significantly to the density field. Consequently, FF no longer acts as a purely volumetric measure but instead reflects the underlying density distribution. A more physical description would therefore require approaches beyond the classical Osterbrock framework and purely statistical PDF-based measures \citep{Spangler2019}, such as fractal analyses or MHD simulations capable of deriving FF directly from the gas dynamics. A detailed exploration of these alternatives is beyond the scope of this work.
\par
One of the most significant results of this study is the linear alignment of all five H\,II regions catalogs in the $\mathrm{log(FF)}$–$\mathrm{log(L_{H\alpha}/R^3)}$ plane (Fig.~\ref{fig:ALLcatalogsFFLR3}). Despite nearly two orders of magnitude difference in radii between \cite{Barnes2026} and the SIGNALS- and MUSE-based catalogs, and despite the varying criteria used to define H\,II region boundaries, all datasets fall on the same relation. This is nontrivial: the scaling relation between $\mathrm{log(FF)}$ and $\mathrm{log(L_{H\alpha}/R^3)}$ does not imply that catalogs defining R via absolute flux thresholds, profile truncation, or HST-based fragmentation would coincide. The usefulness of the $\mathrm{log(FF)}$-$\mathrm{log(L_{H\alpha}/R^3)}$ plane lies in isolating the role of the electron density term. The residual scatter around this relation reflects true variations in $n_e^2$ rather than differences in segmentation. Future work will extend this analysis beyond NGC~628 to a larger sample of galaxies (Aragüete Riesco et al., in prep.). This demonstrates that $\mathrm{L_{H\alpha}/R^3}$, the volumetric H$\alpha$ luminosity density absorbs methodological differences and measures the effective emission measure per unit volume of the emitting clumps. Furthermore, correlations involving $\mathrm{L_{H\alpha}/R^{3}}$ are comparable in all catalogs, provided that the second observable does not contain an explicit dependence on the adopted region size.
\par
As the central cluster’s ionizing output declines, the EW(H$\alpha$) progressively decreases \citep[e.g.][]{Kewley2015, MarmolQueralto2016, Mingozzi2020}. In this context, the observed decrease in EW(H$\alpha$) toward lower FF values may indicate an advanced evolutionary stage. These preliminary results provide a coherent picture linking internal gas structure and evolutionary stage: young compact regions with high EW(H$\alpha$) and high FF; intermediate regions with extended nebulae and moderate EW(H$\alpha$); and older regions with low EW(H$\alpha$) and low FF, with inferred ages spanning from less than 3 to 6.3 Myr. However, given the shared dependence on H$\alpha$ luminosity, the observed FF–EW(H$\alpha$) trend cannot be unambiguously interpreted in terms of evolutionary sequence alone.
\par
However, we caution against a definitive evolutionary interpretation. The classical Strömgren framework assumes uniform density and spherical symmetry, an unrealistic simplification in real H\,II regions \citep{Castaneda1992, Copetti2000}, whereas real H\,II regions are often powered by multiple OB associations at different evolutionary stages. This would introduce additional scatter in any FF--EW(H$\alpha$) relation and further limits its interpretation as a strict evolutionary sequence. In addition, our analysis is exploratory, based on a limited number of H\,II regions with reliable continuum measurements, and relies on EW(H$\alpha$) as an indirect age proxy.
\par
Elevated EW(H$\alpha$) values in bright H\,II regions may also reflect metallicity effects \citep{Leitherer1999} or leakage of ionizing photons \citep{Zurita2002}. A definitive evolutionary interpretation would require independent age constraints beyond the EW(H$\alpha$) proxy, for instance from stellar cluster catalogs. Furthermore, the assumption of a uniform density of ionized gas clumps is an additional simplification that may not hold in turbulent or highly inhomogeneous regions \citep{Copetti2000}.
\par
We examined the relation between FF and PAH emission. The transition in PAH abundance identified in the $\mathrm{log(R_{PAH})}$-$\mathrm{log(L_{H\alpha}/R³)}$ relation corresponds to H\,II regions with volumetric H$\alpha$ luminosity density, $\mathrm{log(L_{H\alpha}/R^3)} \approx$ 31.76-32.11 $\ \mathrm{erg \ s^{-1} \ pc^{-3}}$, ionizing photon rates $\mathrm{log(Q_H)} \approx$ 49.03 - 49.48 $\mathrm{\ s^{-1}}$), and $\mathrm{log(FF)} \approx$ -5.01- -4.02. The consistency of these three independent diagnostics in two catalogs, built from different instruments and radius definition methods, confirms that this physical regime marks the onset of efficient PAH photodissociation in H\,II regions of NGC~628, in agreement with the critical threshold $\mathrm{log(Q_H)} \gtrsim 10^{50}~\mathrm{s^{-1}}$ reported by \citet{Egorov2023b}. 
\par
Within the luminous giant and supergiant H\,II region regime, characterized by high FF values, we find a clear dependence between the PAH-to-dust fraction and FF. PAHs are highly susceptible to Lyman-continuum radiation, undergoing photoionization and photodissociation that progressively reduce their PAH fraction \citep{Dale2025}. High-FF regions, which are typically the most luminous, provide the intense radiation fields required for efficient PAH destruction. In addition, stellar winds and radiation pressure from massive O/B stars can evacuate the central regions and compress dust and gas into shell-like structures where PAH emission is preferentially detected \citep{Watson2008}. Our results suggest that lower-luminosity H\,II regions exhibit higher PAH-to-dust ratios, indicating less efficient destruction of PAHs and carbonaceous grains, consistent with the findings of \citet{Egorov2023b, Egorov2025}. Beyond reduced photodestruction, evolutionary effects may also contribute: the shattering of larger carbonaceous grains and the preferential destruction of VSGs can enhance the relative PAH abundance in older H\,II regions \citep{Jones1996, Galliano2008, Micelotta2010, Seok2014}. Our observations support this scenario in which PAH survival and replenishment become increasingly important as the ionizing radiation field weakens with time.
\par
In the case of very low-luminosity, small H\,II regions, we observe an almost flat dependence of $\mathrm{log(R_{PAH})}$ on $\mathrm{log(FF)}$. This suggests that UV radiation in these regions is insufficient to efficiently destroy PAHs, and that secondary mechanisms unrelated to ionizing radiation may play a more important role in maintaining the PAH population \citep{Reizer2022}. Furthermore, in small H\,II regions, the measured $R_{PAH}$ values are significantly affected by observational biases. As discussed by \cite{Egorov2025}, approximately 45\% of these measurements are overestimated by more than 10\% due to the difficulty of fully isolating the ionized gas from the surrounding diffuse ISM and PDRs at the current resolution. This contamination, compounded by line-of-sight projection effects, may explain why smaller regions often exhibit $R_{PAH}$ values closer to those of the diffuse interstellar medium. However, in NGC~628, the mean $R_{PAH}$ of the diffuse ISM is $3.72^{+4.50}_{-2.84}$ \citep{Sutter2024}. We consistently find higher values in small, low-luminosity H\,II regions, suggesting that additional mechanisms, such as accretion-driven processes leading to the formation of small PAHs, may help sustain the PAH population and partially compensate for reduced radiative destruction. To test the impact of this bias, we propagate the expected overestimation to the median $\mathrm{R_{PAH}}$ for regions with $\mathrm{log(FF)} < -4.4$. If $45\%$ of these regions are overestimated by a factor of 1.10 (\citealt{Egorov2025}), the corrected median becomes $\mathrm{R_{PAH, corr}}$ = $\mathrm{R_{PAH, obs}}$/1.045, implying $\Delta \mathrm{R_{PAH}} \approx -0.02$ dex. This shift is smaller than the intrinsic dispersion of the low-FF subsample ($\sigma_{R_{PAH}}=0.11$~dex in SIGNALS–JWST and $\sigma_{R_{PAH}}=0.22$~dex in PHANGS–JWST) and below the median offset between FF regimes ($\Delta \mathrm{R_{PAH}} \approx 0.18$–$0.25$ dex; Fig.~\ref{fig:histRpahLR3}), indicating that the observational bias cannot explain the elevated $\mathrm{R_{PAH}}$ values in low-FF regions.
\par
Overall, our results show that the FF is a useful diagnostic of the internal structure and evolutionary stage of extragalactic H\,II regions. The apparent slope of the $L_{H\alpha}$–$R$ relation depends on whether the boundary definition is tied to the absolute flux level or to the local profile structure of each region providing a framework for cross-catalog comparisons. The tight alignment of all five catalogs in the $\log(\mathrm{FF})$ --$\log(L_{\mathrm{H}\alpha}/R^3)$ plane indicates that the volumetric luminosity density offers a convenient common representation of the $L_{\mathrm{H}\alpha}$--$R$ coupling when combined with FF estimates, reducing the impact of methodological differences in radius definitions.
\par
The applicability of the FF framework throughout the full luminosity range probed here provides a basis for exploring variations in H\,II region structure and dust properties in diverse extragalactic environments (Aragüete Riesco et al., in prep.).
\section{Conclusions} \label{sec:conclusions}
We presented a detailed census of extragalactic H\,II regions in NGC~628, providing homogeneous measurements of luminosity, emission line fluxes, and radii by combining 475 H\,II regions from the SIGNALS survey and 147 regions from PHANGS-MUSE in NGC~628. We provided measurements of the H$\alpha$ luminosity, emission line fluxes, electron densities, radii, and FF in the two surveys and extended the comparison to three published H\,II region catalogs based on different segmentation methods. Our analysis confirmed a close connection between the FF and several fundamental properties of H\,II regions, highlighting its diagnostic value for probing their internal structure. We also investigated the connection between H\,II regions and the PAH-to-dust ratio, offering new insights into the interplay among star formation, ionized gas, dust, and the interstellar medium. Our analysis yields several key findings that we list below.
\begin{itemize}
	\item The apparent slope of the $L_{H\alpha}$-$R$ relation does not appear to be a universal physical constant, but is an observable that can be affected by the coupling between the boundary definition and the flux distribution of each region. Assuming a Gaussian surface brightness H\,II region profile model, we demonstrated that absolute-threshold methods produce steeper effective slopes than profile-truncated methods. This framework distinguishes physical and methodological effects in H$\alpha$ luminosity--radius relations.	
	\item We showed that the SIGNALS, PHANGS--MUSE, and PHANGS--MUSE/HST H$\alpha$ catalogs agree very well in the $\mathrm{log(FF)}$ --$\mathrm{log}(L_{\mathrm{H}\alpha}/R^3)$ plane. The residual scatter primarily reflects intrinsic variations in $n_e^{2}$ in the H\,II region population. 
	\par
	The volumetric H$\alpha$ luminosity density, $L_{\mathrm{H}\alpha}/R^3$, reduces the covariance between $L_{\mathrm{H}\alpha}$ and $R$ introduced by the adopted region definition, and it provides a consistent way to compare H\,II region properties in different catalogs in slope space. Physically, it traces the effective emission measure per unit volume of the emitting clumps, $L_{\mathrm{H}\alpha}/R^3 \propto n_e^2 \times \mathrm{FF}$.
	In this sense, $L_{\mathrm{H}\alpha}/R^3$ constitutes a useful quantity for cross-catalog comparisons when it is correlated with observables that are independent of the adopted region size.
	\item The FF scaling relations with volume, electron density, and H$\alpha$ luminosity are globally consistent with expectations from a Strömgren-like framework in the luminous regime. However, the scatter around each relation carries physical information about the internal gas structure of H\,II regions and reflects secondary dependences.
	\item Low-luminosity H\,II regions, likely powered by single massive stars, appear to depart from the trends followed by giant and supergiant regions in all three FF scaling relations. These differences are primarily driven by observational sampling and resolution effects and not by a breakdown of the classical FF formulation. The transition occurs over the range $\mathrm{log(L_{H\alpha})} \sim 36.5-37.5~\mathrm{\ erg \ s^{-1}}$, depending on the relation considered.
	\item The negative physical residual $\gamma^{\rm phys}$ in the $\mathrm{log(FF)}$–$\mathrm{log(R^{3})}$ relation reflects that more extended regions host lower $n_e^{2}$, so that the increasing volume-to-luminosity ratio and the dilution of ionized gas act together to reduce FF beyond what is predicted by the definitional and coupling effects alone. The $\mathrm{log(FF)}$–$\mathrm{log(L_{H\alpha})}$ relation indicates that more luminous regions genuinely occupy a larger ionized fraction of their volume.
	\item In an exploratory analysis, the $\log(\mathrm{FF})$--$\log(\mathrm{EW}(\mathrm{H}\alpha))$ relation suggests that regions with high $\log(\mathrm{EW}(\mathrm{H}\alpha))$ (interpreted as younger ages) tend to show higher $\log(\mathrm{FF})$, while intermediate and lower $\log(\mathrm{EW}(\mathrm{H}\alpha))$ regions span progressively lower $\log(\mathrm{FF})$ values. The inferred ages range from $\lesssim 3$ to $\sim 6.3$ Myr, although a robust evolutionary interpretation requires larger samples and independent age constraints beyond the EW(H$\alpha$) proxy. These trends might be consistent with a decrease in FF with H\,II region age.
	\item In NGC~628, the $\log(R_{\rm PAH})$--$\log(L_{\rm H\alpha}/R^{3})$ relation shows a transitional change in the slope at $\log(L_{\rm H\alpha}/R^{3}) \approx 31.76-32.11$ $\mathrm{erg\,s^{-1}\,pc^{-3}}$, below which H\,II regions preserve higher PAH abundances, indicating less efficient photodestruction in more diffuse, evolved environments. Above this, $R_{PAH}$ decreases steeply with increasing $L_{H\alpha}/R^3$, consistent with efficient PAH destruction by the intense radiation fields of luminous H\,II regions. The H\,II regions at the transition are characterized by ionizing photon rates $\mathrm{log(Q_H)} \approx$ 49.03 -- 49.48 $ \mathrm{s^{-1}}$ and $\mathrm{log(FF)} \approx$ -5.01-- -4.02.
\end{itemize}
Taken together, these trends can be indicative of a progressive change in the H\,II region properties, where a lower FF is associated with more extended and diffuse ionized structures that appear to be more favorable to PAH survival. Further investigations extending this analysis to other galaxies and higher spatial resolutions are currently underway. 
\begin{acknowledgements}
	  We thank the anonymous referee for carefully reviewing
	  our manuscript. We truly appreciate the constructive comments provided, which have enabled us to make this paper better.
      L.A.R., J.M.V., S.D.P. acknowledge financial support from the State Agency for Research of the Spanish MCIU through Center of Excellence Severo Ochoa award to the Instituto de Astrofísica de Andalucía CEX2021-001131-S funded by MCIN/AEI/10.13039/ 501100011033, and  from the project grant PID2022-136598NB-C32 Estallidos8. L.A.R., J.M.V. and S.D.P. also acknowledge Laurent Drissen's contribution to the SIGNALS survey. CM acknowledges the support of grant UNAM/DGAPA/PAPIIT IG101223.
      This study was partly based on observations obtained with SITELLE, a joint project of Universit´e Laval, ABB, Universit´e de Montr´eal, and the Canada-France-Hawaii Telescope (CFHT) which is operated by the National Research Council of Canada, the Institut National des Sciences de l’Univers of the Centre National de la Recherche Scientifique of France, and the University of Hawaii. LRN is grateful to the Natural Sciences and Engineering Research Council of Canada NSERC - RGPIN-2023-03487 and the National Science foundation (NSF) - 2109124 for their support.
      The authors wish to recognize and acknowledge the very significant cultural role that the summit of Mauna Kea has always had within the indigenous Hawaiian community. We are most grateful to have the opportunity to conduct observations from this mountain.
      This research has made use of the NASA/IPAC Extragalactic Database (NED), which is operated by the Jet Propulsion Laboratory, California Institute of Technology, under contract with NASA.
      This study is also based on observations collected at the European Southern Observatory under ESO programmes 1100.B-0651, 095.C-0473, and 094.C-0623 (PHANGS–MUSE; PI Schinnerer) and on observations made with the NASA/ESA/CSA JWST. The data were
      obtained from the Mikulski Archive for Space Telescopes at the Space Telescope Science Institute, which is operated by the Association of Universities for Research in Astronomy, Inc., under NASA contract NAS5-03127 for JWST. These observations are associated with program 2107. The specific observations analyzed can be accessed via doi:10.17909/ew88-jt15. 
      Software: This research made use of "Aladin sky atlas" developed at CDS, Strasbourg Observatory, France \citep{Bonnarel2000}, Astropy, a
      community-developed core python (http://www.python.org, \citep{VanRossum2009} package for Astronomy (Astropy Collaboration et al. 2013, 2018, 2022); scikit-learn \citep{Pedregosa2011} and Scikit-Image \citep{Walt2014}, Matplotlib \citep{Hunter2007}; NumPy \citep{Walt2011}; SciPy \citep{Virtanen2020}; Sympy (https://doi.org/10.7717/peerj-cs.103); OpenCV-Python (https://github.com/opencv/opencv-python); and Pandas \citep{McKinney2010}; Photutils \citep{Bradley2019}.
\end{acknowledgements}
\bibliography{bibliography}

@Article{Zurita2026,
  author    = {Zurita, Almudena and Bresolin, Fabio and Florido, Estrella and Verley, Simon and Relaño, Mónica and Beckman, John E},
  journal   = {MNRAS},
  title     = {Electron densities and filling factors of extragalactic H  ii regions: NGC 2403 and NGC 628},
  year      = {2026},
  issn      = {1365-2966},
  month     = Mar,
  number    = {4},
  volume    = {547},
  doi       = {10.1093/mnras/stag480},
  publisher = {Oxford University Press (OUP)},
}

@Article{McClain2026,
  author    = {McClain, Rebecca L. and Leroy, Adam K. and Congiu, Enrico and Barnes, Ashley. T. and Belfiore, Francesco and Egorov, Oleg and Emsellem, Eric and Rosolowsky, Erik and Amiri, Amirnezam and Boquien, Médéric and Chastenet, Jérémy and Chown, Ryan and Dale, Daniel A. and Das, Sanskriti and Glover, Simon C. O. and Grasha, Kathryn and Indebetouw, Rémy and Koch, Eric W. and Mathur, Smita and Méndez-Delgado, J. Eduardo and Oakes, Elias K. and Pan, Hsi-An and Sandstrom, Karin and Sarbadhicary, Sumit K. and Whitmore, Bradley C. and Williams, Thomas G.},
  journal   = {ApJ},
  title     = {Resolved H II Regions in NGC 253: Ionized Gas Structure and Suggestions of a Universal Density–Surface Brightness Relation},
  year      = {2026},
  issn      = {1538-4357},
  month     = Feb,
  number    = {1},
  pages     = {166},
  volume    = {998},
  doi       = {10.3847/1538-4357/ae1f94},
  publisher = {American Astronomical Society},
}

@Article{Barnes2026,
  author    = {Barnes, A. T. and Chandar, R. and Kreckel, K. and Belfiore, F. and Pathak, D. and Thilker, D. and Leroy, A. K. and Groves, B. and Glover, S. C. O. and McClain, R. and Amiri, A. and Bazzi, Z. and Boquien, M. and Congiu, E. and Dale, D. A. and Egorov, O. V. and Emsellem, E. and Grasha, K. and Gonzalez Lobos, J. and Henny, K. and He, H. and Indebetouw, R. and Lee, J. C. and Li, J. and Liang, F.-H. and Larson, K. and Maschmann, D. and Meidt, S. E. and Eduardo Méndez-Delgado, J. and Neumann, J. and Pan, H.-A. and Querejeta, M. and Rosolowsky, E. and Sarbadhicary, S. K. and Scheuermann, F. and Úbeda, L. and Williams, T. G. and Weinbeck, T. D. and Whitmore, B. and Wofford, A. and},
  journal   = {A\&A},
  title     = {The PHANGS-MUSE/HST-H α nebulae catalogue: Parsec-scale resolved structure, physical conditions, and stellar associations across nearby galaxies},
  year      = {2026},
  issn      = {1432-0746},
  month     = Feb,
  pages     = {A95},
  volume    = {706},
  doi       = {10.1051/0004-6361/202555751},
  publisher = {EDP Sciences},
}

@Article{Pathak2025,
  author    = {Pathak, Debosmita and Leroy, Adam K. and Barnes, Ashley. T. and Thompson, Todd A. and Lopez, Laura A. and Sandstrom, Karin M. and Sun, Jiayi and Glover, Simon C. O. and Klessen, Ralf S. and Koch, Eric W. and Larson, Kirsten L. and Lee, Janice and Meidt, Sharon and Sanchez-Blazquez, Patricia and Schinnerer, Eva and Bazzi, Zein and Belfiore, Francesco and Boquien, Médéric and Chown, Ryan and Colombo, Dario and Congiu, Enrico and Egorov, Oleg V. and Eibensteiner, Cosima and Kurapati, Sushma and Querejeta, Miguel and Dale, Daniel A. and Kravtsov, Timo and Padave, Mansi and Pisano, D. J. and Rosolowsky, Erik and Sarbadhicary, Sumit K. and Williams, Thomas G. and Indebetouw, Remy and Pan, Hsi-An and Úbeda, Leonardo and Amiri, Amirnezam and Bigiel, Frank and Blanc, Guillermo A. and Grasha, Kathryn},
  journal   = {ApJL},
  title     = {Masses, Star Formation Efficiencies, and Dynamical Evolution of 18,000 H ii Regions},
  year      = {2025},
  issn      = {2041-8213},
  month     = oct,
  number    = {1},
  pages     = {L20},
  volume    = {993},
  doi       = {10.3847/2041-8213/ae0e70},
  publisher = {American Astronomical Society},
}

@Article{Egorov2025,
  author    = {Egorov, Oleg V. and Leroy, Adam K. and Sandstrom, Karin and Kreckel, Kathryn and Baron, Dalya and Belfiore, Francesco and Chown, Ryan and Sutter, Jessica and Boquien, Médéric and Canal i Saguer, Mar and Congiu, Enrico and Dale, Daniel A. and Egorova, Evgeniya and Huber, Michael and Li, Jing and Williams, Thomas G. and Chastenet, Jérémy and Chiang, I-Da and Gerasimov, Ivan and Hassani, Hamid and Kim, Hwihyun and Koziol, Hannah and Lee, Janice C. and McClain, Rebecca L. and Delgado, José Eduardo Méndez and Pan, Hsi-An and Pathak, Debosmita and Rosolowsky, Erik and Sarbadhicary, Sumit K. and Schinnerer, Eva and Thilker, David and Ubeda, Leonardo and Weinbeck, Tony},
  journal   = {A\&A},
  title     = {Polycyclic aromatic hydrocarbon destruction in star-forming regions across 42 nearby galaxies},
  year      = {2025},
  issn      = {1432-0746},
  month     = nov,
  pages     = {A103},
  volume    = {703},
  doi       = {10.1051/0004-6361/202556427},
  publisher = {EDP Sciences},
}

@Article{Dale2025,
  author    = {Dale, Daniel A. and Graham, Gabrielle B. and Barnes, Ashley T. and Baron, Dalya and Bigiel, Frank and Boquien, Médéric and Chandar, Rupali and Chastenet, Jérémy and Chown, Ryan and Egorov, Oleg V. and Glover, Simon C. O. and Hands, Lindsey and Henny, Kiana F. and Indebetouw, Remy and Klessen, Ralf S. and Larson, Kirsten L. and Lee, Janice C. and Leroy, Adam K. and Maschmann, Daniel and Pathak, Debosmita and Rodríguez, M. Jimena and Rosolowsky, Erik and Sandstrom, Karin and Schinnerer, Eva and Sutter, Jessica and Thilker, David A. and Weinbeck, Tony D. and Whitmore, Bradley C. and Williams, Thomas G. and Wofford, Aida},
  journal   = {AJ},
  title     = {PAH Feature Ratios around Stellar Clusters and Associations in 19 Nearby Galaxies},
  year      = {2025},
  issn      = {1538-3881},
  month     = feb,
  number    = {3},
  pages     = {133},
  volume    = {169},
  doi       = {10.3847/1538-3881/ada89f},
  publisher = {American Astronomical Society},
}

@Article{Baron2025,
  author    = {Baron, Dalya and Sandstrom, Karin M. and Sutter, Jessica and Hassani, Hamid and Groves, Brent and Leroy, Adam K. and Schinnerer, Eva and Boquien, Médéric and Brazzini, Matilde and Chastenet, Jérémy and Dale, Daniel A. and Egorov, Oleg V. and Glover, Simon C. O. and Klessen, Ralf S. and Pathak, Debosmita and Rosolowsky, Erik and Bigiel, Frank and Chevance, Mélanie and Grasha, Kathryn and Hughes, Annie and Méndez-Delgado, J. Eduardo and Pety, Jérôme and Williams, Thomas G. and Hannon, Stephen and Sarbadhicary, Sumit K.},
  journal   = {ApJ},
  title     = {PHANGS-ML: The Universal Relation between PAH Band and Optical Line Ratios across Nearby Star-forming Galaxies},
  year      = {2025},
  issn      = {1538-4357},
  month     = jan,
  number    = {2},
  pages     = {135},
  volume    = {978},
  doi       = {10.3847/1538-4357/ad972a},
  publisher = {American Astronomical Society},
}

@Article{Williams2024,
  author    = {Williams, Thomas G. and Lee, Janice C. and Larson, Kirsten L. and Leroy, Adam K. and Sandstrom, Karin and Schinnerer, Eva and Thilker, David A. and Belfiore, Francesco and Egorov, Oleg V. and Rosolowsky, Erik and Sutter, Jessica and DePasquale, Joseph and Pagan, Alyssa and Berger, Travis A. and Anand, Gagandeep S. and Barnes, Ashley T. and Bigiel, Frank and Boquien, Médéric and Cao, Yixian and Chastenet, Jérémy and Chevance, Mélanie and Chown, Ryan and Dale, Daniel A. and Deger, Sinan and Eibensteiner, Cosima and Emsellem, Eric and Faesi, Christopher M. and Glover, Simon C. O. and Grasha, Kathryn and Hannon, Stephen and Hassani, Hamid and Henshaw, Jonathan D. and Jiménez-Donaire, María J. and Kim, Jaeyeon and Klessen, Ralf S. and Koch, Eric W. and Li, Jing and Liu, Daizhong and Meidt, Sharon E. and Méndez-Delgado, J. Eduardo and Murphy, Eric J. and Neumann, Justus and Neumann, Lukas and Neumayer, Nadine and Oakes, Elias K. and Pathak, Debosmita and Pety, Jérôme and Pinna, Francesca and Querejeta, Miguel and Ramambason, Lise and Romanelli, Andrea and Sormani, Mattia C. and Stuber, Sophia K. and Sun, Jiayi and Teng, Yu-Hsuan and Usero, Antonio and Watkins, Elizabeth J. and Weinbeck, Tony D.},
  journal   = {ApJS},
  title     = {{PHANGS-JWST}: Data-processing Pipeline and First Full Public Data Release},
  year      = {2024},
  issn      = {1538-4365},
  month     = jul,
  number    = {1},
  pages     = {13},
  volume    = {273},
  doi       = {10.3847/1538-4365/ad4be5},
  publisher = {American Astronomical Society},
}

@Article{Ujjwal2024,
  author    = {Ujjwal, Krishnan and Kartha, Sreeja S. and Akhil, Krishna R. and Mathew, Blesson and Subramanian, Smitha and Sudheesh, T. P. and Thomas, Robin},
  journal   = {A\&A},
  title     = {Disentangling the association of PAH molecules with star formation: Insights from the James Webb Space Telescope and from the UltraViolet Imaging Telescope},
  year      = {2024},
  issn      = {1432-0746},
  month     = apr,
  pages     = {A71},
  volume    = {684},
  doi       = {10.1051/0004-6361/202347620},
  publisher = {EDP Sciences},
}

@Article{Sutter2024,
  author    = {Sutter, Jessica and Sandstrom, Karin and Chastenet, Jérémy and Leroy, Adam K. and Koch, Eric W. and Williams, Thomas G. and Chown, Ryan and Belfiore, Francesco and Bigiel, Frank and Boquien, Médéric and Cao, Yixian and Chevance, Mélanie and Dale, Daniel A. and Egorov, Oleg V. and Glover, Simon C. O. and Groves, Brent and Klessen, Ralf S. and Kreckel, Kathryn and Larson, Kirsten L. and Oakes, Elias K. and Pathak, Debosmita and Ramambason, Lise and Rosolowsky, Erik and Watkins, Elizabeth J.},
  journal   = {ApJ},
  title     = {The Fraction of Dust Mass in the Form of Polycyclic Aromatic Hydrocarbons on 10–50 pc Scales in Nearby Galaxies},
  year      = {2024},
  issn      = {1538-4357},
  month     = aug,
  number    = {2},
  pages     = {178},
  volume    = {971},
  doi       = {10.3847/1538-4357/ad54bd},
  publisher = {American Astronomical Society},
}

@Article{Sun2024,
  author    = {Sun, Bingqing and Calzetti, Daniela and Battisti, Andrew J.},
  journal   = {ApJ},
  title     = {The Role of Spiral Arms in Galaxies},
  year      = {2024},
  issn      = {1538-4357},
  month     = sep,
  number    = {2},
  pages     = {137},
  volume    = {973},
  doi       = {10.3847/1538-4357/ad6157},
  publisher = {American Astronomical Society},
}

@Article{Rigopoulou2024,
  author    = {Rigopoulou, D and Donnan, F R and García-Bernete, I and Pereira-Santaella, M and Alonso-Herrero, A and Davies, R and Hunt, L K and Roche, P F and Shimizu, T},
  journal   = {MNRAS},
  title     = {Polycyclic aromatic hydrocarbon emission in galaxies as seen with JWST},
  year      = {2024},
  issn      = {1365-2966},
  month     = jul,
  number    = {2},
  pages     = {1598--1611},
  volume    = {532},
  doi       = {10.1093/mnras/stae1535},
  publisher = {Oxford University Press (OUP)},
}

@Article{Querejeta2024,
  author    = {Querejeta, Miguel and Leroy, Adam K. and Meidt, Sharon E. and Schinnerer, Eva and Belfiore, Francesco and Emsellem, Eric and Klessen, Ralf S. and Sun, Jiayi and Sormani, Mattia and Bešlić, Ivana and Cao, Yixian and Chevance, Mélanie and Colombo, Dario and Dale, Daniel A. and García-Burillo, Santiago and Glover, Simon C. O. and Grasha, Kathryn and Groves, Brent and Koch, Eric. W. and Neumann, Lukas and Pan, Hsi-An and Pessa, Ismael and Pety, Jérôme and Pinna, Francesca and Ramambason, Lise and Razza, Alessandro and Romanelli, Andrea and Rosolowsky, Erik and Ruiz-García, Marina and Sánchez-Blázquez, Patricia and Smith, Rowan and Stuber, Sophia and Ubeda, Leonardo and Usero, Antonio and Williams, Thomas G.},
  journal   = {A\&A},
  title     = {Do spiral arms enhance star formation efficiency?},
  year      = {2024},
  issn      = {1432-0746},
  month     = jul,
  pages     = {A293},
  volume    = {687},
  doi       = {10.1051/0004-6361/202449733},
  publisher = {EDP Sciences},
}

@Article{Li2024,
  author        = {Li, Jing and Kreckel, K. and Sarbadhicary, S. and Egorov, Oleg V. and Groves, B. and Long, K. S. and Congiu, Enrico and Belfiore, Francesco and Glover, Simon C. O. and Barnes, Ashley T. and Bigiel, Frank and Blanc, Guillermo A. and Grasha, Kathryn and Klessen, Ralf S. and Leroy, Adam and Lopez, Laura A. and Méndez-Delgado, J. Eduardo and Neumann, Justus and Schinnerer, Eva and Williams, Thomas G.},
  journal   = {A\&A},
  title         = {Discovery of ~2200 new supernova remnants in 19 nearby star-forming galaxies with MUSE spectroscopy},
  year          = {2024},
  issn          = {1432-0746},
  month         = oct,
  pages         = {A161},
  volume        = {690},
  archiveprefix = {arXiv},
  doi           = {10.1051/0004-6361/202450730},
  eid           = {A161},
  eprint        = {2405.08974},
  primaryclass  = {astro-ph.GA},
  publisher     = {EDP Sciences},
  url           = {https://ui.adsabs.harvard.edu/abs/2024A&A...690A.161L},
}

@Article{Fichtner2024,
  author    = {Fichtner, Yvonne A. and Mackey, Jonathan and Grassitelli, Luca and Romano-Díaz, Emilio and Porciani, Cristiano},
  journal   = {A\&A},
  title     = {Connecting stellar and galactic scales: Energetic feedback from stellar wind bubbles to supernova remnants},
  year      = {2024},
  issn      = {1432-0746},
  month     = sep,
  pages     = {A72},
  volume    = {690},
  doi       = {10.1051/0004-6361/202449638},
  publisher = {EDP Sciences},
}

@Article{Fernandez2024,
  author    = {Fernández, V. and Amorín, R. and Firpo, V. and Morisset, C.},
  journal   = {A\&A},
  title     = {LIME: A LIne MEasuring library for large and complex spectroscopic data sets: I. Implementation of a virtual observatory for JWST spectra},
  year      = {2024},
  issn      = {1432-0746},
  month     = aug,
  pages     = {A69},
  volume    = {688},
  doi       = {10.1051/0004-6361/202449224},
  publisher = {EDP Sciences},
}

@Article{DuartePuertas2024,
  author    = {Duarte Puertas, Salvador and Drissen, Laurent and Robert, Carmelle and Rousseau-Nepton, Laurie and Martin, René Pierre and Amram, Philippe and Martin, Thomas},
  journal   = {MNRAS},
  title     = {Properties of supernova remnants in SIGNALS galaxies – I. NGC 6822 and M33},
  year      = {2024},
  issn      = {1365-2966},
  month     = jul,
  number    = {3},
  pages     = {2677--2704},
  volume    = {533},
  doi       = {10.1093/mnras/stae1641},
  publisher = {Oxford University Press (OUP)},
}

@Misc{Bushouse2024,
  author    = {Bushouse, Howard and Eisenhamer, Jonathan and Dencheva, Nadia and Davies, James and Greenfield, Perry and Morrison, Jane and Hodge, Phil and Simon, Bernie and Grumm, David and Droettboom, Michael and Slavich, Edward and Sosey, Megan and Pauly, Tyler and Miller, Todd and Jedrzejewski, Robert and Hack, Warren and Davis, David and Crawford, Steven and Law, David and Gordon, Karl and Regan, Michael and Cara, Mihai and MacDonald, Ken and Bradley, Larry and Shanahan, Clare and Jamieson, William and Teodoro, Mairan and Williams, Thomas and Pena-Guerrero, Maria},
  title     = {JWST Calibration Pipeline},
  year      = {2024},
  copyright = {Creative Commons Attribution 4.0 International},
  doi       = {10.5281/ZENODO.10463537},
  publisher = {Zenodo},
}

@Article{Thilker2023,
  author    = {Thilker, David A. and Lee, Janice C. and Deger, Sinan and Barnes, Ashley T. and Bigiel, Frank and Boquien, Médéric and Cao, Yixian and Chevance, Mélanie and Dale, Daniel A. and Egorov, Oleg V. and Glover, Simon C. O. and Grasha, Kathryn and Henshaw, Jonathan D. and Klessen, Ralf S. and Koch, Eric and Kruijssen, J. M. Diederik and Leroy, Adam K. and Lessing, Ryan A. and Meidt, Sharon E. and Pinna, Francesca and Querejeta, Miguel and Rosolowsky, Erik and Sandstrom, Karin M. and Schinnerer, Eva and Smith, Rowan J. and Watkins, Elizabeth J. and Williams, Thomas G. and Anand, Gagandeep S. and Belfiore, Francesco and Blanc, Guillermo A. and Chandar, Rupali and Congiu, Enrico and Emsellem, Eric and Groves, Brent and Kreckel, Kathryn and Larson, Kirsten L. and Liu, Daizhong and Pessa, Ismael and Whitmore, Bradley C.},
  journal   = {ApJL},
  title     = {PHANGS–JWST First Results: The Dust Filament Network of NGC 628 and Its Relation to Star Formation Activity},
  year      = {2023},
  issn      = {2041-8213},
  month     = feb,
  number    = {2},
  pages     = {L13},
  volume    = {944},
  doi       = {10.3847/2041-8213/acaeac},
  publisher = {American Astronomical Society},
}

@Article{Teh2023,
  author        = {Teh, Jia Wei and Grasha, Kathryn and Krumholz, Mark R. and Battisti, Andrew J. and Calzetti, Daniela and Rousseau-Nepton, Laurie and Rhea, Carter and Adamo, Angela and Kennicutt, Robert C. and Grebel, Eva K. and Cook, David O. and Combes, Francoise and Messa, Matteo and Linden, Sean T. and Klessen, Ralf S. and Vilchez, Jos{\'e} M. and Fumagalli, Michele and McLeod, Anna and Smith, Linda J. and Chemin, Laurent and Wang, Junfeng and Sabbi, Elena and Sacchi, Elena and Petric, Andreea and Della Bruna, Lorenza and Boselli, Alessandro},
  journal       = {\mnras},
  title         = {Constraining the LyC escape fraction from LEGUS star clusters with SIGNALS H II region observations: a pilot study of NGC 628},
  year          = {2023},
  month         = sep,
  number        = {1},
  pages         = {1191-1210},
  volume        = {524},
  archiveprefix = {arXiv},
  doi           = {10.1093/mnras/stad1780},
  eprint        = {2306.05457},
  primaryclass  = {astro-ph.GA},
  url           = {https://ui.adsabs.harvard.edu/abs/2023MNRAS.524.1191T},
}

@Article{Scheuermann2023,
  author    = {Scheuermann, Fabian and Kreckel, Kathryn and Barnes, Ashley T and Belfiore, Francesco and Groves, Brent and Hannon, Stephen and Lee, Janice C and Minsley, Rebecca and Rosolowsky, Erik and Bigiel, Frank and Blanc, Guillermo A and Boquien, Médéric and Dale, Daniel A and Deger, Sinan and Egorov, Oleg V and Emsellem, Eric and Glover, Simon C O and Grasha, Kathryn and Hassani, Hamid and Jeffreson, Sarah M R and Klessen, Ralf S and Kruijssen, J M Diederik and Larson, Kirsten L and Leroy, Adam K and Lopez, Laura A and Pan, Hsi-An and Sánchez-Blázquez, Patricia and Santoro, Francesco and Schinnerer, Eva and Thilker, David A and Whitmore, Bradley C and Watkins, Elizabeth J and Williams, Thomas G},
  journal   = {MNRAS},
  title     = {Stellar associations powering HII regions – I. Defining an evolutionary sequence},
  year      = {2023},
  issn      = {1365-2966},
  month     = mar,
  number    = {2},
  pages     = {2369--2383},
  volume    = {522},
  doi       = {10.1093/mnras/stad878},
  publisher = {Oxford University Press (OUP)},
}

@Article{Reddy2023a,
  author    = {Reddy, Naveen A. and Sanders, Ryan L. and Shapley, Alice E. and Topping, Michael W. and Kriek, Mariska and Coil, Alison L. and Mobasher, Bahram and Siana, Brian and Rezaee, Saeed},
  journal   = {ApJ},
  title     = {The Impact of Star-formation-rate Surface Density on the Electron Density and Ionization Parameter of High-redshift Galaxies*},
  year      = {2023},
  issn      = {1538-4357},
  month     = jul,
  number    = {1},
  pages     = {56},
  volume    = {951},
  doi       = {10.3847/1538-4357/acd0b1},
  publisher = {American Astronomical Society},
}

@InProceedings{Pineda2023,
  author        = {Pineda, J.~E. and Arzoumanian, D. and Andre, P. and Friesen, R.~K. and Zavagno, A. and Clarke, S.~D. and Inoue, T. and Chen, C. and Lee, Y. and Soler, J.~D. and Kuffmeier, M.},
  booktitle     = {Protostars and Planets VII},
  title         = {From Bubbles and Filaments to Cores and Disks: Gas Gathering and Growth of Structure Leading to the Formation of Stellar Systems},
  year          = {2023},
  editor        = {{Inutsuka}, S. and {Aikawa}, Y. and {Muto}, T. and {Tomida}, K. and {Tamura}, M.},
  month         = jul,
  pages         = {233},
  series        = {Astronomical Society of the Pacific Conference Series},
  volume        = {534},
  archiveprefix = {arXiv},
  doi           = {10.48550/arXiv.2205.03935},
  eprint        = {2205.03935},
  primaryclass  = {astro-ph.GA},
  url           = {https://ui.adsabs.harvard.edu/abs/2023ASPC..534..233P},
}

@Article{Lee2023,
  author        = {Lee, Janice C. and Sandstrom, Karin M. and Leroy, Adam K. and Thilker, David A. and Schinnerer, Eva and Rosolowsky, Erik and Larson, Kirsten L. and Egorov, Oleg V. and Williams, Thomas G. and Schmidt, Judy and Emsellem, Eric and Anand, Gagandeep S. and Barnes, Ashley T. and Belfiore, Francesco and Bešlić, Ivana and Bigiel, Frank and Blanc, Guillermo A. and Bolatto, Alberto D. and Boquien, Médéric and Brok, Jakob den and Cao, Yixian and Chandar, Rupali and Chastenet, Jérémy and Chevance, Mélanie and Chiang 江, I-Da 宜達 and Congiu, Enrico and Dale, Daniel A. and Deger, Sinan and Eibensteiner, Cosima and Faesi, Christopher M. and Glover, Simon C. O. and Grasha, Kathryn and Groves, Brent and Hassani, Hamid and Henny, Kiana F. and Henshaw, Jonathan D. and Hoyer, Nils and Hughes, Annie and Jeffreson, Sarah and Jiménez-Donaire, María J. and Kim, Jaeyeon and Kim, Hwihyun and Klessen, Ralf S. and Koch, Eric W. and Kreckel, Kathryn and Kruijssen, J. M. Diederik and Li, Jing and Liu, Daizhong and Lopez, Laura A. and Maschmann, Daniel and Chen, Ness Mayker and Meidt, Sharon E. and Murphy, Eric J. and Neumann, Justus and Neumayer, Nadine and Pan, Hsi-An and Pessa, Ismael and Pety, Jérôme and Querejeta, Miguel and Pinna, Francesca and Rodríguez, M. Jimena and Saito, Toshiki and Sánchez-Blázquez, Patricia and Santoro, Francesco and Sardone, Amy and Smith, Rowan J. and Sormani, Mattia C. and Scheuermann, Fabian and Stuber, Sophia K. and Sutter, Jessica and Sun 孙, Jiayi 嘉懿 and Teng, Yu-Hsuan and Treß, Robin G. and Usero, Antonio and Watkins, Elizabeth J. and Whitmore, Bradley C. and Razza, Alessandro},
  journal   = {ApJL},
  title         = {The PHANGS–JWST Treasury Survey: Star Formation, Feedback, and Dust Physics at High Angular Resolution in Nearby GalaxieS},
  year          = {2023},
  issn          = {2041-8213},
  month         = feb,
  number        = {2},
  pages         = {L17},
  volume        = {944},
  archiveprefix = {arXiv},
  doi           = {10.3847/2041-8213/acaaae},
  eid           = {L17},
  eprint        = {2212.02667},
  primaryclass  = {astro-ph.GA},
  publisher     = {American Astronomical Society},
  url           = {https://ui.adsabs.harvard.edu/abs/2023ApJ...944L..17L},
}

@Article{Kim2023,
  author    = {Kim, Jaeyeon and Chevance, Mélanie and Kruijssen, J. M. Diederik and Barnes, Ashley. T. and Bigiel, Frank and Blanc, Guillermo A. and Boquien, Médéric and Cao, Yixian and Congiu, Enrico and Dale, Daniel A. and Egorov, Oleg V. and Faesi, Christopher M. and Glover, Simon C. O. and Grasha, Kathryn and Groves, Brent and Hassani, Hamid and Hughes, Annie and Klessen, Ralf S. and Kreckel, Kathryn and Larson, Kirsten L. and Lee, Janice C. and Leroy, Adam K. and Liu, Daizhong and Longmore, Steven N. and Meidt, Sharon E. and Pan, Hsi-An and Pety, Jérôme and Querejeta, Miguel and Rosolowsky, Erik and Saito, Toshiki and Sandstrom, Karin and Schinnerer, Eva and Smith, Rowan J. and Usero, Antonio and Watkins, Elizabeth J. and Williams, Thomas G.},
  journal   = {ApJL},
  title     = {PHANGS–JWST First Results: Duration of the Early Phase of Massive Star Formation in NGC 628},
  year      = {2023},
  issn      = {2041-8213},
  month     = feb,
  number    = {2},
  pages     = {L20},
  volume    = {944},
  doi       = {10.3847/2041-8213/aca90a},
  publisher = {American Astronomical Society},
}

@Article{Hrodmarsson2023,
  author  = {Hrodmarsson, Helgi Rafn and Bouwman, Jordy and Tielens, Alexander G.~G.~M. and Linnartz, Harold},
  journal = {International Journal of Mass Spectrometry},
  title   = {Fragmentation of the PAH cations of Isoviolanthrene and Dicoronylene: A case made for interstellar cyclo[n]carbons as products of universal fragmentation processes},
  year    = {2023},
  month   = mar,
  pages   = {116996},
  volume  = {485},
  doi     = {10.1016/j.ijms.2022.116996},
  url     = {https://ui.adsabs.harvard.edu/abs/2023IJMSp.48516996H},
}

@Article{Groves2023,
  author    = {Groves, B and Kreckel, K and Santoro, F and Belfiore, F and Zavodnik, E and Congiu, E and Egorov, O V and Emsellem, E and Grasha, K and Leroy, A and Scheuermann, F and Schinnerer, E and Watkins, E J and Barnes, A T and Bigiel, F and Dale, D A and Glover, S C O and Pessa, I and Sanchez-Blazquez, P and Williams, T G},
  journal   = {MNRAS},
  title     = {The PHANGS–MUSE nebular catalogue},
  year      = {2023},
  issn      = {1365-2966},
  month     = jan,
  number    = {4},
  pages     = {4902--4952},
  volume    = {520},
  doi       = {10.1093/mnras/stad114},
  publisher = {Oxford University Press (OUP)},
}

@Article{Egorov2023b,
  author        = {Egorov, Oleg V. and Kreckel, Kathryn and Sandstrom, Karin M. and Leroy, Adam K. and Glover, Simon C. O. and Groves, Brent and Kruijssen, J. M. Diederik and Barnes, Ashley. T. and Belfiore, Francesco and Bigiel, F. and Blanc, Guillermo A. and Boquien, Médéric and Cao, Yixian and Chastenet, Jérémy and Chevance, Mélanie and Congiu, Enrico and Dale, Daniel A. and Emsellem, Eric and Grasha, Kathryn and Klessen, Ralf S. and Larson, Kirsten L. and Liu, Daizhong and Murphy, Eric J. and Pan, Hsi-An and Pessa, Ismael and Pety, Jérôme and Rosolowsky, Erik and Scheuermann, Fabian and Schinnerer, Eva and Sutter, Jessica and Thilker, David A. and Watkins, Elizabeth J. and Williams, Thomas G.},
  journal   = {ApJL},
  title         = {PHANGS–JWST First Results: Destruction of the PAH Molecules in {HII} Regions Probed by JWST and MUSE},
  year          = {2023},
  issn          = {2041-8213},
  month         = feb,
  number        = {2},
  pages         = {L16},
  volume        = {944},
  archiveprefix = {arXiv},
  doi           = {10.3847/2041-8213/acac92},
  eid           = {L16},
  eprint        = {2212.09159},
  primaryclass  = {astro-ph.GA},
  publisher     = {American Astronomical Society},
  url           = {https://ui.adsabs.harvard.edu/abs/2023ApJ...944L..16E},
}

@Article{Egorov2023a,
  author    = {Egorov, Oleg V. and Kreckel, Kathryn and Glover, Simon C. O. and Groves, Brent and Belfiore, Francesco and Emsellem, Eric and Klessen, Ralf S. and Leroy, Adam K. and Meidt, Sharon E. and Sarbadhicary, Sumit K. and Schinnerer, Eva and Watkins, Elizabeth J. and Whitmore, Brad C. and Barnes, Ashley T. and Congiu, Enrico and Dale, Daniel A. and Grasha, Kathryn and Larson, Kirsten L. and Lee, Janice C. and Méndez-Delgado, J. Eduardo and Thilker, David A. and Williams, Thomas G.},
  journal   = {A\&A},
  title     = {Quantifying the energy balance between the turbulent ionised gas and young stars},
  year      = {2023},
  issn      = {1432-0746},
  month     = oct,
  pages     = {A153},
  volume    = {678},
  doi       = {10.1051/0004-6361/202346919},
  publisher = {EDP Sciences},
}

@Article{Easeman2023,
  author    = {Easeman, Bethan and Schady, Patricia and Wuyts, Stijn and Yates, Robert M},
  journal   = {MNRAS},
  title     = {Optimal metallicity diagnostics for MUSE observations of low-z galaxies},
  year      = {2023},
  issn      = {1365-2966},
  month     = nov,
  number    = {3},
  pages     = {5484--5502},
  volume    = {527},
  doi       = {10.1093/mnras/stad3464},
  publisher = {Oxford University Press (OUP)},
}

@Article{Congiu2023,
  author    = {Congiu, Enrico and Blanc, Guillermo A. and Belfiore, Francesco and Santoro, Francesco and Scheuermann, Fabian and Kreckel, Kathryn and Emsellem, Eric and Groves, Brent and Pan, Hsi-An and Bigiel, Frank and Dale, Daniel A. and Glover, Simon C. O. and Grasha, Kathryn and Egorov, Oleg V. and Leroy, Adam and Schinnerer, Eva and Watkins, Elizabeth J. and Williams, Thomas G.},
  journal   = {A\&A},
  title     = {PHANGS-MUSE: Detection and Bayesian classification of ~40 000 ionised nebulae in nearby spiral galaxies},
  year      = {2023},
  issn      = {1432-0746},
  month     = apr,
  pages     = {A148},
  volume    = {672},
  doi       = {10.1051/0004-6361/202245153},
  publisher = {EDP Sciences},
}

@Article{Chastenet2023a,
  author        = {Chastenet, Jérémy and Sutter, Jessica and Sandstrom, Karin and Belfiore, Francesco and Egorov, Oleg V. and Larson, Kirsten L. and Leroy, Adam K. and Liu, Daizhong and Rosolowsky, Erik and Thilker, David A. and Watkins, Elizabeth J. and Williams, Thomas G. and Barnes, Ashley. T. and Bigiel, Frank and Boquien, Médéric and Chevance, Mélanie and Chiang 江, I-Da 宜 達 and Dale, Daniel A. and Kruijssen, J. M. Diederik and Emsellem, Eric and Grasha, Kathryn and Groves, Brent and Hassani, Hamid and Hughes, Annie and Kreckel, Kathryn and Meidt, Sharon E. and Rickards Vaught, Ryan J. and Sardone, Amy and Schinnerer, Eva},
  journal   = {ApJL},
  title         = {PHANGS–JWST First Results: Variations in PAH Fraction as a Function of ISM Phase and Metallicity},
  year          = {2023},
  issn          = {2041-8213},
  month         = feb,
  number        = {2},
  pages         = {L11},
  volume        = {944},
  archiveprefix = {arXiv},
  doi           = {10.3847/2041-8213/acadd7},
  eid           = {L11},
  eprint        = {2301.00578},
  primaryclass  = {astro-ph.GA},
  publisher     = {American Astronomical Society},
  url           = {https://ui.adsabs.harvard.edu/abs/2023ApJ...944L..11C},
}

@Article{Cappellari2023,
  author        = {Cappellari, Michele},
  journal       = {\mnras},
  title         = {Full spectrum fitting with photometry in PPXF: stellar population versus dynamical masses, non-parametric star formation history and metallicity for 3200 LEGA-C galaxies at redshift z {\ensuremath{\approx}} 0.8},
  year          = {2023},
  month         = dec,
  number        = {3},
  pages         = {3273-3300},
  volume        = {526},
  archiveprefix = {arXiv},
  doi           = {10.1093/mnras/stad2597},
  eprint        = {2208.14974},
  primaryclass  = {astro-ph.GA},
  url           = {https://ui.adsabs.harvard.edu/abs/2023MNRAS.526.3273C},
}

@Article{Santoro2022,
  author    = {Santoro, Francesco and Kreckel, Kathryn and Belfiore, Francesco and Groves, Brent and Congiu, Enrico and Thilker, David A. and Blanc, Guillermo A. and Schinnerer, Eva and Ho, I-Ting and Diederik Kruijssen, J. M. and Meidt, Sharon and Klessen, Ralf S. and Schruba, Andreas and Querejeta, Miguel and Pessa, Ismael and Chevance, Mélanie and Kim, Jaeyeon and Emsellem, Eric and McElroy, Rebecca and Barnes, Ashley T. and Bigiel, Frank and Boquien, Médéric and Dale, Daniel A. and Glover, Simon C. O. and Grasha, Kathryn and Lee, Janice and Leroy, Adam K. and Pan, Hsi-An and Rosolowsky, Erik and Saito, Toshiki and Sanchez-Blazquez, Patricia and Watkins, Elizabeth J. and Williams, Thomas G.},
  journal   = {A\&A},
  title     = {PHANGS–MUSE: The H II region luminosity function of local star-forming galaxies},
  year      = {2022},
  issn      = {1432-0746},
  month     = feb,
  pages     = {A188},
  volume    = {658},
  doi       = {10.1051/0004-6361/202141907},
  publisher = {EDP Sciences},
}

@Article{Reizer2022,
  author    = {Reizer, Edina and Viskolcz, Béla and Fiser, Béla},
  journal   = {Chemosphere},
  title     = {Formation and growth mechanisms of polycyclic aromatic hydrocarbons: A mini-review},
  year      = {2022},
  issn      = {0045-6535},
  month     = mar,
  pages     = {132793},
  volume    = {291},
  doi       = {10.1016/j.chemosphere.2021.132793},
  publisher = {Elsevier BV},
}

@Article{Lee2022,
  author    = {Lee, Janice C. and Whitmore, Bradley C. and Thilker, David A. and Deger, Sinan and Larson, Kirsten L. and Ubeda, Leonardo and Anand, Gagandeep S. and Boquien, Médéric and Chandar, Rupali and Dale, Daniel A. and Emsellem, Eric and Leroy, Adam K. and Rosolowsky, Erik and Schinnerer, Eva and Schmidt, Judy and Lilly, James and Turner, Jordan and Van Dyk, Schuyler and White, Richard L. and Barnes, Ashley T. and Belfiore, Francesco and Bigiel, Frank and Blanc, Guillermo A. and Cao, Yixian and Chevance, Melanie and Congiu, Enrico and Egorov, Oleg V. and Glover, Simon C. O. and Grasha, Kathryn and Groves, Brent and Henshaw, Jonathan D. and Hughes, Annie and Klessen, Ralf S. and Koch, Eric and Kreckel, Kathryn and Kruijssen, J. M. Diederik and Liu, Daizhong and Lopez, Laura A. and Mayker, Ness and Meidt, Sharon E. and Murphy, Eric J. and Pan, Hsi-An and Pety, Jérôme and Querejeta, Miguel and Razza, Alessandro and Saito, Toshiki and Sánchez-Blázquez, Patricia and Santoro, Francesco and Sardone, Amy and Scheuermann, Fabian and Schruba, Andreas and Sun, Jiayi and Usero, Antonio and Watkins, E. and Williams, Thomas G.},
  journal   = {ApJS},
  title     = {The PHANGS-HST Survey: Physics at High Angular Resolution in Nearby Galaxies with the Hubble Space Telescope},
  year      = {2022},
  issn      = {1538-4365},
  month     = jan,
  number    = {1},
  pages     = {10},
  volume    = {258},
  doi       = {10.3847/1538-4365/ac1fe5},
  publisher = {American Astronomical Society},
}

@Article{Herrero2022,
  author    = {Herrero, Víctor J. and Jiménez-Redondo, Miguel and Peláez, Ramón J. and Maté, Belén and Tanarro, Isabel},
  journal   = {Front. Astron. Space Sci.},
  title     = {Structure and evolution of interstellar carbonaceous dust. Insights from the laboratory},
  year      = {2022},
  issn      = {2296-987X},
  month     = dec,
  volume    = {9},
  doi       = {10.3389/fspas.2022.1083288},
  publisher = {Frontiers Media SA},
}

@Article{GarciaBernete2022,
  author        = {Garc{\'\i}a-Bernete, I. and Rigopoulou, D. and Alonso-Herrero, A. and Pereira-Santaella, M. and Roche, P.~F. and Kerkeni, B.},
  journal       = {\mnras},
  title         = {Polycyclic aromatic hydrocarbons in Seyfert and star-forming galaxies},
  year          = {2022},
  month         = jan,
  number        = {3},
  pages         = {4256-4275},
  volume        = {509},
  archiveprefix = {arXiv},
  doi           = {10.1093/mnras/stab3127},
  eprint        = {2011.10882},
  primaryclass  = {astro-ph.GA},
  url           = {https://ui.adsabs.harvard.edu/abs/2022MNRAS.509.4256G},
}

@Article{Emsellem2022,
  author    = {Emsellem, Eric and Schinnerer, Eva and Santoro, Francesco and Belfiore, Francesco and Pessa, Ismael and McElroy, Rebecca and Blanc, Guillermo A. and Congiu, Enrico and Groves, Brent and Ho, I-Ting and Kreckel, Kathryn and Razza, Alessandro and Sanchez-Blazquez, Patricia and Egorov, Oleg and Faesi, Chris and Klessen, Ralf S. and Leroy, Adam K. and Meidt, Sharon and Querejeta, Miguel and Rosolowsky, Erik and Scheuermann, Fabian and Anand, Gagandeep S. and Barnes, Ashley T. and Bešlić, Ivana and Bigiel, Frank and Boquien, Médéric and Cao, Yixian and Chevance, Mélanie and Dale, Daniel A. and Eibensteiner, Cosima and Glover, Simon C. O. and Grasha, Kathryn and Henshaw, Jonathan D. and Hughes, Annie and Koch, Eric W. and Kruijssen, J. M. Diederik and Lee, Janice and Liu, Daizhong and Pan, Hsi-An and Pety, Jérôme and Saito, Toshiki and Sandstrom, Karin M. and Schruba, Andreas and Sun, Jiayi and Thilker, David A. and Usero, Antonio and Watkins, Elizabeth J. and Williams, Thomas G.},
  journal   = {A\&A},
  title     = {The PHANGS-MUSE survey: Probing the chemo-dynamical evolution of disc galaxies},
  year      = {2022},
  issn      = {1432-0746},
  month     = mar,
  pages     = {A191},
  volume    = {659},
  doi       = {10.1051/0004-6361/202141727},
  publisher = {EDP Sciences},
}

@Article{Belfiore2022,
  author        = {Belfiore, F. and Santoro, F. and Groves, B. and Schinnerer, E. and Kreckel, K. and Glover, S. C. O. and Klessen, R. S. and Emsellem, E. and Blanc, G. A. and Congiu, E. and Barnes, A. T. and Boquien, M. and Chevance, M. and Dale, D. A. and Diederik Kruijssen, J. M. and Leroy, A. K. and Pan, H.-A. and Pessa, I. and Schruba, A. and Williams, T. G.},
  journal   = {A\&A},
  title         = {A tale of two DIGs: The relative role of H II regions and low-mass hot evolved stars in powering the diffuse ionised gas (DIG) in PHANGS–MUSE galaxies},
  year          = {2022},
  issn          = {1432-0746},
  month         = mar,
  pages         = {A26},
  volume        = {659},
  archiveprefix = {arXiv},
  doi           = {10.1051/0004-6361/202141859},
  eid           = {A26},
  eprint        = {2111.14876},
  primaryclass  = {astro-ph.GA},
  publisher     = {EDP Sciences},
  url           = {https://ui.adsabs.harvard.edu/abs/2022A&A...659A..26B},
}

@Article{Rigopoulou2021,
  author    = {Rigopoulou, D and Barale, M and Clary, D C and Shan, X and Alonso-Herrero, A and García-Bernete, I and Hunt, L and Kerkeni, B and Pereira-Santaella, M and Roche, P F},
  journal   = {MNRAS},
  title     = {The properties of polycyclic aromatic hydrocarbons in galaxies: constraints on PAH sizes, charge and radiation fields},
  year      = {2021},
  issn      = {1365-2966},
  month     = apr,
  number    = {4},
  pages     = {5287--5300},
  volume    = {504},
  doi       = {10.1093/mnras/stab959},
  publisher = {Oxford University Press (OUP)},
}

@Article{Omont2021,
  author    = {Omont, A. and Bettinger, H. F.},
  journal   = {A\&A},
  title     = {Intermediate-size fullerenes as degradation products of interstellar polycyclic aromatic hydrocarbons},
  year      = {2021},
  issn      = {1432-0746},
  month     = jun,
  pages     = {A193},
  volume    = {650},
  doi       = {10.1051/0004-6361/202140675},
  publisher = {EDP Sciences},
}

@Article{Leroy2021,
  author        = {Leroy, Adam K. and Schinnerer, Eva and Hughes, Annie and Rosolowsky, Erik and Pety, J{\'e}r{\^o}me and Schruba, Andreas and Usero, Antonio and Blanc, Guillermo A. and Chevance, M{\'e}lanie and Emsellem, Eric and Faesi, Christopher M. and Herrera, Cinthya N. and Liu, Daizhong and Meidt, Sharon E. and Querejeta, Miguel and Saito, Toshiki and Sandstrom, Karin M. and Sun, Jiayi and Williams, Thomas G. and Anand, Gagandeep S. and Barnes, Ashley T. and Behrens, Erica A. and Belfiore, Francesco and Benincasa, Samantha M. and Be{\v{s}}li{\'c}, Ivana and Bigiel, Frank and Bolatto, Alberto D. and den Brok, Jakob S. and Cao, Yixian and Chandar, Rupali and Chastenet, J{\'e}r{\'e}my and Chiang, I.-Da and Congiu, Enrico and Dale, Daniel A. and Deger, Sinan and Eibensteiner, Cosima and Egorov, Oleg V. and Garc{\'\i}a-Rodr{\'\i}guez, Axel and Glover, Simon C.~O. and Grasha, Kathryn and Henshaw, Jonathan D. and Ho, I. Ting and Kepley, Amanda A. and Kim, Jaeyeon and Klessen, Ralf S. and Kreckel, Kathryn and Koch, Eric W. and Kruijssen, J.~M. Diederik and Larson, Kirsten L. and Lee, Janice C. and Lopez, Laura A. and Machado, Josh and Mayker, Ness and McElroy, Rebecca and Murphy, Eric J. and Ostriker, Eve C. and Pan, Hsi-An and Pessa, Ismael and Puschnig, Johannes and Razza, Alessandro and S{\'a}nchez-Bl{\'a}zquez, Patricia and Santoro, Francesco and Sardone, Amy and Scheuermann, Fabian and Sliwa, Kazimierz and Sormani, Mattia C. and Stuber, Sophia K. and Thilker, David A. and Turner, Jordan A. and Utomo, Dyas and Watkins, Elizabeth J. and Whitmore, Bradley},
  journal       = {\apjs},
  title         = {PHANGS-ALMA: Arcsecond CO(2-1) Imaging of Nearby Star-forming Galaxies},
  year          = {2021},
  issn          = {1538-4365},
  month         = dec,
  number        = {2},
  pages         = {43},
  volume        = {257},
  archiveprefix = {arXiv},
  doi           = {10.3847/1538-4365/ac17f3},
  eid           = {43},
  eprint        = {2104.07739},
  primaryclass  = {astro-ph.GA},
  publisher     = {American Astronomical Society},
  url           = {https://ui.adsabs.harvard.edu/abs/2021ApJS..257...43L},
}

@Article{Draine2021,
  author    = {Draine, B. T. and Li, Aigen and Hensley, Brandon S. and Hunt, L. K. and Sandstrom, K. and Smith, J.-D. T.},
  journal   = {ApJ},
  title     = {Excitation of Polycyclic Aromatic Hydrocarbon Emission: Dependence on Size Distribution, Ionization, and Starlight Spectrum and Intensity},
  year      = {2021},
  issn      = {1538-4357},
  month     = aug,
  number    = {1},
  pages     = {3},
  volume    = {917},
  doi       = {10.3847/1538-4357/abff51},
  publisher = {American Astronomical Society},
}

@Article{Anand2021,
  author        = {Anand, Gagandeep S. and Lee, Janice C. and Van Dyk, Schuyler D. and Leroy, Adam K. and Rosolowsky, Erik and Schinnerer, Eva and Larson, Kirsten and Kourkchi, Ehsan and Kreckel, Kathryn and Scheuermann, Fabian and Rizzi, Luca and Thilker, David and Tully, R. Brent and Bigiel, Frank and Blanc, Guillermo A. and Boquien, M{\'e}d{\'e}ric and Chandar, Rupali and Dale, Daniel and Emsellem, Eric and Deger, Sinan and Glover, Simon C.~O. and Grasha, Kathryn and Groves, Brent and S. Klessen, Ralf and Kruijssen, J.~M. Diederik and Querejeta, Miguel and S{\'a}nchez-Bl{\'a}zquez, Patricia and Schruba, Andreas and Turner, Jordan and Ubeda, Leonardo and Williams, Thomas G. and Whitmore, Brad},
  journal       = {\mnras},
  title         = {Distances to PHANGS galaxies: New tip of the red giant branch measurements and adopted distances},
  year          = {2021},
  month         = mar,
  number        = {3},
  pages         = {3621-3639},
  volume        = {501},
  archiveprefix = {arXiv},
  doi           = {10.1093/mnras/staa3668},
  eprint        = {2012.00757},
  primaryclass  = {astro-ph.GA},
  url           = {https://ui.adsabs.harvard.edu/abs/2021MNRAS.501.3621A},
}

@Article{Virtanen2020,
  author    = {Virtanen, Pauli and Gommers, Ralf and Oliphant, Travis E. and Haberland, Matt and Reddy, Tyler and Cournapeau, David and Burovski, Evgeni and Peterson, Pearu and Weckesser, Warren and Bright, Jonathan and van der Walt, Stéfan J. and Brett, Matthew and Wilson, Joshua and Millman, K. Jarrod and Mayorov, Nikolay and Nelson, Andrew R. J. and Jones, Eric and Kern, Robert and Larson, Eric and Carey, C J and Polat, İlhan and Feng, Yu and Moore, Eric W. and VanderPlas, Jake and Laxalde, Denis and Perktold, Josef and Cimrman, Robert and Henriksen, Ian and Quintero, E. A. and Harris, Charles R. and Archibald, Anne M. and Ribeiro, Antônio H. and Pedregosa, Fabian and van Mulbregt, Paul and Vijaykumar, Aditya and Bardelli, Alessandro Pietro and Rothberg, Alex and Hilboll, Andreas and Kloeckner, Andreas and Scopatz, Anthony and Lee, Antony and Rokem, Ariel and Woods, C. Nathan and Fulton, Chad and Masson, Charles and Häggström, Christian and Fitzgerald, Clark and Nicholson, David A. and Hagen, David R. and Pasechnik, Dmitrii V. and Olivetti, Emanuele and Martin, Eric and Wieser, Eric and Silva, Fabrice and Lenders, Felix and Wilhelm, Florian and Young, G. and Price, Gavin A. and Ingold, Gert-Ludwig and Allen, Gregory E. and Lee, Gregory R. and Audren, Hervé and Probst, Irvin and Dietrich, Jörg P. and Silterra, Jacob and Webber, James T and Slavič, Janko and Nothman, Joel and Buchner, Johannes and Kulick, Johannes and Schönberger, Johannes L. and de Miranda Cardoso, José Vinícius and Reimer, Joscha and Harrington, Joseph and Rodríguez, Juan Luis Cano and Nunez-Iglesias, Juan and Kuczynski, Justin and Tritz, Kevin and Thoma, Martin and Newville, Matthew and Kümmerer, Matthias and Bolingbroke, Maximilian and Tartre, Michael and Pak, Mikhail and Smith, Nathaniel J. and Nowaczyk, Nikolai and Shebanov, Nikolay and Pavlyk, Oleksandr and Brodtkorb, Per A. and Lee, Perry and McGibbon, Robert T. and Feldbauer, Roman and Lewis, Sam and Tygier, Sam and Sievert, Scott and Vigna, Sebastiano and Peterson, Stefan and More, Surhud and Pudlik, Tadeusz and Oshima, Takuya and Pingel, Thomas J. and Robitaille, Thomas P. and Spura, Thomas and Jones, Thouis R. and Cera, Tim and Leslie, Tim and Zito, Tiziano and Krauss, Tom and Upadhyay, Utkarsh and Halchenko, Yaroslav O. and Vázquez-Baeza, Yoshiki},
  journal   = {Nat. Methods},
  title     = {SciPy 1.0: fundamental algorithms for scientific computing in Python},
  year      = {2020},
  issn      = {1548-7105},
  month     = feb,
  number    = {3},
  pages     = {261--272},
  volume    = {17},
  doi       = {10.1038/s41592-019-0686-2},
  publisher = {Springer Science and Business Media LLC},
}

@Article{Vazdekis2020,
  author        = {Vazdekis, A. and Cervi{\~n}o, M. and Montes, M. and Mart{\'\i}n-Navarro, I. and Beasley, M.~A.},
  journal       = {\mnras},
  title         = {Surface brightness fluctuation spectra to constrain stellar population properties},
  year          = {2020},
  month         = apr,
  number        = {4},
  pages         = {5131-5152},
  volume        = {493},
  archiveprefix = {arXiv},
  doi           = {10.1093/mnras/staa629},
  eprint        = {2003.02563},
  primaryclass  = {astro-ph.GA},
  url           = {https://ui.adsabs.harvard.edu/abs/2020MNRAS.493.5131V},
}

@Article{Mingozzi2020,
  author    = {Mingozzi, M. and Belfiore, F. and Cresci, G. and Bundy, K. and Bershady, M. and Bizyaev, D. and Blanc, G. and Boquien, M. and Drory, N. and Fu, H. and Maiolino, R. and Riffel, R. and Schaefer, A. and Storchi-Bergmann, T. and Telles, E. and Tremonti, C. and Zakamska, N. and Zhang, K.},
  journal   = {A\&A},
  title     = {SDSS IV MaNGA: Metallicity and ionisation parameter in local star-forming galaxies from Bayesian fitting to photoionisation models},
  year      = {2020},
  issn      = {1432-0746},
  month     = apr,
  pages     = {A42},
  volume    = {636},
  doi       = {10.1051/0004-6361/201937203},
  publisher = {EDP Sciences},
}

@Article{Lang2020,
  author        = {Lang, Philipp and Meidt, Sharon E. and Rosolowsky, Erik and Nofech, Joseph and Schinnerer, Eva and Leroy, Adam K. and Emsellem, Eric and Pessa, Ismael and Glover, Simon C.~O. and Groves, Brent and Hughes, Annie and Kruijssen, J.~M. Diederik and Querejeta, Miguel and Schruba, Andreas and Bigiel, Frank and Blanc, Guillermo A. and Chevance, M{\'e}lanie and Colombo, Dario and Faesi, Christopher and Henshaw, Jonathan D. and Herrera, Cinthya N. and Liu, Daizhong and Pety, J{\'e}r{\^o}me and Puschnig, Johannes and Saito, Toshiki and Sun, Jiayi and Usero, Antonio},
  journal       = {\apj},
  title         = {PHANGS CO Kinematics: Disk Orientations and Rotation Curves at 150 pc Resolution},
  year          = {2020},
  issn          = {1538-4357},
  month         = jul,
  number        = {2},
  pages         = {122},
  volume        = {897},
  archiveprefix = {arXiv},
  doi           = {10.3847/1538-4357/ab9953},
  eid           = {122},
  eprint        = {2005.11709},
  primaryclass  = {astro-ph.GA},
  publisher     = {American Astronomical Society},
  url           = {https://ui.adsabs.harvard.edu/abs/2020ApJ...897..122L},
}

@Article{Kumar2020,
  author    = {Kumar, M. S. N. and Palmeirim, P. and Arzoumanian, D. and Inutsuka, S. I.},
  journal   = {A\&A},
  title     = {Unifying low- and high-mass star formation through density-amplified hubs of filaments: The highest mass stars ($>$100M⊙) form only in hubs},
  year      = {2020},
  issn      = {1432-0746},
  month     = oct,
  pages     = {A87},
  volume    = {642},
  doi       = {10.1051/0004-6361/202038232},
  publisher = {EDP Sciences},
}

@Article{DellaBruna2020,
  author        = {Della Bruna, Lorenza and Adamo, Angela and Bik, Arjan and Fumagalli, Michele and Walterbos, Rene and Östlin, Göran and Bruzual, Gustavo and Calzetti, Daniela and Charlot, Stephane and Grasha, Kathryn and Smith, Linda J. and Thilker, David and Wofford, Aida},
  journal   = {A\&A},
  title         = {Studying the ISM at ∼10 pc scale in NGC 7793 with MUSE: I. Data description and properties of the ionised gas},
  year          = {2020},
  issn          = {1432-0746},
  month         = mar,
  pages         = {A134},
  volume        = {635},
  archiveprefix = {arXiv},
  doi           = {10.1051/0004-6361/201937173},
  eid           = {A134},
  eprint        = {2002.08966},
  primaryclass  = {astro-ph.GA},
  publisher     = {EDP Sciences},
  url           = {https://ui.adsabs.harvard.edu/abs/2020A&A...635A.134D},
}

@Article{Spangler2019,
  author        = {Spangler, Steven R. and Bergerud, Brandon M. and Beauchamp, Kara M.},
  journal       = {arXiv e-prints},
  title         = {Analytic Estimates of the Effect of Plasma Density Fluctuations on HII Region Density Diagnostics},
  year          = {2019},
  month         = oct,
  pages         = {arXiv:1910.08466},
  archiveprefix = {arXiv},
  doi           = {10.48550/arXiv.1910.08466},
  eid           = {arXiv:1910.08466},
  eprint        = {1910.08466},
  primaryclass  = {astro-ph.SR},
  url           = {https://ui.adsabs.harvard.edu/abs/2019arXiv191008466S},
}

@Article{RousseauNepton2019,
  author        = {Rousseau-Nepton, L. and Martin, R.~P. and Robert, C. and Drissen, L. and Amram, P. and Prunet, S. and Martin, T. and Moumen, I. and Adamo, A. and Alarie, A. and Barmby, P. and Boselli, A. and Bresolin, F. and Bureau, M. and Chemin, L. and Fernandes, R.~C. and Combes, F. and Crowder, C. and Della Bruna, L. and Duarte Puertas, S. and Egusa, F. and Epinat, B. and Ksoll, V.~F. and Girard, M. and G{\'o}mez Llanos, V. and Gouliermis, D. and Grasha, K. and Higgs, C. and Hlavacek-Larrondo, J. and Ho, I. T. and Iglesias-P{\'a}ramo, J. and Joncas, G. and Kam, Z.~S. and Karera, P. and Kennicutt, R.~C. and Klessen, R.~S. and Lianou, S. and Liu, L. and Liu, Q. and de Amorim, A. Luiz and Lyman, J.~D. and Martel, H. and Mazzilli-Ciraulo, B. and McLeod, A.~F. and Melchior, A. L. and Millan, I. and Moll{\'a}, M. and Momose, R. and Morisset, C. and Pan, H. A. and Pati, A.~K. and Pellerin, A. and Pellegrini, E. and P{\'e}rez, I. and Petric, A. and Plana, H. and Rahner, D. and Ruiz Lara, T. and S{\'a}nchez-Menguiano, L. and Spekkens, K. and Stasi{\'n}ska, G. and Takamiya, M. and Vale Asari, N. and V{\'\i}lchez, J.~M.},
  journal       = {\mnras},
  title         = {SIGNALS: I. Survey description},
  year          = {2019},
  month         = nov,
  number        = {4},
  pages         = {5530-5546},
  volume        = {489},
  archiveprefix = {arXiv},
  doi           = {10.1093/mnras/stz2455},
  eprint        = {1908.09017},
  primaryclass  = {astro-ph.GA},
  url           = {https://ui.adsabs.harvard.edu/abs/2019MNRAS.489.5530R},
}

@Article{Points2019,
  author    = {Points, Sean D. and Long, Knox S. and Winkler, P. Frank and Blair, William P.},
  journal   = {ApJ},
  title     = {Kinematics: A Clean Diagnostic for Separating Supernova Remnants from H ii Regions in Nearby Galaxies},
  year      = {2019},
  issn      = {1538-4357},
  month     = dec,
  number    = {1},
  pages     = {66},
  volume    = {887},
  doi       = {10.3847/1538-4357/ab4e98},
  publisher = {American Astronomical Society},
}

@Article{Moumen2019,
  author        = {Moumen, I. and Robert, C. and Devost, D. and Martin, R.~P. and Rousseau-Nepton, L. and Drissen, L. and Martin, T.},
  journal       = {\mnras},
  title         = {3D optical spectroscopic study of NGC 3344 with SITELLE: I. Identification and confirmation of supernova remnants},
  year          = {2019},
  month         = sep,
  number        = {1},
  pages         = {803-829},
  volume        = {488},
  archiveprefix = {arXiv},
  doi           = {10.1093/mnras/stz1734},
  eprint        = {1906.10021},
  primaryclass  = {astro-ph.GA},
  url           = {https://ui.adsabs.harvard.edu/abs/2019MNRAS.488..803M},
}

@Article{Drissen2019,
  author    = {Drissen, Laurent and Martin, Thomas and Rousseau-Nepton, Laurie and Robert, Carmelle and Martin, R Pierre and Baril, Marc and Prunet, Simon and Joncas, Gilles and Thibault, Simon and Brousseau, Denis and Mandar, Julie and Grandmont, Frédéric and Yee, Howard and Simard, Luc},
  journal   = {MNRAS},
  title     = {SITELLE: an Imaging Fourier Transform Spectrometer for the Canada–France–Hawaii Telescope},
  year      = {2019},
  issn      = {1365-2966},
  month     = mar,
  number    = {3},
  pages     = {3930--3946},
  volume    = {485},
  doi       = {10.1093/mnras/stz627},
  publisher = {Oxford University Press (OUP)},
}

@Misc{Bradley2019,
  author    = {Bradley, Larry and Sipocz, Brigitta and Robitaille, Thomas and Tollerud, Erik and {Zé Vinícius} and Deil, Christoph and Barbary, Kyle and Günther, Hans Moritz and Cara, Mihai and Busko, Ivo and Conseil, Simon and Droettboom, Michael and {Azalee Bostroem} and {E. M. Bray} and Bratholm, Lars Andersen and Wilson, Tom and Craig, Matt and Barentsen, Geert and Pascual, Sergio and Donath, Axel and Greco, Johnny and Perren, Gabriel and {P. L. Lim} and Kerzendorf, Wolfgang},
  title     = {astropy/photutils: v0.6},
  year      = {2019},
  copyright = {Open Access},
  doi       = {10.5281/ZENODO.2533376},
  publisher = {Zenodo},
}

@Article{Bergerud2019,
  author    = {Bergerud, Brandon M and Spangler, Steven R and Beauchamp, Kara M},
  journal   = {MNRAS},
  title     = {Realistic models for filling and abundance discrepancy factors in photoionized nebulae},
  year      = {2019},
  issn      = {1365-2966},
  month     = dec,
  number    = {1},
  pages     = {1142--1153},
  volume    = {492},
  doi       = {10.1093/mnras/stz3515},
  publisher = {Oxford University Press (OUP)},
}

@Article{RousseauNepton2018,
  author        = {Rousseau-Nepton, L and Robert, C and Martin, R P and Drissen, L and Martin, T},
  journal   = {MNRAS},
  title         = {NGC628 with SITELLE: I. Imaging spectroscopy of 4285 {HII} region candidates},
  year          = {2018},
  issn          = {1365-2966},
  month         = feb,
  number        = {3},
  pages         = {4152--4186},
  volume        = {477},
  archiveprefix = {arXiv},
  doi           = {10.1093/mnras/sty477},
  eprint        = {1704.05121},
  primaryclass  = {astro-ph.GA},
  publisher     = {Oxford University Press (OUP)},
  url           = {https://ui.adsabs.harvard.edu/abs/2018MNRAS.477.4152R},
}

@Article{Ragan2018,
  author    = {Ragan, S E and Moore, T J T and Eden, D J and Hoare, M G and Urquhart, J S and Elia, D and Molinari, S},
  journal   = {MNRAS},
  title     = {The role of spiral arms in Milky Way star formation},
  year      = {2018},
  issn      = {1365-2966},
  month     = jun,
  number    = {2},
  pages     = {2361--2373},
  volume    = {479},
  doi       = {10.1093/mnras/sty1672},
  publisher = {Oxford University Press (OUP)},
}

@Article{Galliano2018,
  author    = {Galliano, Frédéric and Galametz, Maud and Jones, Anthony P.},
  journal   = {ARA\&A},
  title     = {The Interstellar Dust Properties of Nearby Galaxies},
  year      = {2018},
  issn      = {1545-4282},
  month     = sep,
  number    = {1},
  pages     = {673--713},
  volume    = {56},
  doi       = {10.1146/annurev-astro-081817-051900},
  publisher = {Annual Reviews},
}

@Article{Winkler2017,
  author    = {Winkler, P. Frank and Blair, William P. and Long, Knox S.},
  journal   = {ApJ},
  title     = {A Spectroscopic Study of the Rich Supernova Remnant Population in M83},
  year      = {2017},
  issn      = {1538-4357},
  month     = apr,
  number    = {2},
  pages     = {83},
  volume    = {839},
  doi       = {10.3847/1538-4357/aa683d},
  publisher = {American Astronomical Society},
}

@Article{Oey2017,
  author    = {Oey, M. S. and López-Hernández, J. and Kellar, J. A. and Pellegrini, E. W. and Gordon, K. D. and Jameson, K. E. and Li, A. and Madden, S. C. and Meixner, M. and Roman-Duval, J. and Bot, C. and Rubio, M. and Tielens, A. G. G. M.},
  journal   = {ApJ},
  title     = {Dust Emission at 8 and 24 μm as Diagnostics of {HII} Region Radiative Transfer},
  year      = {2017},
  issn      = {1538-4357},
  month     = jul,
  number    = {1},
  pages     = {63},
  volume    = {844},
  doi       = {10.3847/1538-4357/aa73d3},
  publisher = {American Astronomical Society},
}

@Article{Bacon2017,
  author        = {Bacon, Roland and Conseil, Simon and Mary, David and Brinchmann, Jarle and Shepherd, Martin and Akhlaghi, Mohammad and Weilbacher, Peter M. and Piqueras, Laure and Wisotzki, Lutz and Lagattuta, David and Epinat, Benoit and Guerou, Adrien and Inami, Hanae and Cantalupo, Sebastiano and Courbot, Jean Baptiste and Contini, Thierry and Richard, Johan and Maseda, Michael and Bouwens, Rychard and Bouché, Nicolas and Kollatschny, Wolfram and Schaye, Joop and Marino, Raffaella Anna and Pello, Roser and Herenz, Christian and Guiderdoni, Bruno and Carollo, Marcella},
  journal   = {A\&A},
  title         = {The MUSE Hubble Ultra Deep Field Survey: I. Survey description, data reduction, and source detection},
  year          = {2017},
  issn          = {1432-0746},
  month         = nov,
  pages         = {A1},
  volume        = {608},
  archiveprefix = {arXiv},
  doi           = {10.1051/0004-6361/201730833},
  eid           = {A1},
  eprint        = {1710.03002},
  primaryclass  = {astro-ph.GA},
  publisher     = {EDP Sciences},
  url           = {https://ui.adsabs.harvard.edu/abs/2017A&A...608A...1B},
}

@Article{Shipley2016,
  author    = {Shipley, Heath V. and Papovich, Casey and Rieke, George H. and Brown, Michael J. I. and Moustakas, John},
  journal   = {ApJ},
  title     = {A NEW STAR FORMATION RATE CALIBRATION FROM POLYCYCLIC AROMATIC HYDROCARBON EMISSION FEATURES AND APPLICATION TO HIGH-REDSHIFT GALAXIES},
  year      = {2016},
  issn      = {1538-4357},
  month     = feb,
  number    = {1},
  pages     = {60},
  volume    = {818},
  doi       = {10.3847/0004-637x/818/1/60},
  publisher = {American Astronomical Society},
}

@Article{MarmolQueralto2016,
  author    = {Mármol-Queraltó, E. and McLure, R. J. and Cullen, F. and Dunlop, J. S. and Fontana, A. and McLeod, D. J.},
  journal   = {MNRAS},
  title     = {The evolution of the equivalent width of the Hα emission line and specific star formation rate in star-forming galaxies at 1 $<$ z $<$ 5},
  year      = {2016},
  issn      = {1365-2966},
  month     = may,
  number    = {4},
  pages     = {3587--3597},
  volume    = {460},
  doi       = {10.1093/mnras/stw1212},
  publisher = {Oxford University Press (OUP)},
}

@Article{Smith2015,
  author    = {Smith, Rowan J. and Glover, Simon C. O. and Klessen, Ralf S. and Fuller, Gary A.},
  journal   = {MNRAS},
  title     = {On the nature of star-forming filaments – II. Subfilaments and velocities},
  year      = {2015},
  issn      = {1365-2966},
  month     = dec,
  number    = {4},
  pages     = {3640--3655},
  volume    = {455},
  doi       = {10.1093/mnras/stv2559},
  publisher = {Oxford University Press (OUP)},
}

@InProceedings{Martin2015,
  author    = {Martin, T. and Drissen, L. and Joncas, G.},
  booktitle = {Astronomical Data Analysis Software an Systems XXIV (ADASS XXIV)},
  title     = {ORBS, ORCS, OACS, a Software Suite for Data Reduction and Analysis of the Hyperspectral Imagers SITELLE and SpIOMM},
  year      = {2015},
  editor    = {{Taylor}, A.~R. and {Rosolowsky}, E.},
  month     = sep,
  pages     = {327},
  series    = {Astronomical Society of the Pacific Conference Series},
  volume    = {495},
  url       = {https://ui.adsabs.harvard.edu/abs/2015ASPC..495..327M},
}

@Article{Kewley2015,
  author    = {Kewley, Lisa J. and Zahid, H. Jabran and Geller, Margaret J. and Dopita, Michael A. and Hwang, Ho Seong and Fabricant, Dan},
  journal   = {ApJ},
  title     = {A RISE IN THE IONIZING PHOTONS IN STAR-FORMING GALAXIES OVER THE PAST 8 BILLION YEARS},
  year      = {2015},
  issn      = {2041-8213},
  month     = oct,
  number    = {2},
  pages     = {L20},
  volume    = {812},
  doi       = {10.1088/2041-8205/812/2/l20},
  publisher = {American Astronomical Society},
}

@Article{Walt2014,
  author    = {van der Walt, Stéfan and Schönberger, Johannes L. and Nunez-Iglesias, Juan and Boulogne, François and Warner, Joshua D. and Yager, Neil and Gouillart, Emmanuelle and Yu, Tony},
  journal   = {PeerJ},
  title     = {scikit-image: image processing in Python},
  year      = {2014},
  issn      = {2167-8359},
  month     = jun,
  pages     = {e453},
  volume    = {2},
  doi       = {10.7717/peerj.453},
  publisher = {PeerJ},
}

@Article{Seok2014,
  author    = {Seok, Ji Yeon and Hirashita, Hiroyuki and Asano, Ryosuke S.},
  journal   = {MNRAS},
  title     = {Formation history of polycyclic aromatic hydrocarbons in galaxies},
  year      = {2014},
  issn      = {1365-2966},
  month     = feb,
  number    = {2},
  pages     = {2186--2196},
  volume    = {439},
  doi       = {10.1093/mnras/stu120},
  publisher = {Oxford University Press (OUP)},
}

@Article{Makarov2014,
  author    = {Makarov, Dmitry and Prugniel, Philippe and Terekhova, Nataliya and Courtois, Hélène and Vauglin, Isabelle},
  journal   = {A\&A},
  title     = {HyperLEDA. III. The catalogue of extragalactic distances},
  year      = {2014},
  issn      = {1432-0746},
  month     = oct,
  pages     = {A13},
  volume    = {570},
  doi       = {10.1051/0004-6361/201423496},
  publisher = {EDP Sciences},
}

@Article{Madau2014,
  author    = {Madau, Piero and Dickinson, Mark},
  journal   = {ARA\&A},
  title     = {Cosmic Star-Formation History},
  year      = {2014},
  issn      = {1545-4282},
  month     = aug,
  number    = {1},
  pages     = {415--486},
  volume    = {52},
  doi       = {10.1146/annurev-astro-081811-125615},
  publisher = {Annual Reviews},
}

@InBook{Tielens2013,
  author    = {Tielens, A. G. G. M.},
  pages     = {499--548},
  publisher = {Springer Netherlands},
  title     = {Interstellar PAHs and Dust},
  year      = {2013},
  isbn      = {9789400756120},
  booktitle = {Planets, Stars and Stellar Systems},
  doi       = {10.1007/978-94-007-5612-0_10},
}

@Article{Louie2013,
  author    = {Louie, Melissa and Koda, Jin and Egusa, Fumi},
  journal   = {ApJ},
  title     = {GEOMETRIC OFFSETS ACROSS SPIRAL ARMS IN M51: NATURE OF GAS AND STAR FORMATION TRACERS},
  year      = {2013},
  issn      = {1538-4357},
  month     = jan,
  number    = {2},
  pages     = {94},
  volume    = {763},
  doi       = {10.1088/0004-637x/763/2/94},
  publisher = {American Astronomical Society},
}

@Article{Cedres2013,
  author        = {Cedr{\'e}s, Bernab{\'e} and Beckman, John E. and Bongiovanni, {\'A}ngel and Cepa, Jordi and Asensio Ramos, Andr{\'e}s and Giammanco, Corrado and Cabrera-Lavers, Antonio and Alfaro, Emilio J.},
  journal       = {\apjl},
  title         = {The Filling Factor-Radius Relation for 58 {HII} Regions across the Disk of NGC 6946},
  year          = {2013},
  month         = mar,
  number        = {1},
  pages         = {L24},
  volume        = {765},
  archiveprefix = {arXiv},
  doi           = {10.1088/2041-8205/765/1/L24},
  eid           = {L24},
  eprint        = {1302.1009},
  primaryclass  = {astro-ph.CO},
  url           = {https://ui.adsabs.harvard.edu/abs/2013ApJ...765L..24C},
}

@Article{Berg2013,
  author        = {Berg, Danielle A. and Skillman, Evan D. and Garnett, Donald R. and Croxall, Kevin V. and Marble, Andrew R. and Smith, J.~D. and Gordon, Karl and Kennicutt, Jr., Robert C.},
  journal       = {\apj},
  title         = {New Radial Abundance Gradients for NGC 628 and NGC 2403},
  year          = {2013},
  month         = oct,
  number        = {2},
  pages         = {128},
  volume        = {775},
  archiveprefix = {arXiv},
  doi           = {10.1088/0004-637X/775/2/128},
  eid           = {128},
  eprint        = {1309.0584},
  primaryclass  = {astro-ph.CO},
  url           = {https://ui.adsabs.harvard.edu/abs/2013ApJ...775..128B},
}

@Article{Wisnioski2012,
  author    = {Wisnioski, Emily and Glazebrook, Karl and Blake, Chris and Poole, Gregory B. and Green, Andrew W. and Wyder, Ted and Martin, Chris},
  journal   = {MNRAS},
  title     = {Scaling relations of star-forming regions: from kpc-sized clumps to H ii regions: Scaling relations of star-forming regions},
  year      = {2012},
  issn      = {0035-8711},
  month     = apr,
  number    = {4},
  pages     = {3339--3355},
  volume    = {422},
  doi       = {10.1111/j.1365-2966.2012.20850.x},
  publisher = {Oxford University Press (OUP)},
}

@Article{Vazdekis2012,
  author    = {Vazdekis, A. and Ricciardelli, E. and Cenarro, A. J. and Rivero-González, J. G. and Díaz-García, L. A. and Falcón-Barroso, J.},
  journal   = {MNRAS},
  title     = {MIUSCAT: extended MILES spectral coverage - I. Stellar population synthesis models: Extended MILES stellar population models},
  year      = {2012},
  issn      = {0035-8711},
  month     = jun,
  number    = {1},
  pages     = {157--171},
  volume    = {424},
  doi       = {10.1111/j.1365-2966.2012.21179.x},
  publisher = {Oxford University Press (OUP)},
}

@InProceedings{Martin2012,
  author    = {Martin, T. and Drissen, L. and Joncas, G.},
  booktitle = {Software and Cyberinfrastructure for Astronomy II},
  title     = {ORBS: A data reduction software for the imaging Fourier transform spectrometers SpIOMM and SITELLE},
  year      = {2012},
  editor    = {Radziwill, Nicole M. and Chiozzi, Gianluca},
  month     = sep,
  pages     = {84513K},
  publisher = {SPIE},
  volume    = {8451},
  doi       = {10.1117/12.925420},
  issn      = {0277-786X},
}

@Article{Leonidaki2012,
  author        = {Leonidaki, I. and Boumis, P. and Zezas, A.},
  journal   = {MNRAS},
  title         = {A multiwavelength study of supernova remnants in six nearby galaxies – II. New optically selected supernova remnants},
  year          = {2012},
  issn          = {0035-8711},
  month         = dec,
  number        = {1},
  pages         = {189--220},
  volume        = {429},
  archiveprefix = {arXiv},
  doi           = {10.1093/mnras/sts324},
  eprint        = {1211.6746},
  primaryclass  = {astro-ph.CO},
  publisher     = {Oxford University Press (OUP)},
  url           = {https://ui.adsabs.harvard.edu/abs/2013MNRAS.429..189L},
}

@Article{Kennicutt2012,
  author        = {Kennicutt, Robert C. and Evans, Neal J.},
  journal       = {\araa},
  title         = {Star Formation in the Milky Way and Nearby Galaxies},
  year          = {2012},
  issn          = {1545-4282},
  month         = sep,
  number        = {1},
  pages         = {531-608},
  volume        = {50},
  archiveprefix = {arXiv},
  doi           = {10.1146/annurev-astro-081811-125610},
  eprint        = {1204.3552},
  primaryclass  = {astro-ph.GA},
  publisher     = {Annual Reviews},
  url           = {https://ui.adsabs.harvard.edu/abs/2012ARA&A..50..531K},
}

@Article{Walt2011,
  author    = {van der Walt, Stéfan and Colbert, S Chris and Varoquaux, Gaël},
  journal   = {Computing in Science \& Engineering},
  title     = {The NumPy Array: A Structure for Efficient Numerical Computation},
  year      = {2011},
  issn      = {1521-9615},
  month     = mar,
  number    = {2},
  pages     = {22--30},
  volume    = {13},
  doi       = {10.1109/mcse.2011.37},
  publisher = {Institute of Electrical and Electronics Engineers (IEEE)},
}

@Article{SimonDiaz2011,
  author    = {Simón-Díaz, S. and García-Rojas, J. and Esteban, C. and Stasińska, G. and López-Sánchez, A. R. and Morisset, C.},
  journal   = {A\&A},
  title     = {A detailed study of the {HII} region M 43 and its ionizing star: I. Stellar parameters and nebular empirical analysis},
  year      = {2011},
  issn      = {1432-0746},
  month     = may,
  pages     = {A57},
  volume    = {530},
  doi       = {10.1051/0004-6361/201116608},
  publisher = {EDP Sciences},
}

@Article{Pedregosa2011,
  author  = {Fabian Pedregosa and Ga{{\"e}}l Varoquaux and Alexandre Gramfort and Vincent Michel and Bertrand Thirion and Olivier Grisel and Mathieu Blondel and Peter Prettenhofer and Ron Weiss and Vincent Dubourg and Jake Vanderplas and Alexandre Passos and David Cournapeau and Matthieu Brucher and Matthieu Perrot and {{\'E}}douard Duchesnay},
  journal = {Journal of Machine Learning Research},
  title   = {Scikit-learn: Machine Learning in Python},
  year    = {2011},
  number  = {85},
  pages   = {2825--2830},
  volume  = {12},
  url     = {http://jmlr.org/papers/v12/pedregosa11a.html},
}

@Article{Lee2011,
  author    = {Lee, Jong Hwan and Hwang, Narae and Lee, Myung Gyoon},
  journal   = {ApJ},
  title     = {H II REGION LUMINOSITY FUNCTION OF THE INTERACTING GALAXY M51},
  year      = {2011},
  issn      = {1538-4357},
  month     = jun,
  number    = {2},
  pages     = {75},
  volume    = {735},
  doi       = {10.1088/0004-637x/735/2/75},
  publisher = {American Astronomical Society},
}

@Article{Aniano2011,
  author    = {Aniano, G. and Draine, B. T. and Gordon, K. D. and Sandstrom, K.},
  journal   = {PASP},
  title     = {Common-Resolution Convolution Kernels for Space- and Ground-Based Telescopes},
  year      = {2011},
  issn      = {1538-3873},
  month     = oct,
  number    = {908},
  pages     = {1218--1236},
  volume    = {123},
  doi       = {10.1086/662219},
  publisher = {IOP Publishing},
}

@Article{Micelotta2010,
  author    = {Micelotta, E. R. and Jones, A. P. and Tielens, A. G. G. M.},
  journal   = {A\&A},
  title     = {Polycyclic aromatic hydrocarbon processing in interstellar shocks},
  year      = {2010},
  issn      = {1432-0746},
  month     = feb,
  pages     = {A36},
  volume    = {510},
  doi       = {10.1051/0004-6361/200911682},
  publisher = {EDP Sciences},
}

@InProceedings{McKinney2010,
  author     = {McKinney, Wes},
  booktitle  = {Proceedings of the 9th Python in Science Conference},
  title      = {Data Structures for Statistical Computing in Python},
  year       = {2010},
  pages      = {56--61},
  publisher  = {SciPy},
  series     = {SciPy},
  collection = {SciPy},
  doi        = {10.25080/majora-92bf1922-00a},
  issn       = {2575-9752},
}

@Article{Gutierrez2010,
  author    = {Gutiérrez, Leonel and Beckman, John E.},
  journal   = {ApJ},
  title     = {THE GALAXY-WIDE DISTRIBUTIONS OF MEAN ELECTRON DENSITY IN THE H II REGIONS OF M51 AND NGC 4449},
  year      = {2010},
  issn      = {2041-8213},
  month     = jan,
  number    = {1},
  pages     = {L44--L48},
  volume    = {710},
  doi       = {10.1088/2041-8205/710/1/l44},
  publisher = {American Astronomical Society},
}

@InProceedings{Bacon2010,
  author    = {Bacon, R. and Accardo, M. and Adjali, L. and Anwand, H. and Bauer, S. and Biswas, I. and Blaizot, J. and Boudon, D. and Brau-Nogue, S. and Brinchmann, J. and Caillier, P. and Capoani, L. and Carollo, C. M. and Contini, T. and Couderc, P. and Daguisé, E. and Deiries, S. and Delabre, B. and Dreizler, S. and Dubois, J. and Dupieux, M. and Dupuy, C. and Emsellem, E. and Fechner, T. and Fleischmann, A. and François, M. and Gallou, G. and Gharsa, T. and Glindemann, A. and Gojak, D. and Guiderdoni, B. and Hansali, G. and Hahn, T. and Jarno, A. and Kelz, A. and Koehler, C. and Kosmalski, J. and Laurent, F. and Le Floch, M. and Lilly, S. J. and Lizon, J.-L. and Loupias, M. and Manescau, A. and Monstein, C. and Nicklas, H. and Olaya, J.-C. and Pares, L. and Pasquini, L. and Pécontal-Rousset, A. and Pelló, R. and Petit, C. and Popow, E. and Reiss, R. and Remillieux, A. and Renault, E. and Roth, M. and Rupprecht, G. and Serre, D. and Schaye, J. and Soucail, G. and Steinmetz, M. and Streicher, O. and Stuik, R. and Valentin, H and Vernet, J. and Weilbacher, P. and Wisotzki, L. and Yerle, N.},
  booktitle = {Ground-based and Airborne Instrumentation for Astronomy III},
  title     = {The MUSE second-generation VLT instrument},
  year      = {2010},
  editor    = {McLean, Ian S. and Ramsay, Suzanne K. and Takami, Hideki},
  month     = jul,
  publisher = {SPIE},
  doi       = {10.1117/12.856027},
  issn      = {0277-786X},
}

@Book{VanRossum2009,
  author    = {Van Rossum, Guido and Drake, Fred L.},
  publisher = {CreateSpace},
  title     = {Python 3 Reference Manual},
  year      = {2009},
  address   = {Scotts Valley, CA},
  isbn      = {1441412697},
  doi       = {10.5555/1593511},
}

@Article{Hunt2009,
  author    = {Hunt, L. K. and Hirashita, H.},
  journal   = {A\&A},
  title     = {The size-density relation of extragalactic {HII} regions},
  year      = {2009},
  issn      = {1432-0746},
  month     = oct,
  number    = {3},
  pages     = {1327--1343},
  volume    = {507},
  doi       = {10.1051/0004-6361/200912020},
  publisher = {EDP Sciences},
}

@Article{Haffner2009,
  author        = {Haffner, L.~M. and Dettmar, R. J. and Beckman, J.~E. and Wood, K. and Slavin, J.~D. and Giammanco, C. and Madsen, G.~J. and Zurita, A. and Reynolds, R.~J.},
  journal   = {Rev. Mod. Phys.},
  title         = {The warm ionized medium in spiral galaxies},
  year          = {2009},
  month         = jul,
  number        = {3},
  pages         = {969-997},
  volume        = {81},
  archiveprefix = {arXiv},
  doi           = {10.1103/RevModPhys.81.969},
  eprint        = {0901.0941},
  primaryclass  = {astro-ph.GA},
  url           = {https://ui.adsabs.harvard.edu/abs/2009RvMP...81..969H},
}

@Article{FloresFajardo2009,
  author        = {Flores-Fajardo, N. and Morisset, C. and Binette, L.},
  journal       = {\rmxaa},
  title         = {Analysis of the diffuse ionized gas database: DIGEDA},
  year          = {2009},
  month         = oct,
  pages         = {261-272},
  volume        = {45},
  archiveprefix = {arXiv},
  doi           = {10.48550/arXiv.0910.2261},
  eprint        = {0910.2261},
  primaryclass  = {astro-ph.GA},
  url           = {https://ui.adsabs.harvard.edu/abs/2009RMxAA..45..261F},
}

@Article{Cajko2009,
  author        = {Cajko, K.~O. and Crawford, E.~J. and Filipovic, M.~D.},
  journal   = {Serb. Astron. J.},
  title         = {Multifrequency Observations of One of the Largest Supernova Remnants in the Local Group of Galaxies, LMC - SNR J0450-709},
  year          = {2009},
  month         = dec,
  pages         = {55-60},
  volume        = {179},
  archiveprefix = {arXiv},
  doi           = {10.2298/SAJ0979055C},
  eprint        = {0909.0310},
  primaryclass  = {astro-ph.CO},
  url           = {https://ui.adsabs.harvard.edu/abs/2009SerAJ.179...55C},
}

@Article{Watson2008,
  author        = {Watson, C. and Povich, M.~S. and Churchwell, E.~B. and Babler, B.~L. and Chunev, G. and Hoare, M. and Indebetouw, R. and Meade, M.~R. and Robitaille, T.~P. and Whitney, B.~A.},
  journal       = {\apj},
  title         = {Infrared Dust Bubbles: Probing the Detailed Structure and Young Massive Stellar Populations of Galactic H II Regions},
  year          = {2008},
  month         = jul,
  number        = {2},
  pages         = {1341-1355},
  volume        = {681},
  archiveprefix = {arXiv},
  doi           = {10.1086/588005},
  eprint        = {0806.0609},
  primaryclass  = {astro-ph},
  url           = {https://ui.adsabs.harvard.edu/abs/2008ApJ...681.1341W},
}

@Article{Phillips2008,
  author    = {Phillips, J. P. and Ramos-Larios, G.},
  journal   = {MNRAS},
  title     = {Observations of 58 compact {HII} regions at mid-infrared wavelengths},
  year      = {2008},
  issn      = {1365-2966},
  month     = dec,
  number    = {4},
  pages     = {1527--1544},
  volume    = {391},
  doi       = {10.1111/j.1365-2966.2008.13952.x},
  publisher = {Oxford University Press (OUP)},
}

@Article{O’Dell2008,
  author    = {O’Dell, C. R. and Henney, W. J. and Abel, N. P. and Ferland, G. J. and Arthur, S. J.},
  journal   = {AJ},
  title     = {THE THREE-DIMENSIONAL DYNAMIC STRUCTURE OF THE INNER ORION NEBULA},
  year      = {2008},
  issn      = {1538-3881},
  month     = dec,
  number    = {1},
  pages     = {367--382},
  volume    = {137},
  doi       = {10.1088/0004-6256/137/1/367},
  publisher = {American Astronomical Society},
}

@Article{Galliano2008,
  author    = {Galliano, Frédéric and Madden, Suzanne C. and Tielens, Alexander G. G. M. and Peeters, Els and Jones, Anthony P.},
  journal   = {ApJ},
  title     = {Variations of the Mid‐IR Aromatic Features inside and among Galaxies},
  year      = {2008},
  issn      = {1538-4357},
  month     = may,
  number    = {1},
  pages     = {310--345},
  volume    = {679},
  doi       = {10.1086/587051},
  publisher = {American Astronomical Society},
}

@Article{Thilker2007,
  author   = {Thilker, David A. and Boissier, Samuel and Bianchi, Luciana and Calzetti, Daniela and Boselli, Alessandro and Dale, Daniel A. and Seibert, Mark and Braun, Robert and Burgarella, Denis and Gil de Paz, Armando and Helou, George and Walter, Fabian and Kennicutt, Jr., R.~C. and Madore, Barry F. and Martin, D. Christopher and Barlow, Tom A. and Forster, Karl and Friedman, Peter G. and Morrissey, Patrick and Neff, Susan G. and Schiminovich, David and Small, Todd and Wyder, Ted K. and Donas, Jos{\'e} and Heckman, Timothy M. and Lee, Young-Wook and Milliard, Bruno and Rich, R. Michael and Szalay, Alex S. and Welsh, Barry Y. and Yi, Sukyoung K.},
  journal  = {\apjs},
  title    = {Ultraviolet and Infrared Diagnostics of Star Formation and Dust in NGC 7331},
  year     = {2007},
  month    = dec,
  number   = {2},
  pages    = {572-596},
  volume   = {173},
  doi      = {10.1086/516646},
  url      = {https://ui.adsabs.harvard.edu/abs/2007ApJS..173..572T},
}

@Article{Ramya2007,
  author        = {Ramya, S. and Sahu, D.~K. and Prabhu, T.~P.},
  journal       = {\mnras},
  title         = {Study of star formation in NGC 1084},
  year          = {2007},
  month         = oct,
  number        = {2},
  pages         = {511-524},
  volume        = {381},
  archiveprefix = {arXiv},
  doi           = {10.1111/j.1365-2966.2007.12232.x},
  eprint        = {0707.2366},
  primaryclass  = {astro-ph},
  url           = {https://ui.adsabs.harvard.edu/abs/2007MNRAS.381..511R},
}

@Article{Hunter2007,
  author    = {Hunter, John D.},
  journal   = {Computing in Science \& Engineering},
  title     = {Matplotlib: A 2D Graphics Environment},
  year      = {2007},
  issn      = {1521-9615},
  number    = {3},
  pages     = {90--95},
  volume    = {9},
  doi       = {10.1109/mcse.2007.55},
  publisher = {Institute of Electrical and Electronics Engineers (IEEE)},
}

@Article{Draine2007a,
  author    = {Draine, B. T. and Li, Aigen},
  journal   = {ApJ},
  title     = {Infrared Emission from Interstellar Dust. IV. The Silicate‐Graphite‐PAH Model in the Post‐SpitzerEra},
  year      = {2007},
  issn      = {1538-4357},
  month     = mar,
  number    = {2},
  pages     = {810--837},
  volume    = {657},
  doi       = {10.1086/511055},
  publisher = {American Astronomical Society},
}

@Book{Osterbrock2006,
  author = {Osterbrock, Donald E. and Ferland, Gary J.},
  title  = {Astrophysics of gaseous nebulae and active galactic nuclei},
  year   = {2006},
  url    = {https://ui.adsabs.harvard.edu/abs/2006agna.book.....O},
}

@Article{Madsen2006,
  author        = {Madsen, G.~J. and Reynolds, R.~J. and Haffner, L.~M.},
  journal       = {\apj},
  title         = {A Multiwavelength Optical Emission Line Survey of Warm Ionized Gas in the Galaxy},
  year          = {2006},
  month         = nov,
  number        = {1},
  pages         = {401-425},
  volume        = {652},
  archiveprefix = {arXiv},
  doi           = {10.1086/508441},
  eprint        = {astro-ph/0609558},
  primaryclass  = {astro-ph},
  url           = {https://ui.adsabs.harvard.edu/abs/2006ApJ...652..401M},
}

@Article{Kewley2006,
  author    = {Kewley, L. J. and Groves, B. and Kauffmann, G. and Heckman, T.},
  journal   = {MNRAS},
  title     = {The host galaxies and classification of active galactic nuclei},
  year      = {2006},
  issn      = {1365-2966},
  month     = nov,
  number    = {3},
  pages     = {961--976},
  volume    = {372},
  doi       = {10.1111/j.1365-2966.2006.10859.x},
  publisher = {Oxford University Press (OUP)},
}

@Article{Mathis2005,
  author    = {Mathis, John S. and Wood, Kenneth},
  journal   = {MNRAS},
  title     = {The effects of clumping on derived abundances in H ii regions},
  year      = {2005},
  issn      = {1365-2966},
  month     = jun,
  number    = {1},
  pages     = {227--235},
  volume    = {360},
  doi       = {10.1111/j.1365-2966.2005.09029.x},
  publisher = {Oxford University Press (OUP)},
}

@Article{Elwert2005,
  author        = {Elwert, T. and Dettmar, R. J.},
  journal       = {\apj},
  title         = {Constraining the Extra Heating of the Diffuse Ionized Gas in the Milky Way},
  year          = {2005},
  month         = oct,
  number        = {1},
  pages         = {277-282},
  volume        = {632},
  archiveprefix = {arXiv},
  doi           = {10.1086/432899},
  eprint        = {astro-ph/0506601},
  primaryclass  = {astro-ph},
  url           = {https://ui.adsabs.harvard.edu/abs/2005ApJ...632..277E},
}

@Article{Rapacioli2004,
  author    = {Rapacioli, M. and Joblin, C. and Boissel, P.},
  journal   = {A\&A},
  title     = {Spectroscopy of polycyclic aromatic hydrocarbons and very small grains in photodissociation regions},
  year      = {2004},
  issn      = {1432-0746},
  month     = dec,
  number    = {1},
  pages     = {193--204},
  volume    = {429},
  doi       = {10.1051/0004-6361:20041247},
  publisher = {EDP Sciences},
}

@Article{Giammanco2004,
  author    = {Giammanco, C. and Beckman, J. E. and Zurita, A. and Relaño, M.},
  journal   = {A\&A},
  title     = {Propagation of ionizing radiation in {HII} regions: The effects of optically thick density fluctuations},
  year      = {2004},
  issn      = {1432-0746},
  month     = sep,
  number    = {3},
  pages     = {877--885},
  volume    = {424},
  doi       = {10.1051/0004-6361:20040468},
  publisher = {EDP Sciences},
}

@Article{Kauffmann2003,
  author        = {Kauffmann, Guinevere and Heckman, Timothy M. and Simon White, D. M. and Charlot, Stéphane and Tremonti, Christy and Brinchmann, Jarle and Bruzual, Gustavo and Peng, Eric W. and Seibert, Mark and Bernardi, Mariangela and Blanton, Michael and Brinkmann, Jon and Castander, Francisco and Csábai, Istvan and Fukugita, Masataka and Ivezic, Zeljko and Munn, Jeffrey A. and Nichol, Robert C. and Padmanabhan, Nikhil and Thakar, Aniruddha R. and Weinberg, David H. and York, Donald},
  journal   = {MNRAS},
  title         = {Stellar masses and star formation histories for {10⁵} galaxies from the Sloan Digital Sky Survey},
  year          = {2003},
  issn          = {1365-2966},
  month         = may,
  number        = {1},
  pages         = {33--53},
  volume        = {341},
  archiveprefix = {arXiv},
  doi           = {10.1046/j.1365-8711.2003.06291.x},
  eprint        = {astro-ph/0204055},
  primaryclass  = {astro-ph},
  publisher     = {Oxford University Press (OUP)},
  url           = {https://ui.adsabs.harvard.edu/abs/2003MNRAS.341...33K},
}

@Article{Chabrier2003,
  author    = {Chabrier, Gilles},
  journal   = {PASP},
  title     = {Galactic Stellar and Substellar Initial Mass Function},
  year      = {2003},
  issn      = {1538-3873},
  month     = jul,
  number    = {809},
  pages     = {763--795},
  volume    = {115},
  doi       = {10.1086/376392},
  publisher = {IOP Publishing},
}

@Article{Zurita2002,
  author    = {Zurita, A. and Beckman, J. E. and Rozas, M. and Ryder, S.},
  journal   = {A\&A},
  title     = {The origin of the ionization of the diffuse ionized gas in spirals: II. Modelling the distribution of ionizing radiation in NGC 157},
  year      = {2002},
  issn      = {1432-0746},
  month     = may,
  number    = {3},
  pages     = {801--815},
  volume    = {386},
  doi       = {10.1051/0004-6361:20020212},
  publisher = {EDP Sciences},
}

@Article{Kewley2001,
  author        = {Kewley, L. J. and Dopita, M. A. and Sutherland, R. S. and Heisler, C. A. and Trevena, J.},
  journal   = {ApJ},
  title         = {Theoretical Modeling of Starburst Galaxies},
  year          = {2001},
  issn          = {1538-4357},
  month         = jul,
  number        = {1},
  pages         = {121--140},
  volume        = {556},
  archiveprefix = {arXiv},
  doi           = {10.1086/321545},
  eprint        = {astro-ph/0106324},
  primaryclass  = {astro-ph},
  publisher     = {American Astronomical Society},
  url           = {https://ui.adsabs.harvard.edu/abs/2001ApJ...556..121K},
}

@Article{Ferriere2001,
  author        = {Ferri{\`e}re, Katia M.},
  journal   = {Rev. Mod. Phys.},
  title         = {The interstellar environment of our galaxy},
  year          = {2001},
  month         = oct,
  number        = {4},
  pages         = {1031-1066},
  volume        = {73},
  archiveprefix = {arXiv},
  doi           = {10.1103/RevModPhys.73.1031},
  eprint        = {astro-ph/0106359},
  primaryclass  = {astro-ph},
  url           = {https://ui.adsabs.harvard.edu/abs/2001RvMP...73.1031F},
}

@Article{Thilker2000,
  author    = {Thilker, David A. and Braun, Robert and Walterbos, René A. M.},
  journal   = {AJ},
  title     = {HIIphot: Automated Photometry of HII Regions Applied to M51},
  year      = {2000},
  issn      = {0004-6256},
  month     = dec,
  number    = {6},
  pages     = {3070--3087},
  volume    = {120},
  doi       = {10.1086/316852},
  publisher = {American Astronomical Society},
}

@Article{Copetti2000,
  author   = {Copetti, M.~V.~F. and Mallmann, J.~A.~H. and Schmidt, A.~A. and Casta{\~n}eda, H.~O.},
  journal  = {\aap},
  title    = {Internal variation of electron density in galactic Hbt II regions},
  year     = {2000},
  month    = may,
  pages    = {621-636},
  volume   = {357},
  url      = {https://ui.adsabs.harvard.edu/abs/2000A&A...357..621C},
}

@Article{Bonnarel2000,
  author   = {Bonnarel, F. and Fernique, P. and Bienaym{\'e}, O. and Egret, D. and Genova, F. and Louys, M. and Ochsenbein, F. and Wenger, M. and Bartlett, J.~G.},
  journal  = {\aaps},
  title    = {The ALADIN interactive sky atlas. A reference tool for identification of astronomical sources},
  year     = {2000},
  month    = apr,
  pages    = {33-40},
  volume   = {143},
  doi      = {10.1051/aas:2000331},
  url      = {https://ui.adsabs.harvard.edu/abs/2000A&AS..143...33B},
}

@Misc{Williams1999,
  author    = {Williams, Jonathan P. and Blitz, Leo and McKee, Christopher F.},
  title     = {The Structure and Evolution of Molecular Clouds: from Clumps to Cores to the IMF},
  year      = {1999},
  copyright = {Assumed arXiv.org perpetual, non-exclusive license to distribute this article for submissions made before January 2004},
  doi       = {10.48550/ARXIV.ASTRO-PH/9902246},
  publisher = {arXiv},
}

@Article{Leitherer1999,
  author    = {Leitherer, Claus and Schaerer, Daniel and Goldader, Jeffrey D. and Delgado, Rosa M. Gonzalez and Robert, Carmelle and Kune, Denis Foo and de Mello, Duilia F. and Devost, Daniel and Heckman, Timothy M.},
  journal   = {ApJS},
  title     = {Starburst99: Synthesis Models for Galaxies with Active Star Formation},
  year      = {1999},
  issn      = {1538-4365},
  month     = jul,
  number    = {1},
  pages     = {3--40},
  volume    = {123},
  doi       = {10.1086/313233},
  publisher = {American Astronomical Society},
}

@Article{Vacca1996,
  author   = {Vacca, William D. and Garmany, Catharine D. and Shull, J. Michael},
  journal  = {\apj},
  title    = {The Lyman-Continuum Fluxes and Stellar Parameters of O and Early B-Type Stars},
  year     = {1996},
  month    = apr,
  pages    = {914},
  volume   = {460},
  doi      = {10.1086/177020},
  url      = {https://ui.adsabs.harvard.edu/abs/1996ApJ...460..914V},
}

@Article{Jones1996,
  author   = {Jones, A.~P. and Tielens, A.~G.~G.~M. and Hollenbach, D.~J.},
  journal  = {\apj},
  title    = {Grain Shattering in Shocks: The Interstellar Grain Size Distribution},
  year     = {1996},
  month    = oct,
  pages    = {740},
  volume   = {469},
  doi      = {10.1086/177823},
  url      = {https://ui.adsabs.harvard.edu/abs/1996ApJ...469..740J},
}

@Article{Zaritsky1994,
  author   = {Zaritsky, Dennis and Kennicutt, Jr., Robert C. and Huchra, John P.},
  journal  = {\apj},
  title    = {HII Regions and the Abundance Properties of Spiral Galaxies},
  year     = {1994},
  month    = jan,
  pages    = {87},
  volume   = {420},
  doi      = {10.1086/173544},
  url      = {https://ui.adsabs.harvard.edu/abs/1994ApJ...420...87Z},
}

@Article{Williams1994,
  author    = {Williams, Jonathan P. and de Geus, Eugene J. and Blitz, Leo},
  journal   = {ApJ},
  title     = {Determining structure in molecular clouds},
  year      = {1994},
  issn      = {1538-4357},
  month     = jun,
  pages     = {693},
  volume    = {428},
  doi       = {10.1086/174279},
  publisher = {American Astronomical Society},
}

@Article{ODonnell1994,
  author    = {O’Donnell, James E.},
  journal   = {ApJ},
  title     = {R[SUB]nu[/SUB]-dependent optical and near-ultraviolet extinction},
  year      = {1994},
  issn      = {1538-4357},
  month     = feb,
  pages     = {158},
  volume    = {422},
  doi       = {10.1086/173713},
  publisher = {American Astronomical Society},
}

@Article{Castaneda1992,
  author   = {Castaneda, H.~O. and Vilchez, J.~M. and Copetti, M.~V.~F.},
  journal  = {\aap},
  title    = {Density studies on giant extragalactic HII regions.},
  year     = {1992},
  month    = jul,
  pages    = {370-380},
  volume   = {260},
  url      = {https://ui.adsabs.harvard.edu/abs/1992A&A...260..370C},
}

@Article{Rubin1991,
  author   = {Rubin, R.~H. and Simpson, J.~P. and Haas, M.~R. and Erickson, E.~F.},
  journal  = {\apj},
  title    = {Axisymmetric Model of the Ionized Gas in the Orion Nebula},
  year     = {1991},
  month    = jun,
  pages    = {564},
  volume   = {374},
  doi      = {10.1086/170145},
  url      = {https://ui.adsabs.harvard.edu/abs/1991ApJ...374..564R},
}

@Book{Osterbrock1989,
  author = {Osterbrock, Donald E.},
  title  = {Astrophysics of gaseous nebulae and active galactic nuclei},
  year   = {1989},
  url    = {https://ui.adsabs.harvard.edu/abs/1989agna.book.....O},
}

@Article{Cardelli1989,
  author    = {Cardelli, Jason A. and Clayton, Geoffrey C. and Mathis, John S.},
  journal   = {ApJ},
  title     = {The relationship between infrared, optical, and ultraviolet extinction},
  year      = {1989},
  issn      = {1538-4357},
  month     = oct,
  pages     = {245},
  volume    = {345},
  doi       = {10.1086/167900},
  publisher = {American Astronomical Society},
}

@Article{Hartigan1987,
  author   = {Hartigan, Patrick and Raymond, John and Hartmann, Lee},
  journal  = {\apj},
  title    = {Radiative Bow Shock Models of Herbig-Haro Objects},
  year     = {1987},
  month    = may,
  pages    = {323},
  volume   = {316},
  doi      = {10.1086/165204},
  url      = {https://ui.adsabs.harvard.edu/abs/1987ApJ...316..323H},
}

@Article{Elmegreen1987,
  author   = {Elmegreen, Debra Meloy and Elmegreen, Bruce G.},
  journal  = {\apj},
  title    = {Arm Classifications for Spiral Galaxies},
  year     = {1987},
  month    = mar,
  pages    = {3},
  volume   = {314},
  doi      = {10.1086/165034},
  url      = {https://ui.adsabs.harvard.edu/abs/1987ApJ...314....3E},
}

@Article{Cox1985,
  author   = {Cox, D.~P. and Raymond, J.~C.},
  journal  = {\apj},
  title    = {Preionization-dependent families of radiative shock waves.},
  year     = {1985},
  month    = nov,
  pages    = {651-659},
  volume   = {298},
  doi      = {10.1086/163649},
  url      = {https://ui.adsabs.harvard.edu/abs/1985ApJ...298..651C},
}

@Article{McCall1984,
  author   = {McCall, M.~L.},
  journal  = {\mnras},
  title    = {Emission coefficients for gaseous nebulae : three-level atom approximations.},
  year     = {1984},
  month    = may,
  pages    = {253-259},
  volume   = {208},
  doi      = {10.1093/mnras/208.2.253},
  url      = {https://ui.adsabs.harvard.edu/abs/1984MNRAS.208..253M},
}

@Article{Kennicutt1984,
  author   = {Kennicutt, Jr., R.~C.},
  journal  = {\apj},
  title    = {Structural properties of giant H II regions in nearby galaxies.},
  year     = {1984},
  month    = dec,
  pages    = {116-130},
  volume   = {287},
  doi      = {10.1086/162669},
  url      = {https://ui.adsabs.harvard.edu/abs/1984ApJ...287..116K},
}

@Article{Baldwin1981,
  author    = {Baldwin, J. A. and Phillips, M. M. and Terlevich, R.},
  journal   = {PASP},
  title     = {Classification parameters for the emission-line spectra of extragalactic objects},
  year      = {1981},
  issn      = {1538-3873},
  month     = feb,
  pages     = {5},
  volume    = {93},
  doi       = {10.1086/130766},
  publisher = {IOP Publishing},
}

@Article{Sabbadin1977,
  author   = {Sabbadin, F. and Minello, S. and Bianchini, A.},
  journal  = {\aap},
  title    = {Sharpless 176: a large, nearby planetary nebula.},
  year     = {1977},
  month    = aug,
  pages    = {147-149},
  volume   = {60},
  url      = {https://ui.adsabs.harvard.edu/abs/1977A&A....60..147S},
}

@Article{Mathewson1973,
  author   = {Mathewson, D.~S. and Clarke, J.~N.},
  journal  = {\apj},
  title    = {Supernova remnants in the Large Magellanic Cloud.},
  year     = {1973},
  month    = mar,
  pages    = {725-738},
  volume   = {180},
  doi      = {10.1086/152002},
  url      = {https://ui.adsabs.harvard.edu/abs/1973ApJ...180..725M},
}

@Article{Osterbrock1959,
  author    = {Osterbrock, Donald and Flather, Edith},
  journal   = {ApJ},
  title     = {Electron Densities in the Orion NEBULA.II.},
  year      = {1959},
  issn      = {1538-4357},
  month     = jan,
  pages     = {26},
  volume    = {129},
  doi       = {10.1086/146592},
  publisher = {American Astronomical Society},
}
\begin{appendix}
\section{Analytical derivation of the $\mathrm{log(L_{H\alpha})}$-$\mathrm{log(R)}$ relation}\label{app:LR_derivation}
\subsection{Absolute-threshold radius definition methods}
The H$\alpha$ surface brightness $\Sigma_{H\alpha}$ is related to the emission measure $\mathrm{EM} = \int n_e^2\, dl$ through:
\begin{equation} \label{eq:SigmaEM}
	\Sigma_{H\alpha} = \frac{\alpha_{H\alpha}^{\rm eff}\, h\nu_{H\alpha}}{4\pi}\,
	\mathrm{EM} = 1.09\times10^{-6}\,\mathrm{EM}
	\quad (\mathrm{erg\,s^{-1}\,cm^{-2}\,sr^{-1}})
\end{equation}
\noindent where $\alpha_{H\alpha}^{\rm eff} = 1.17\times10^{-13}$~cm$^3$~s$^{-1}$ at $T_e = 10^4$~K (Case~B; \citealt{Osterbrock2006}) and EM is expressed in pc~cm$^{-6}$. This relation is exact under uniform $T_e$, which is implicit in adopting a fixed $\mathrm{EM_{min}}$ threshold for all regions \citep{Groves2023}.
\par
We can model each H\,II region as having a Gaussian surface brightness profile:
\begin{equation}  \label{eq:GaussProfile}
	\Sigma(r) = \Sigma_0\,\exp\!\left(-\frac{r^2}{2\sigma^2}\right)
\end{equation}
\noindent where an absolute flux threshold $\Sigma_{\min}$ truncates the profile at the radius where $\Sigma(R_{\rm limit}) = \Sigma_{\min}$:
\begin{equation} \label{eq:Rlimit}
	R_{\rm limit} = \sigma\sqrt{2\ln\!\left(\frac{\Sigma_0}{\Sigma_{\min}}\right)}
\end{equation}
Since $\Sigma_{\min}$ is fixed and $\Sigma_0$ varies, $R_{\rm limit}$ grows with the peak surface brightness of each region. The total luminosity integrated up to $R_{\rm limit}$ is:
\begin{equation}  \label{eq:LRlimit}
	L_{H\alpha} = 2\pi\sigma^2\Sigma_0
	\left[1 - \exp\!\left(-\frac{R_{\rm limit}^2}{2\sigma^2}\right)\right]
\end{equation}
Substituting the threshold condition, we get:
\begin{equation} \label{eq:LRlimit2}
	\mathrm{L_{H\alpha}} \approx 2\pi\sigma^2\Sigma_{\min}
	\exp\!\left(\frac{R_{\rm limit}^2}{2\sigma^2}\right)
\end{equation}
Differentiating in log space yields the local slope of the $\mathrm{log(L_{H\alpha})}$ on $\mathrm{log(R_{limit})}$ relation is:
\begin{equation}  \label{eq:slope}
	\kappa \equiv \frac{\mathrm{d\log(L_{H\alpha})}}{\mathrm{d\log(R_{\rm limit})}} \propto \frac{R_{\rm limit}^2}{\sigma^2}
\end{equation}
Consequently, the slope $\kappa$ depends on $R_{\rm limit}/\sigma$, which increases with the ratio $\Sigma_0/\Sigma_{\min}$. Brighter regions have larger radii before reaching the threshold. Table~\ref{tab:slopes} shows this dependence for some values of $\Sigma_0/\Sigma_{\min}$:
\begin{table}[H]
	\centering
	\caption{$R_{\rm limit}/\sigma$ and slope $\kappa$ for a Gaussian profile as a function of $\Sigma_0/\Sigma_{\min}$.}
	\label{tab:slopes}
	\begin{tabular}{cccc}
		\hline\hline
		$\Sigma_0/\Sigma_{\min}$ & $R_{\rm limit}/\sigma$ & $\kappa$ & Regime \\
		\hline
		5    & 1.79 & 3.2  & Very faint regions    \\
		10   & 2.15 & 4.6 & Typical regions   \\
		50  & 2.80 & 7.8 & Bright regions   \\
		\hline
	\end{tabular}
\end{table}
For bright regions ($\Sigma_0/\Sigma_{\min} \sim 50$), $R_{\rm limit}/\sigma \approx 2.80$, while for typical regions ($\Sigma_0/\Sigma_{\min} \sim 10$) it is $\sim 4.6$. Fitting a single power law across a population spanning roughly 2~dex in $\Sigma_0/\Sigma_{\min}$ averages over the intrinsic curvature of the Gaussian profiles, yielding an effective slope above three, consistent with the values reported by \cite{Groves2023} and \cite{Congiu2023}.
\par
The resulting slopes for \cite{RousseauNepton2018} and this work evaluated at the 5th quantile, $\kappa_{05} \sim= 0.83$ and $1.66$, are consistent with a weak decrease of $\Sigma_0$ with increasing $R$ because the method isolates the core of each region. A further consequence is that absolute-threshold methods compress the scatter perpendicular to the $L_{H\alpha}$-$R$ relation. Regions with identical intrinsic $\sigma$ but different H$\alpha$ are assigned larger radii, which shifts them primarily along the correlation. This explains why the \cite{Groves2023, Congiu2023} catalogs produce a narrow distribution.
\par
Two catalogs in our comparison achieve $\kappa_{05}$ values between those of the two pure categories. In \cite{Barnes2026} the coupling between $R_{\rm cir}$ and $L_{H\alpha}$, yields $\kappa_{05} \sim 1.95$. Here, the coupling set by the absolute threshold is overcome because the flux extension is limited within each subregion of the MUSE parent mask. The resulting method is a hybrid of the two major categories of pure methods. In \cite{Zurita2026}, H\,II regions are detected above $3\sigma_{\rm noise}$ in a background-subtracted H$\alpha$ image, where the background traces the spatially variable DIG emission. This local background subtraction reduces the dynamic range of the residual image relative to the original, partially suppressing the dependence of $R_{\rm limit}$ on $\Sigma_0$ that drives $\kappa_{05}$ to high values in pure absolute-threshold methods. The resulting $L_{H\alpha}$–$R$ slope is $\kappa \sim 2.69$ for NGC~628, close to the physical expectation of $\kappa = 3$ for uniform-density regions.
\par
Although the two methods yield intermediate $\kappa_{05}$, the mechanisms differ: in \citet{Barnes2026} suppression is topological, while in \citet{Zurita2026} it is photometric. As a result, the two approaches respond differently to DIG morphology or resolution, and neither guarantees $\kappa \approx 3$ in general.
\subsection{Profile-truncated radius definition methods}  
In contrast, profile-truncated radii definition methods apply a relative truncation criterion that fixes $R_{\rm limit} \approx k\sigma$ independently of $\Sigma_0$, where $k$ depends on the profile shape but not on the absolute flux level. Substituting $R_{\rm limit} = k\sigma$ into Eq.~\ref{eq:LRlimit}:
\begin{equation}  \label{eq:Lrelative}
	L \approx 2\pi k^2 \sigma^2 \Sigma_0 = 2\pi R_{\rm cut}^2 \Sigma_0
\end{equation}
Since $R_{\rm limit}$ is now decoupled from $\Sigma_0$, the observed slope reflects the intrinsic variation of $\Sigma_0$ with $R$ across the population rather than the influence of the threshold:
\begin{equation}  \label{eq:slopeRelative}
	\kappa = \frac{\mathrm{d\log(L_{H\alpha})}}{\mathrm{d\log R_{\rm limit}} }
	= 2 + \frac{\mathrm{d\log\Sigma_0}}{\mathrm{d\log R_{\rm limit}}}
\end{equation}
As a consequence of that, \cite{RousseauNepton2018} and this work show a pronounced triangular "Christmas-tree" distribution with $\sim 3$~dex of vertical scatter at fixed $R$. This scatter reflects physical variations in local electron density and FF. In contrast to absolute-threshold methods, where regions with identical intrinsic $\sigma$ but different H$\alpha$ luminosities are assigned different radii which mainly moves the regions along the correlation rather than across it, as mentioned above.
\par
Overall, the slope $\kappa_{05}$ primarily reflects the radius-definition method when comparing different pipelines within a single galaxy. When the same method is applied across multiple galaxies, variations in $\kappa_{05}$ can also reflect genuine physical differences among H\,II region populations.
\section{Analytical derivation of the $\mathrm{log(FF)}-\mathrm{log(R^{3}})$ slope}\label{app:gamma_derivation} 
The observed slope $\gamma_R$ of the $\mathrm{log(FF)}$-$\mathrm{log}(R)$ relation can be derived from the classical FF definition Eq.~\ref{ec:linealff} and the $\mathrm{L_{H\alpha}}$-$\mathrm{R}$ coupling effect. We confirm empirically that $n_e$ shows no significant correlation with R across the catalogs, or it is negligible. Substituting the empirical power-law relation between $\mathrm{L_{H\alpha}}$ and $R$: $\mathrm{log(L_{H\alpha})} = \kappa\,\mathrm{log(R)} + \xi$ into Eq.~\ref{ec:linealff} yields:
\begin{equation}
	\mathrm{log(FF)} = \kappa\,\mathrm{log(R)} + \xi - 3\,\mathrm{log(R)} + \mathrm{const} = (\kappa - 3)\,\mathrm{log(R)} + \mathrm{const}.
	\label{ec:FF_R_derived}
\end{equation}
\noindent The slope of the $\mathrm{log(FF)}$-$\mathrm{log(R)}$ relation is therefore:
\begin{equation}
	\gamma_R = \kappa - 3,
	\label{ec:gammaR}
\end{equation}
\noindent where the factor of three difference with respect to the volume arises from the change of variable $\mathrm{log(R^{3})} = 3\,\mathrm{log(R)}$. Consequently, the slope of the $\mathrm{log(FF)}$-$\mathrm{log(R^{3})}$ relation is:
\begin{equation}
	\gamma = \frac{\kappa - 3}{3}.
	\label{ec:gamma}
\end{equation}
This result demonstrates that $\gamma$ is influenced by the $\mathrm{L_{H\alpha}}$-R slope $\kappa$, which depends on the radii definition criterion rather than on the physical properties of the ionized gas (see Sect.~\ref{subsubsec:FFR}). The physical component of the observed slope is:
\begin{equation}
	\gamma^{\rm phys}_R = \gamma^{\rm obs}_R 
	- (\kappa - 3)
	\label{ec:gamma_R_phys}
\end{equation}
A negative $\gamma^{\rm phys}_R$ indicates that larger H\,II regions genuinely have lower mean rms densities, consistent with an evolutionary scenario in which more extended regions represent later stages of expansion and dispersal of the ionized gas (Sect.~\ref{sec:discussion}). A value of $\gamma^{\rm phys}_R \approx 0$ indicates that the observed trend is entirely explained by the $L_{H\alpha}$-$R$ coupling.
\section{Analytical derivation of the 
	$\mathrm{log(FF)}$-$\mathrm{log(L_{H\alpha})}$ slope}
\label{app:kappa_L_derivation}
The observed slope $\eta$ of the $\mathrm{log(FF)}$-$\mathrm{log(L_{H\alpha})}$ relation can be derived from the classical FF definition (Eq.~\ref{ec:linealff}) and the $\mathrm{L_{H\alpha}}$-R coupling. Substituting the empirical power-law relation $\mathrm{log(L_{H\alpha})} = \kappa\,\mathrm{log(R)} + \xi$, which gives 
$\mathrm{log}(R) = \frac{1}{\kappa}([\mathrm{log}(L_{H\alpha}) - \xi])$, into Eq.~\ref{ec:linealff} and substituting $\mathrm{log(R)} = 
\frac{1}{\kappa}\,\mathrm{log(L_{H\alpha})} + \mathrm{const}$:
\begin{equation}
	\mathrm{log(FF)} = \mathrm{log(L_{H\alpha})}
	- \frac{3}{\kappa}\,\mathrm{log(L_{H\alpha})} 
	+ \mathrm{const}
	\propto \left(1 - \frac{3}{\kappa}\right)
	\mathrm{log(L_{H\alpha})} 
	\label{ec:FF_L_derived}
\end{equation}
The slope of the $\mathrm{log(FF)}$-$\mathrm{log(L_{H\alpha})}$ relation is therefore:
\begin{equation}
	{\eta = 1 - \frac{3}{\kappa} 
		= \frac{\kappa - 3}{\kappa}}
	\label{ec:etaL}
\end{equation}
The physical component of the observed slope is:
\begin{equation}
	\eta^{\rm phys}_R = \eta^{\rm obs}_R 
	- \frac{\kappa - 3}{\kappa}
	\label{ec:eta_R_phys}
\end{equation}
This result demonstrates that $\eta$ is also influenced by the $\mathrm{L_{H\alpha}}$-$R$ slope $\kappa$, and therefore by the radii definition criterion. Three limiting cases can be considered:
\begin{itemize}
	\item $\kappa = 3$: $\eta = 0$. When the radii definition recovers the physical scaling $\mathrm{L_{H\alpha}} \propto R^3$, the FF shows no methodological dependence on the  luminosity-size coupling, and the full observed slope is physical.
	\item $\kappa < 3$: $\eta < 0$. For profile-truncated methods ($\kappa \sim 1.55$), the predicted methodological slope is $\eta \sim -0.93$. Since the observed slope is positive ($\eta^{\rm obs} \sim +1.2$), the physical component $\eta^{\rm phys} = \eta^{\rm obs} - \eta^{\rm pred} \sim +2.1$ is large and positive, reflecting the true increase of FF with luminosity in the 
	H\,II region population.
	\item $\kappa > 3$: $\eta > 0$. For absolute-threshold methods ($\kappa \sim 4.4$-$4.9$), the methodological contribution is $\eta \sim +0.32$-$+0.39$, already positive, so part of the observed positive FF-$\mathrm{L_{H\alpha}}$ correlation is a radii definition effect rather than a physical signal.
\end{itemize}
Comparing Eqs.~\ref{ec:gammaR} and~\ref{ec:etaL}, the three methodological slopes are related by:
\begin{equation}
	\eta = \frac{\gamma_R}{\kappa} 
	= \frac{\kappa - 3}{\kappa}
	\label{ec:eta_gamma_relation}
\end{equation}
\noindent which shows that all three slopes above mentioned are influenced by the single parameter $\kappa$. The method-independent observable in slope space that absorbs this dependence is the volumetric luminosity density $\mathrm{L_{H\alpha}/R^3}$,  whose relation with FF is invariant under changes in $\kappa$, making it the method-independent observable to the induced covariance between $\mathrm{L_{H\alpha}}$ and R, that absorbs the methodological variance introduced by different radii definitions across catalogs.
\begin{figure*}
	\begin{center}
		\includegraphics[width=\linewidth]{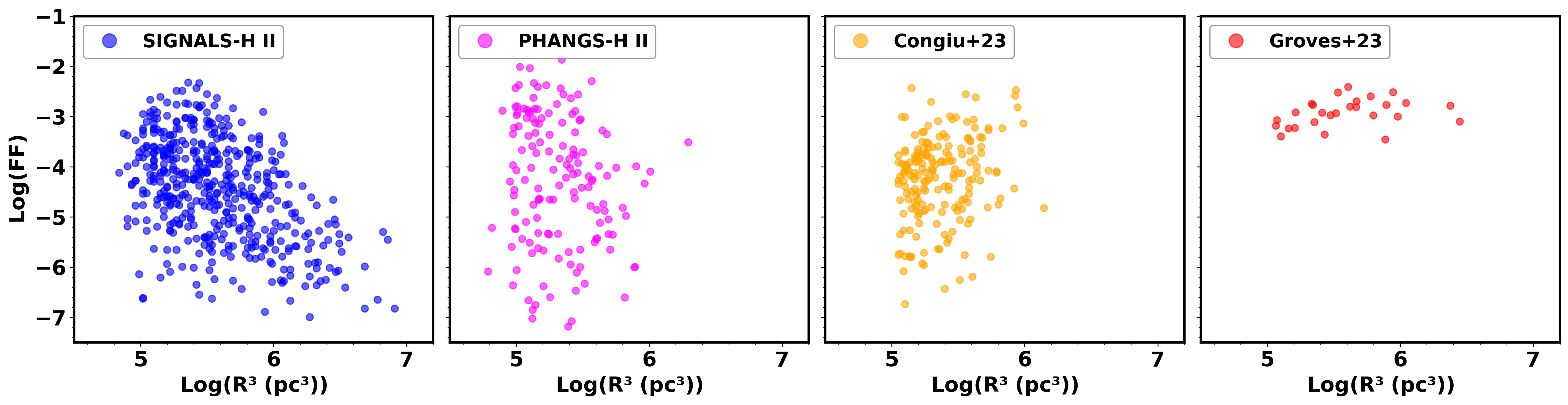}
		\caption[]{$\mathrm{Log(FF)}$ as a function of $\mathrm{log(R^{3})}$ for the SIGNALS-H\,II (blue), PHANGS-H\,II (magenta), \citet{Congiu2023} (orange), and \citet{Groves2023} (red) catalogs. Catalogs adopting truncated-profile radius definitions (blue and magenta dots) show a negative trend with $R$, whereas absolute-threshold catalogs (orange and red dots) tend to display an elongated flatter relation, reflecting the artificial covariance between $\mathrm{L_{H\alpha}}$ and $R$ introduced by the threshold-based radii definition methods.}
		\label{fig:FF_4plots}
	\end{center}
\end{figure*}
\begin{table}
	\centering
	\caption{Analytical expressions for the predicted methodological contributions to the main scaling relations. The physical component of an observed slope is obtained as
		$\tau^{\rm phys}=\tau^{\rm obs}-\tau^{\rm pred}$.}
	\label{tab:predicted_slopes}
	\setlength{\tabcolsep}{6pt}
	\renewcommand{\arraystretch}{1.2}
	\begin{tabular}{lc}
		\hline\hline
		Relation &
		$\tau^{\rm pred}$ \\
		\hline
		$\mathrm{log(FF)}-\mathrm{log(R^{3})}$
		&
		$\displaystyle \frac{\kappa-3}{3}$
		\\[6pt]
		$\mathrm{log(FF)}-\mathrm{log(n_e)}$
		&
		$-2^{\dagger}$
		\\[3pt]
		$\mathrm{log(FF)}-\mathrm{log(L_{H\alpha})}$
		&
		$\displaystyle \frac{\kappa-3}{\kappa}$
		\\[6pt]
		$\mathrm{log(FF)}-\mathrm{log(L_{H\alpha}/R^{3})}$
		&
		$\displaystyle 1-2B^{\ddagger}$
		\\[3pt]
		\hline
	\end{tabular}
	\tablefoot{
		$^{\dagger}$ The predicted slope of the
		$\mathrm{log(FF)}-\mathrm{log(n_e)}$ relation follows directly from the definitional identity (Eq.~\ref{ec:linealff}) and is independent of $\kappa$.
		$^{\ddagger}$ Here,	$B$ is the slope of the empirical relation
		$\mathrm{log(n_e)}$ versus $\mathrm{log}(L_{H\alpha}/R^{3})$
		(Eq.~\ref{ec:ne_beta}; Appendix~\ref{app:LR3_derivation}).
	}
\end{table}
\section{Analytical derivation of the 
	\texorpdfstring{$\mathrm{log(FF)}$-$\mathrm{log(L_{H\alpha}/R^3)}$}{Log(FF)-Log(L_Hα/R^3)}}
\label{app:LR3_derivation}
Starting from the definition of the FF and the empirical $\mathrm{log(L_{H\alpha})}$--$\mathrm{log(R)}$ relation with slope $\kappa$, we substitute $\mathrm{log(L_{H\alpha})} = \kappa\,\mathrm{log(R)} + \xi$ into Eq.~\ref{ec:linealff}, obtaining:
\begin{equation}
	\mathrm{log(FF)} = (\kappa - 3)\mathrm{log(R)} + \xi - 2\mathrm{log(n_e)} - 31.6 
	\label{ec:D.1}
\end{equation}
We define the logarithm of the volumetric H$\alpha$ luminosity density as:
\begin{equation}
	\mathcal{E} \equiv \mathrm{log(L_{H\alpha}/R^{3})} 
	= \mathrm{log(L_{H\alpha})} - 3\mathrm{log(R)}
	\label{ec:D.2}
\end{equation}
\noindent which leads to:
\begin{equation}
	\mathcal{E} = (\kappa - 3)\mathrm{log(R)} + \xi 
	\label{ec:D.3}
\end{equation}
Combining Eqs.~\ref{ec:D.1} and \ref{ec:D.3}, the dependence on $(\kappa - 3)$ cancels explicitly, yielding:
\begin{equation}
	\mathrm{log(FF)} = \mathcal{E} - 2\mathrm{log(n_e)} - 31.6
	= \mathrm{log\!\left(\frac{L_{H\alpha}}{R^3}\right)} - 2\mathrm{log(n_e)} - 31.6
	\label{ec:D.4}
\end{equation}
At fixed $n_e$, this implies:
\begin{equation}
	\frac{\mathrm{d\log(FF)}}{\mathrm{d\log(L_{H\alpha}/R^3)}}\bigg|_{n_e} = +1
	\quad (\kappa \neq 3)
	\label{ec:D.5}
\end{equation}
Physically, $\kappa = 3$ corresponds to the idealized case in which the mean local electron density is the same for all H\,II regions regardless of size, consistent with a pure Strömgren sphere model with uniform $n_e$ across the population. In this limiting case, $\mathrm{log(L_{H\alpha}/R^{3})}$ carries no variance across the population, and the slope of the $\mathrm{log(FF)}$-$\mathrm{log(L_{H\alpha}/R^{3})}$ relation becomes observationally indeterminate: Eq.~\ref{ec:D.5} remains valid, but it cannot be measured as a regression slope because the independent variable has zero dispersion. 
\par
When also accounting for the dependence of $n_e$ on $\mathrm{L_{H\alpha}/R^{3}}$ across the H\,II region population, we proceed via partial differentiation. If $\log(n_e)$ correlates with $\mathrm{L_{H\alpha}/R^{3}}$ with slope $B$:
\begin{equation}
	B \equiv \frac{\partial \log(n_e)}{\partial \log\!\left(\frac{L_{H\alpha}}{R^3}\right)}
	\label{ec:ne_beta}
\end{equation}
The analytical expressions describing the methodological contributions to the observed slopes are summarized in Table~\ref{tab:predicted_slopes}. The observed, predicted, and physical slopes recovered for each catalog are listed in Table~\ref{tab:slopes_allcatalogs}.
\subsection{Slope Conditions for Method-Independent Correlations with $\mathrm{log(L_{H\alpha}/R^{3})}$}
\label{app:LR3_cross_conditions}
The $\mathrm{L_{H\alpha}/R^{3}}$ is a method-independent observable for characterizing the ionized gas structure across catalogs with fundamentally different radii definitions. However, this independence does not automatically extend to cross-correlations between $L_{H\alpha}/R^{3}$ and arbitrary photometric or physical observables. For a general observable $\Theta$, the correlation $\mathrm{log(\Theta)}$-$\mathrm{log(L_{H\alpha}/R^{3})}$ is independent of $\kappa$ if and only if $\Theta$ does not depend on the region radius. 
\begin{equation}
	\left.\frac{\partial \ Log (\Theta)}{\partial \ Log(R)}\right| = 0
	\label{ec:param_condition}
\end{equation}
If this condition holds, the observed correlation is purely physical and comparable across catalogs. If it does not hold, the observed slope contains a methodological contribution that scales with $\kappa$ and varies between catalogs.
\par
Observables that satisfy this condition include emission-line ratios, band ratios ($\mathrm{R_{\rm PAH}}$) and equivalent widths for example. 
\par
Observables that do not satisfy this condition include absolute luminosities in any band ($L_{\rm IR}$, $L_{\rm UV}$, $L_{\rm CO}$), masses derived from integrated luminosities, the ionizing photons rate, the ionization parameter and the surface brightnesses for example. In the present work, the PAH-to-dust ratio $R_{\rm PAH}$ satisfies condition because it is a ratio of fluxes measured within apertures matched to each H\,II region. Its value does not depend on the H\,II radius, assuming that the spatial distributions of PAH emission and warm dust emission are approximately uniform within the integration aperture at the working resolution. 
\subsection{Comparison across catalogs}
\label{app:LR3_verification}
\begin{table*}[!t]
	\centering
	\caption{Observed ($\tau^{\rm obs}$), predicted ($\tau^{\rm pred}$), and physical ($\tau^{\rm phys}=\tau^{\rm obs}-\tau^{\rm pred}$) slopes for the main scaling relations in the five H\,II region catalogs.}
	\label{tab:slopes_allcatalogs}
	\setlength{\tabcolsep}{3pt}
	\renewcommand{\arraystretch}{1.25}
	\begin{tabular}{l c
			ccc
			ccc
			cccc}
		\hline\hline
		Catalog
		&
		$\kappa_{05}$
		& \multicolumn{3}{c}
		{$\mathrm{log(FF)}-\mathrm{log(R^{3})}$}
		& \multicolumn{3}{c}
		{$\mathrm{log(FF)}-\mathrm{log(L_{H\alpha})}$}
		& \multicolumn{4}{c}
		{$\mathrm{log(FF)}-\mathrm{log(L_{H\alpha}/R^{3})}$}
		\\
		\cmidrule(lr){3-5}
		\cmidrule(lr){6-8}
		\cmidrule(lr){9-12}
		&
		&
		obs & pred & phys
		&
		obs & pred & phys
		&
		obs & $B^{\dagger}$ & pred$^{\dagger\dagger}$ & phys
		\\
		\hline
		SIGNALS-MUSE
		&
		0.83
		&
		$-1.13$ & $-0.72$ & $-0.41$
		&
		$1.60$ & $-2.61$ & $4.21$
		&
		$1.294$ & $-0.138$ & $1$ & $0.294$
		\\
		PHANGS-H\,II
		&
		1.66
		&
		$-1.18$ & $-0.45$ & $-0.73$
		&
		$1.22$ & $-0.81$ & $2.03$
		&
		$1.303$ & $-0.160$ & $1$ & $0.303$
		\\
		Barnes et al. (2026)
		&
		1.95
		&
		$-0.32$ & $-0.35$ & $+0.03$
		&
		$-0.006$ & $-0.54$ & $+0.53$
		&
		$0.978$ & $0.217$ & $1$ & $-0.022$
		\\
		Congiu et al. (2023)
		&
		4.53
		&
		$0.77$ & $+0.51$ & $+0.26$
		&
		$1.10$ & $+0.34$ & $+0.76$
		&
		$1.296$ & $-0.148$ & $1$ & $0.296$
		\\
		Groves et al. (2023)
		&
		3.95
		&
		$0.23$ & $+0.32$ & $-0.09$
		&
		$0.31$ & $+0.24$ & $+0.07$
		&
		$0.931$ & $0.035$ & $1$ & $-0.069$
		\\
		\hline
	\end{tabular}
	\tablefoot{
		Observed slopes of the $\mathrm{log(FF)}-\mathrm{log(R^{3})}$ and $\mathrm{log(FF)}-\mathrm{log(L_{H\alpha})}$
		relations are described in Appendices~\ref{app:gamma_derivation}	and	\ref{app:kappa_L_derivation}, respectively.
		$^{\dagger}$ B is the slope of the empirical relation $\mathrm{log(n_e)}$ versus $\mathrm{L_{H\alpha}/R^{3}}$ measured independently for each catalog (Appendix~\ref{app:LR3_derivation}), and is listed for reference only. 
		$^{\dagger\dagger}$ For the relation 
		$\mathrm{log(FF)}$-$\mathrm{L_{H\alpha}/R^{3}}$, the predicted slope is $1$ without accounting for the dependence of $n_e$ on $\mathrm{L_{H\alpha}/R^{3}}$ (Appendix~\ref{app:LR3_derivation}).	Predicted slopes are computed from the analytical expressions summarized in Table~\ref{tab:predicted_slopes}.
	}
\end{table*}
For each catalog, B is measured from a linear regression of $\mathrm{log(n_e)}$ on $\mathrm{log(L_{H\alpha}/R^3)}$. The results are summarized in the last columns of Table~\ref{tab:slopes_allcatalogs}. The values reveal that the profile-truncated catalogs (SIGNALS-H\,II and PHANGS-H\,II) yield B $<0$, indicating a genuine anti-correlation between local electron density and volumetric emissivity. Thus H\,II regions with higher $\mathrm{L_{H\alpha}/R^{3}}$ tend to have slightly lower $n_e$, consistent with the luminous regions hosting large ionized volumes at moderate density rather than compact high-density gas. In contrast, \cite{Congiu2023, Groves2023} and \cite{Barnes2026} yield positive and B $\sim 0$. As noted in Subsubsect.~\ref{subsubsec:localdensity}, \cite{Barnes2026} derive electron densities from the \cite{Groves2023} catalog, so the two catalogs share the same $n_e$ distribution. The compressed dynamic range of $n_e$ in these samples (Fig.~\ref{fig:comparne}) suppresses the statistical detectability of the physical anti-correlation, driving B $\sim 0$ regardless of whether the physical relation exists. \cite{Congiu2023}, which also uses an absolute-threshold segmentation method but a wider range of electron densities, yields B much lower than zero because B depends on the dynamic range of $n_e$ in each catalog.
\par
The physical component of the observed slope from the $\mathrm{log(FF)}$-$\mathrm{log(L_{H\alpha}/R^3)}$ relation is:
\begin{equation}
	\eta^{\rm phys}_{L/R^3} = \eta^{\rm obs}_{L/R^3} 
	- 1
	\label{ec:eta_LR3_phys}
\end{equation}
Across all three catalogs with large dynamic range of $n_e$ (SIGNALS-H\,II, PHANGS-H\,II and \cite{Congiu2023}), the physical slopes in the last "phys"' column of Table~\ref{tab:slopes_allcatalogs}, once the predicted slope of $+1$ is accounted for, reflect the true physical signal whereby more luminous H\,II regions are associated with higher values of $\log(\mathrm{FF})$. This confirms that $\mathrm{log(L_{H\alpha}/R^3)}$ provides a complete and quantitative account of the slope variations across catalogs with fundamentally different radius definition methods.
\par
Note that the observational slopes derived here have been obtained from a univariate analysis. To account for the nonzero values of $B$, a multivariate regression of $\mathrm{log(FF)}$ on $\mathrm{log(L_{H\alpha}/R^3)}$ and $\log(n_e)$ simultaneously would be required.
\section{Effect of the galactocentric luminosity correction } \label{app:galactocentric_correction}
The radial luminosity correction was calculated but only appears as significant for the SIGNALS catalog which removes the decrease of $L_{\mathrm{H}\alpha}$ with galactocentric radius, characterized by a gradient of $\lambda=-0.56\pm0.03$, supported by the regression analysis and Spearman's rank correlation ($\rho=-0.38$, $p< 0.001$). 
\par
To assess the impact of this procedure, we repeated the analysis using the corrected luminosities (Table~\ref{tab:rg_correction}). The correction has a negligible effect on the $\log(\mathrm{FF})$--$\log(R^3)$ and $\log(\mathrm{FF})$--$\log(n_{\mathrm e})$ relations in SIGNALS-H\,II, with median differences of only 0.05 and 0.01 dex, respectively, well within the corresponding $1\sigma$ uncertainties. A somewhat larger difference of 0.12 dex is found for the $\log(\mathrm{FF})$--$\log(L_{\mathrm{H}\alpha})$ relation, which is expected because $L_{\mathrm{H}\alpha}$ enters explicitly as the independent variable.
\begin{table}[H]
	\caption{Effect of the galactocentric luminosity correction on
		the slopes of the FF scaling relations for the SIGNALS-H\,II sample.}
	\label{tab:rg_correction}
	\centering
	\begin{tabular}{lccc}
		\hline\hline
		Relation & With correction & Without correction & $\Delta$ \\
		\hline
		$\log(\mathrm{FF})$--$\log(R^3)$            & $-1.18 \pm 0.10$ & $-1.13 \pm 0.10$ & 0.05 \\
		$\log(\mathrm{FF})$--$\log(n_e)$            & $-2.64 \pm 0.10$ & $-2.65 \pm 0.10$ & 0.01 \\
		$\log(\mathrm{FF})$--$\log(L_{\mathrm{H}\alpha})$ & $1.72 \pm 0.06$  & $1.60 \pm 0.06$  & 0.12 \\
		\hline
	\end{tabular}
	\tablefoot{Slopes and $1\sigma$ uncertainties from ordinary least-squares
		regression.}
\end{table}
Importantly, the galactocentric correction does not modify the inferred $L_{\mathrm{H}\alpha}$--$R$ scaling exponents in SIGNALS, including the lower-envelope estimator $\kappa_{05}$ used throughout this work. 
\section{Selection criteria for H\,II regions} \label{app:selection_criteria}
In Table~\ref{tab:HIIcriteria}, we summarize the selection criteria applied to identify the final H\,II regions in the SIGNALS-H\,II and PHANGS-H\,II catalogs. These criteria include constraints on emission line fluxes, signal-to-noise ratios, pseudo-Voigt profile parameters, and classical BPT diagnostics. 
\begin{table*}
	\centering
	\caption[]{\label{tab:HIIcriteria}Criteria for H\,II region selection}
	\begin{adjustbox}{width=\linewidth}
		\begin{tabular}{lccc}
			\hline 
			\noalign{\smallskip}
			Column &
			Unit &
			Description &
			Criteria \\
			\hline \hline
			\noalign{\smallskip}
			\multicolumn{4}{c}{SIGNALS survey} \\
			\noalign{\smallskip}
			\hline
			\noalign{\smallskip}
			$\mathrm{R_{g}}$    & kpc & Galactocentric radius & deprojected  with \\
			&     &                       &$i$ of \cite{Lang2020} \\
			$\mathrm{L_{H\alpha}}$   & $10^{-7} \ erg \ s^{-1}$ & H$\alpha$ total luminosity corrected for extinction  &  $\neq$ 0    \\
			I0p      & $mW/m^2$ & pseudo-Voigt profile parameter I0 & $\neq$ 0\\
			sigp     & pc & pseudo-Voigt profile parameter $\sigma$ &  $\neq$ 0\\
			R2p       &     & pseudo-Voigt profile correlation coefficient $ R^2$ & 0.65 \\
			$\mathrm{log(NII/H\alpha)}$ & Log & Log line ratio $\mathrm{[NII]6583/H\alpha}$  & BPT diagnosis criteria\\
			SNR1      &     & line ratio $\mathrm{[NII]6583/H\alpha}$ best S/N & >4 \\
			$\mathrm{log([OIII]/H\beta)}$ & Log & Log line ratio [$\mathrm{OIII]5007/H\beta}$ & BPT diagnosis criteria \\
			SNR4      &     & line ratio $\mathrm{[OIII]5007/H\beta}$ best S/N & >4 \\
			$\mathrm{log([SII]16/[SII]31)}$ & Log & Log line ratio $\mathrm{[SII]6716/[SII]6731}$ & $ Log([\mathrm{S\,II}]\,6716/6731)$ $\in$ (-0.36, 0.137)\\
			SNR10     &     & line ratio $\mathrm{[SII]6716/[SII]6731}$ best S/N & > 4  \\
			$\mathrm{log([SII]6716+6731/H\alpha})$ &   & Log line ratio $\mathrm{[SII]6716+6731/H\alpha}$ &  $\geq$ 0.4 \\
			SNR2     &      & line ratio $\mathrm{[SII]6716+6731/H\alpha}$ best S/N & > 4 \\
			\noalign{\smallskip}
			\hline
			\noalign{\smallskip}
			\multicolumn{4}{c}{PHANGS-MUSE Ionized Regions Catalog} \\
			\noalign{\smallskip}
			\hline
			\noalign{\smallskip}
			pseudo-Voigt R2       &     & pseudo-Voigt profile correlation coefficient $ R^{2}$ & 0.65 \\
			Emission line fluxes    &     &    &  >0\\
			$\mathrm{EW}$ H1 6563A    &  & $\mathrm{H\alpha}$  equivalent width  & < 0 \\
			$\mathrm{S/N}$ H1 6563A   &  & line ratio $\mathrm{H\alpha}$ best S/N & >4 \\
			$\mathrm{S/N}$ H1 4861A   &  & line ratio $\mathrm{H\beta}$ best S/N & >4 \\
			$\mathrm{S/N}$ N2 6583A   &  & line ratio $\mathrm{[NII]6583/H\alpha}$ best S/N & >4 \\
			$\mathrm{S/N}$ O3 5007A   &  & line ratio $\mathrm{[OIII]5007/H\beta}$ best S/N & >4 \\
			$\mathrm{S/N}$ S2 6716A   &  & line ratio $\mathrm{[SII]6716/}$ best S/N & >4 \\
			$\mathrm{S/N}$ S2 6731A   &  & line ratio $\mathrm{[SII]6731/}$ best S/N & >4 \\
			$\mathrm{log([SII]16/[SII]31)}$ & & Log line ratio $\mathrm{[SII]6716/[SII]6731}$ & $ Log([\mathrm{S\,II}]\,6716/6731)$ $\in$ (-0.36, 0.14)\\
			\noalign{\smallskip}
			\hline
		\end{tabular}
	\end{adjustbox}
\end{table*}
\section{Additional figures}\label{app:Additional figures}
\begin{figure*}
	\begin{center}
		\includegraphics[width=0.245\linewidth]{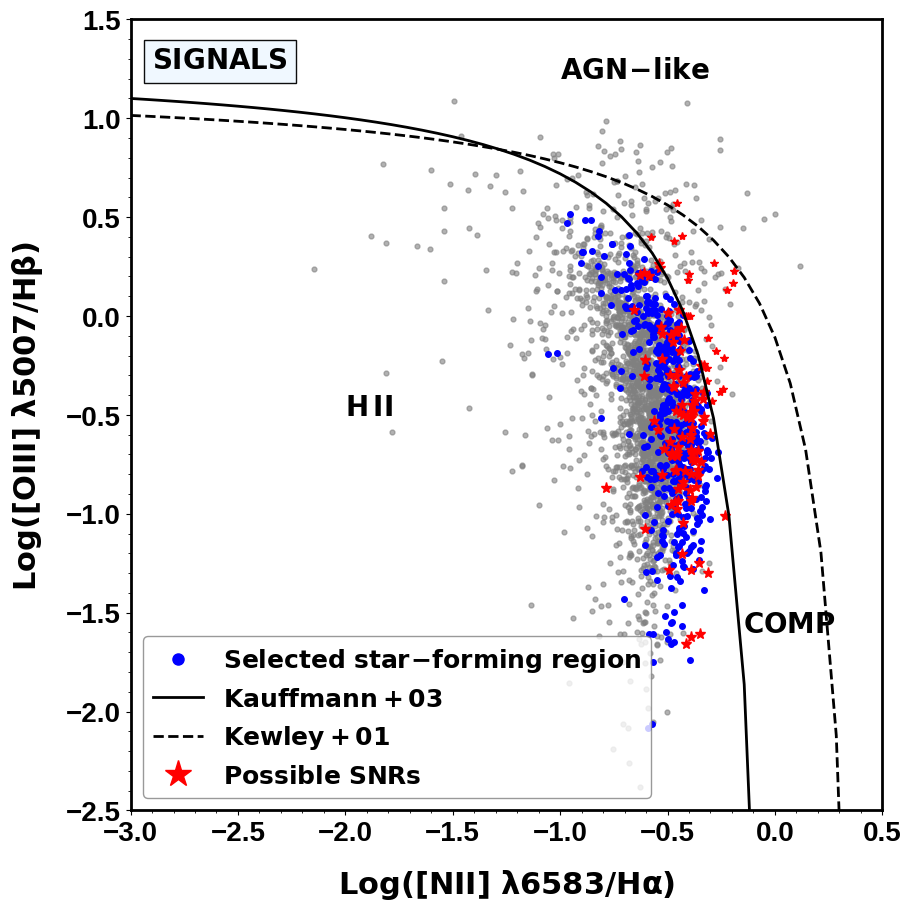} \includegraphics[width=0.245\linewidth]{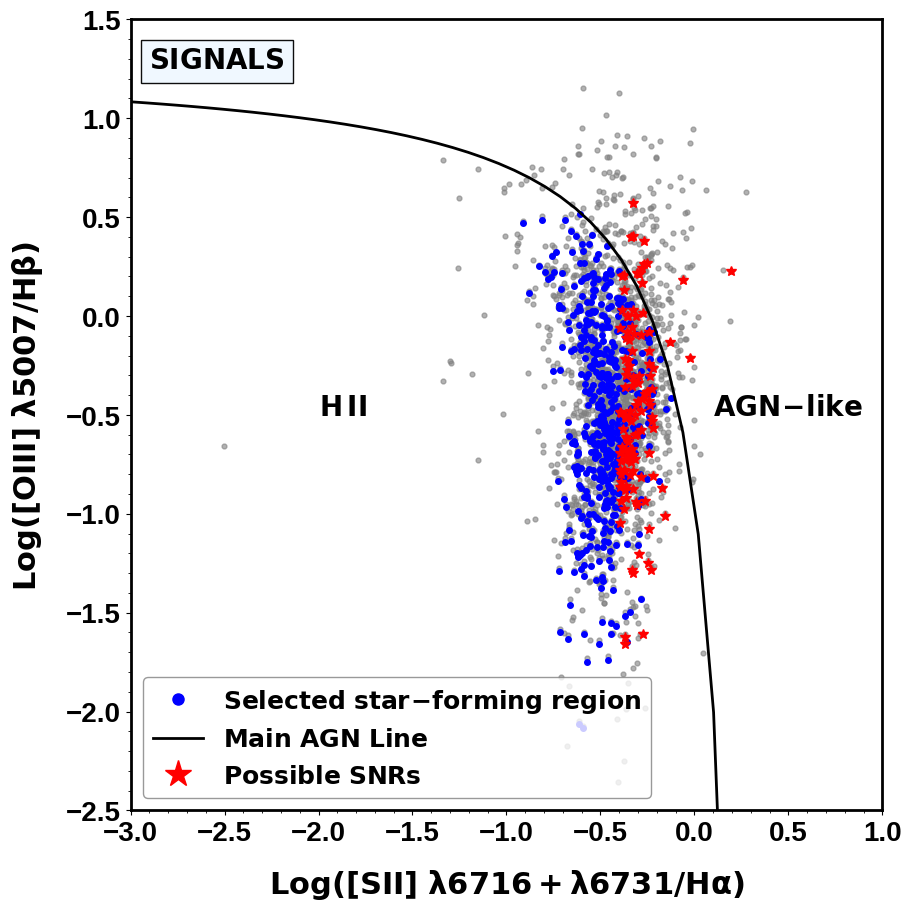} \includegraphics[width=0.245\linewidth]
		{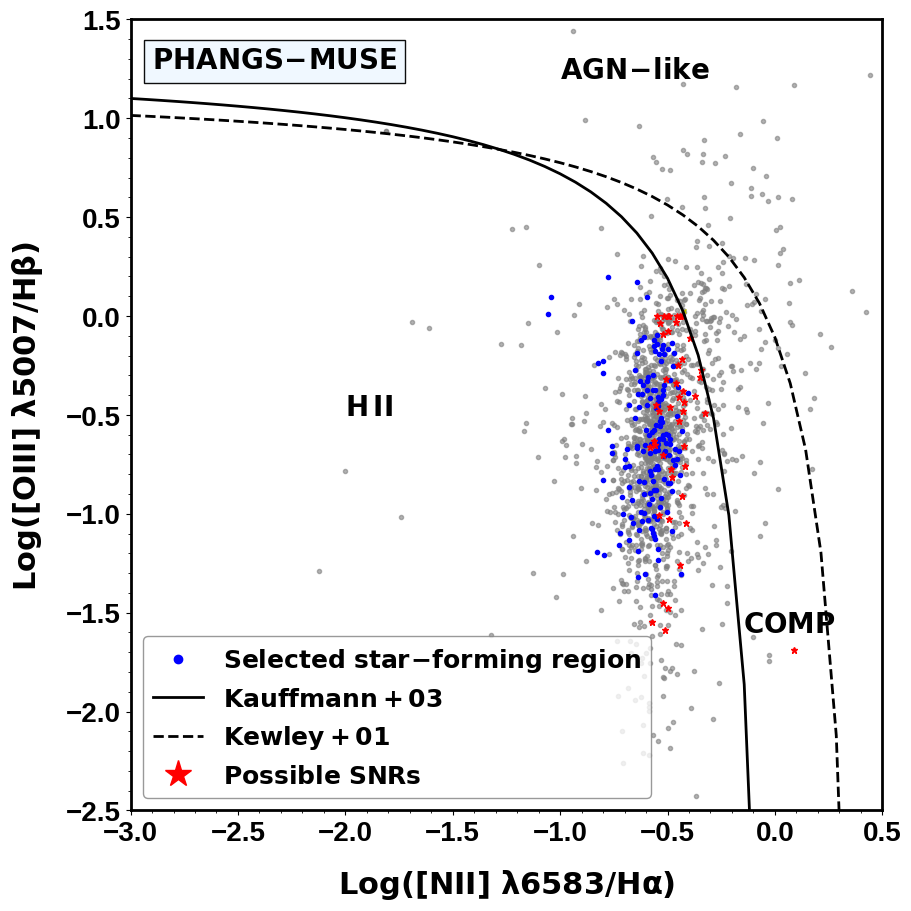}
		\includegraphics[width=0.245\linewidth]
		{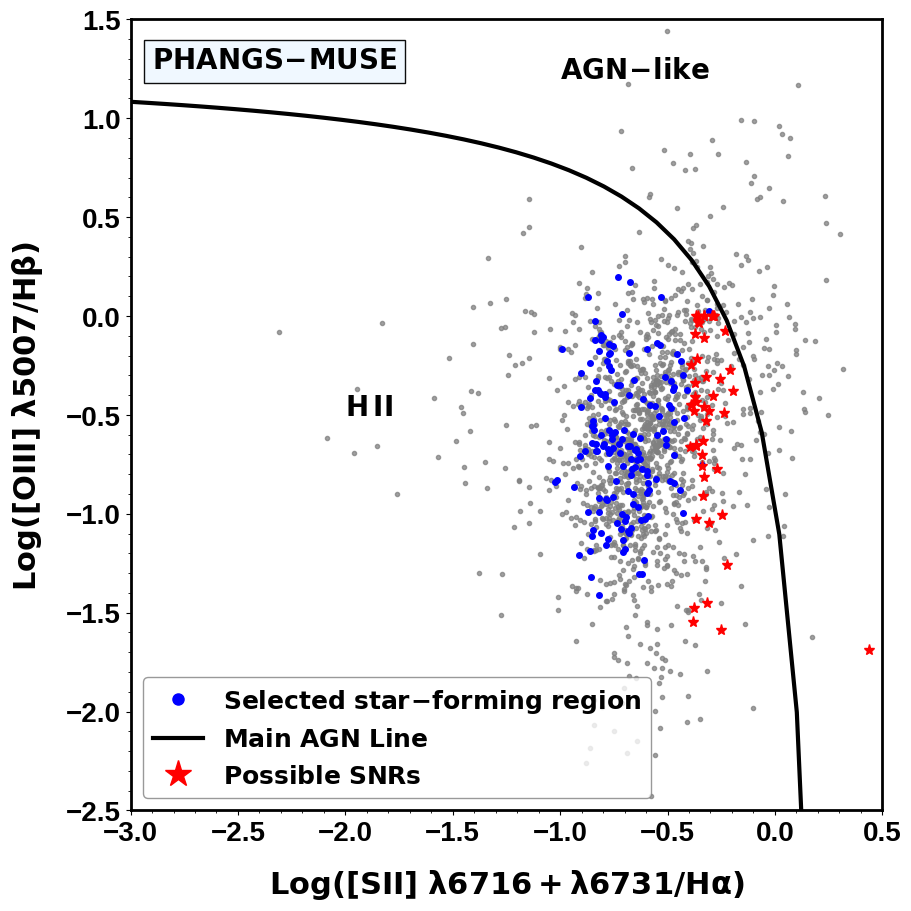}
		\caption[]{BPT-NII and BPT-SII diagrams of the SIGNALS and PHANGS sample in NGC~628. In both panels, the solid and dashed black lines indicate the theoretical boundaries separating the domain of the star-forming, composite and AGN-like regions as prescribed by \cite{Kewley2001} and \cite{Kauffmann2003}. The upper and lower panels show the SIGNAL-H\,II and PHANGS-H\,II regions, respectively, represented as blue points. Regions identified as possible supernova remnants (SNR) are included as red stars. Gray points indicate regions with $R^{2}_{fit} > 0.65$.}  
		\label{fig:BPT}
	\end{center}
\end{figure*}
\begin{figure*}
	\centering
	\begin{minipage}{0.495\textwidth}
		\centering
		\vspace{1em}
		\includegraphics[width=\linewidth]{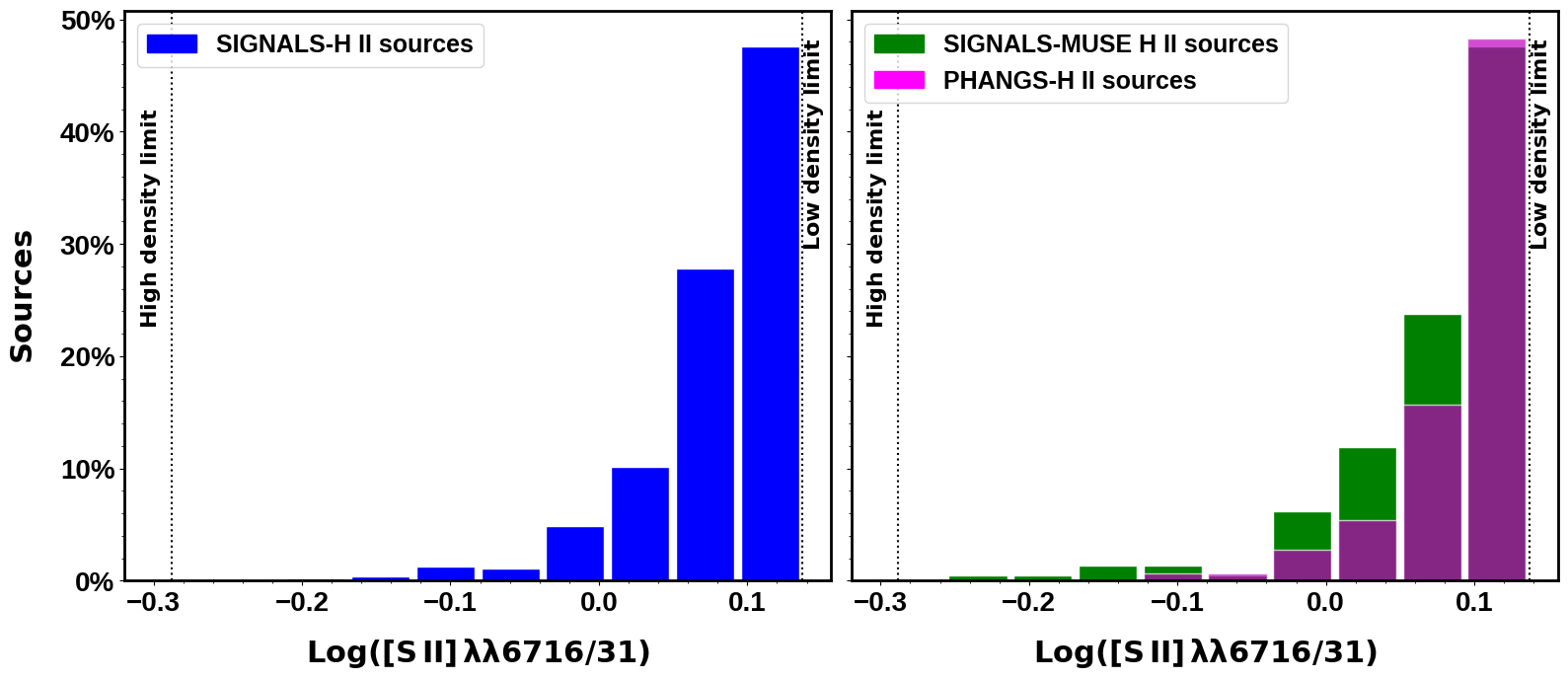}
		\captionof{figure}{In situ electron density, $\mathrm{n_e}$, against the ratio $\mathrm{[S\,II]6716/[S\,II]6731}$ in NGC~628 for the SIGNALS-H\,II subsample (left panel). In the right panel, we compared PHANGS-H\,II regions with SIGNALS-MUSE, a subsample from SIGNALS-H\,II within the MUSE mask FoV. Vertical dashed lines indicate, from left to right, the high and low electron density limits based on our criteria outlined in Subsubsect.~\ref{subsubsec:localdensity}.}
		\label{fig:ratioSII}
	\end{minipage}
	\hfill
	\begin{minipage}{0.495\textwidth}
		\centering
		\includegraphics[width=\linewidth]{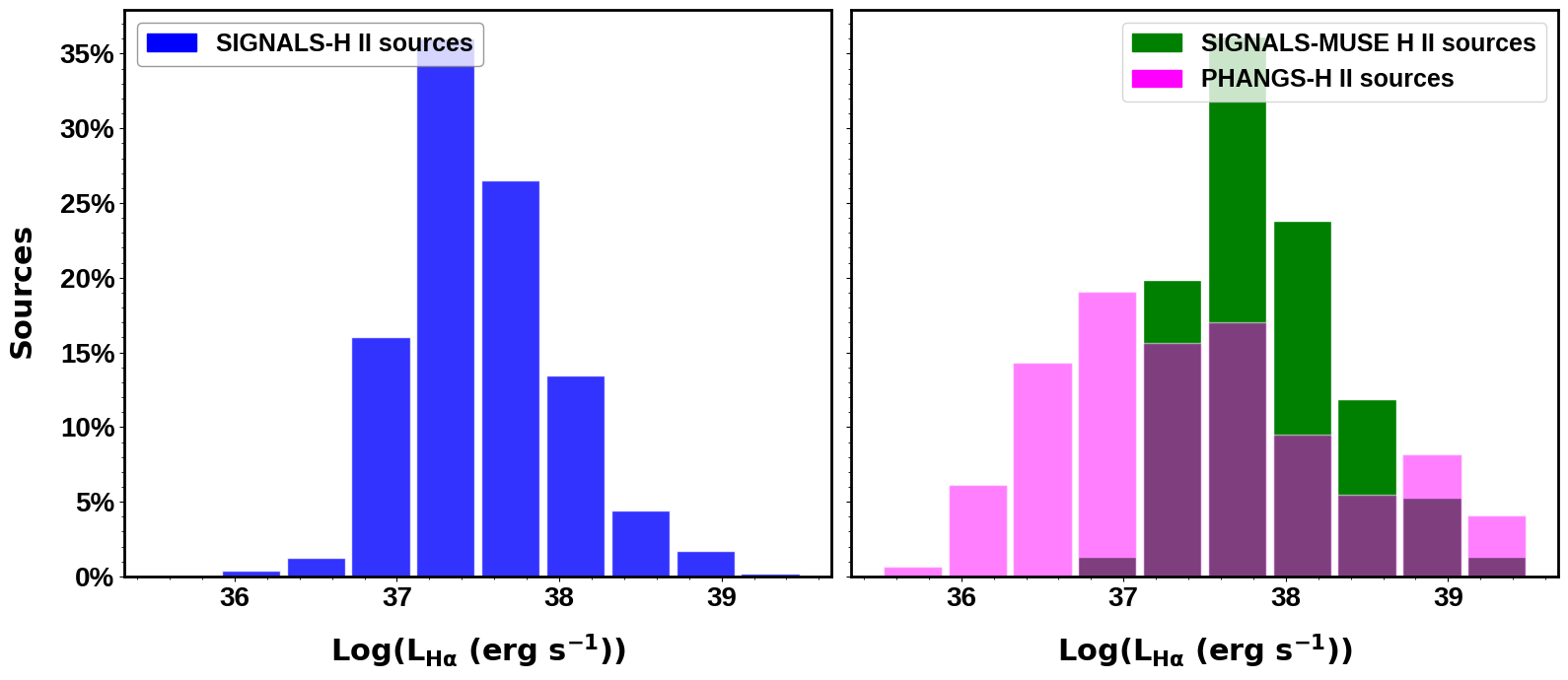}
		\captionof{figure}{H$\alpha$ luminosity distribution of the H\,II regions selected in our final sample. The left panel focuses only on the SIGNALS-H\,II subsample. In the right panel, we compared the PHANGS-H\,II subsample (magenta) with the restricted SIGNALS-MUSE sample (cyan). The vertical axis represents the total number of regions in percentage.}
		\label{fig:signalsphangsLHa}
	\end{minipage}
	\vspace{1.5em}	
	\begin{minipage}{0.495\textwidth}
		\centering
		\vspace{2em}
		\includegraphics[width=\linewidth]{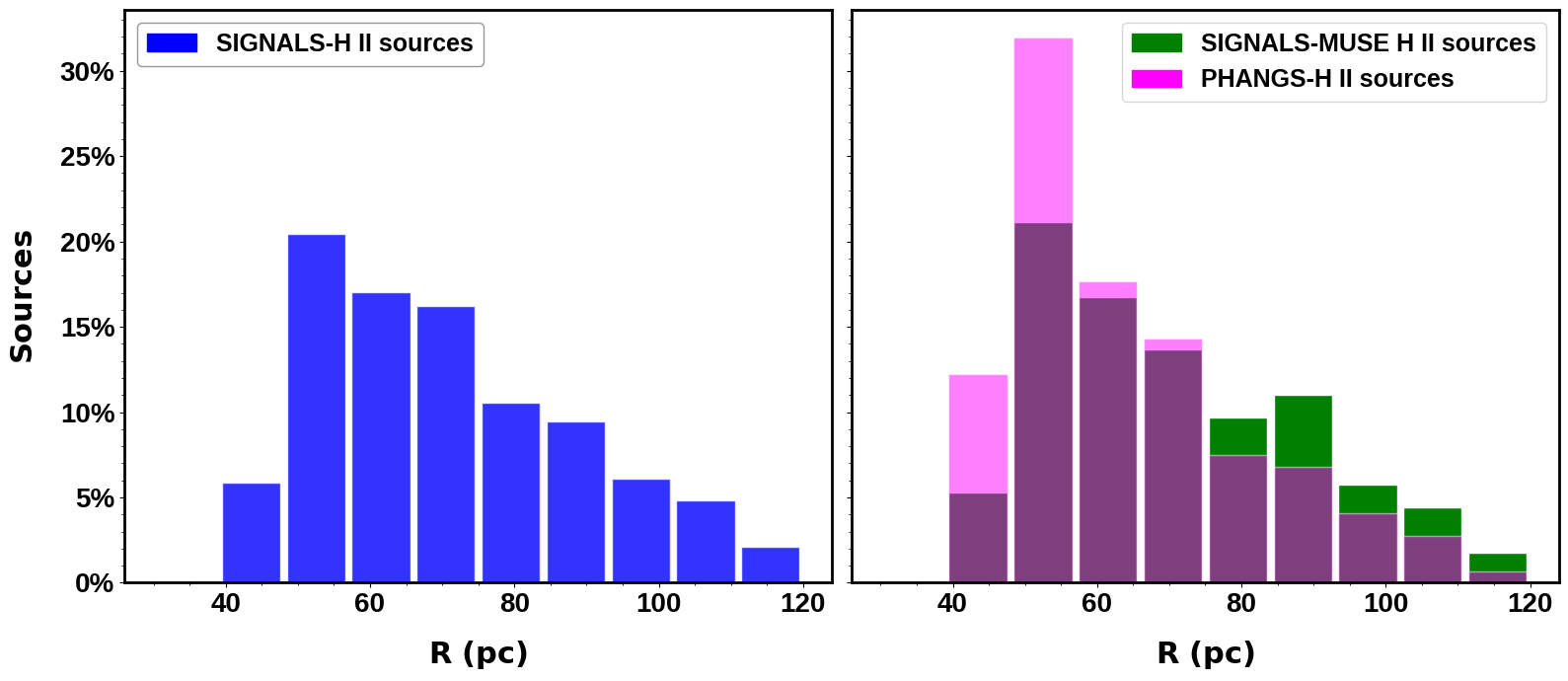}
		\captionof{figure}{Size distribution of the radii of the H\,II regions in our final sample considering only the SIGNALS-H\,II subsample (left panel). In the right panel, we compared the PHANGS-H\,II subsample (magenta) with the restricted SIGNALS-MUSE sample (cyan). The equivalent radii of the PHANGS-H\,II subsample are the radii obtained by circularizing their ellipsoidal volume. The vertical axis represents the total number of regions in percentage.}
		\label{fig:signalsphangsSize}
	\end{minipage}
	\hfill
	\begin{minipage}{0.495\textwidth}
		\centering
		\includegraphics[width=\linewidth]{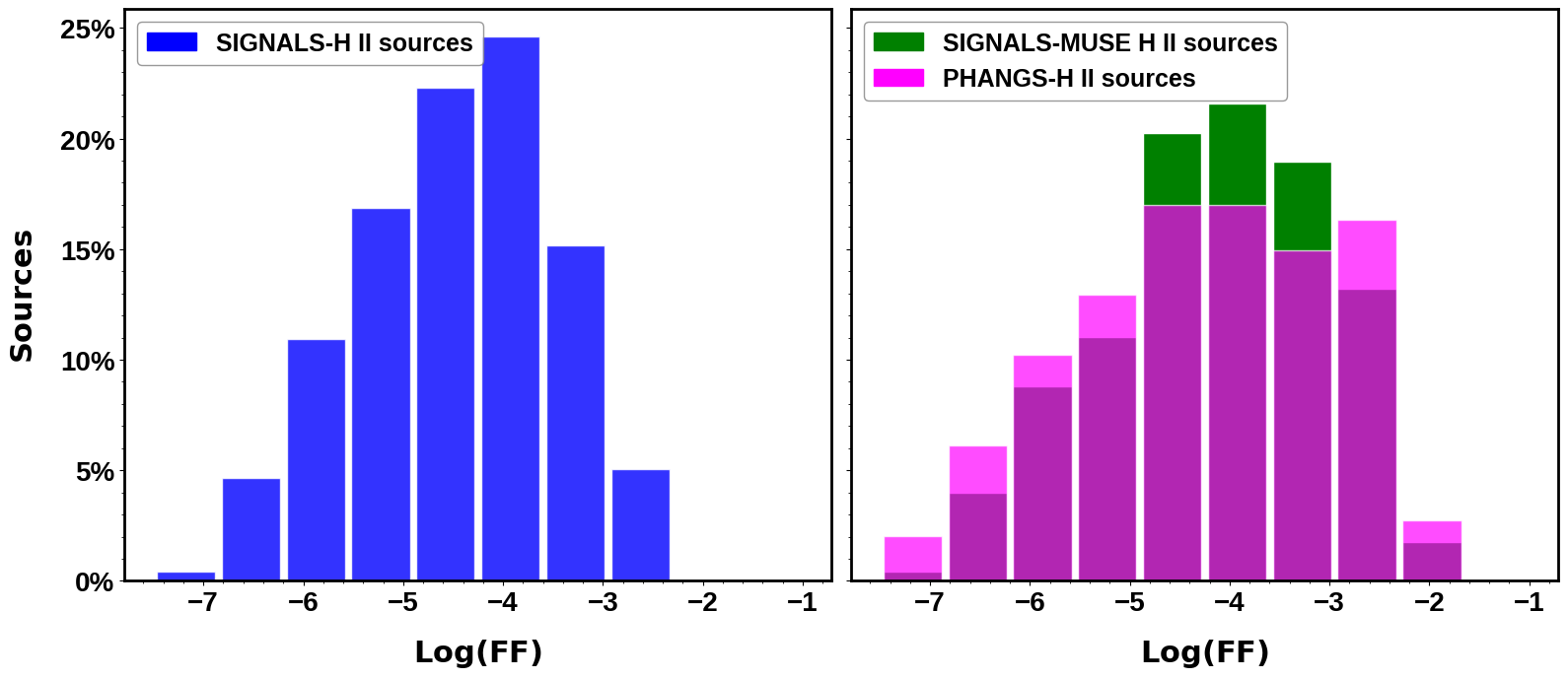}
		\captionof{figure}{FF distribution in our final sample considering the SIGNALS-H\,II subsample (left panel), and compared the PHANGS-H\,II subsample (magenta) with the restricted SIGNALS-MUSE sample (green) in the right panel. The vertical axis represents the total number of regions in percentage.}
		\label{fig:signalsphangsff}
	\end{minipage}
\end{figure*}
\vspace{1.5em}
\begin{figure*}
	\centering
	\includegraphics[width=0.32\linewidth]{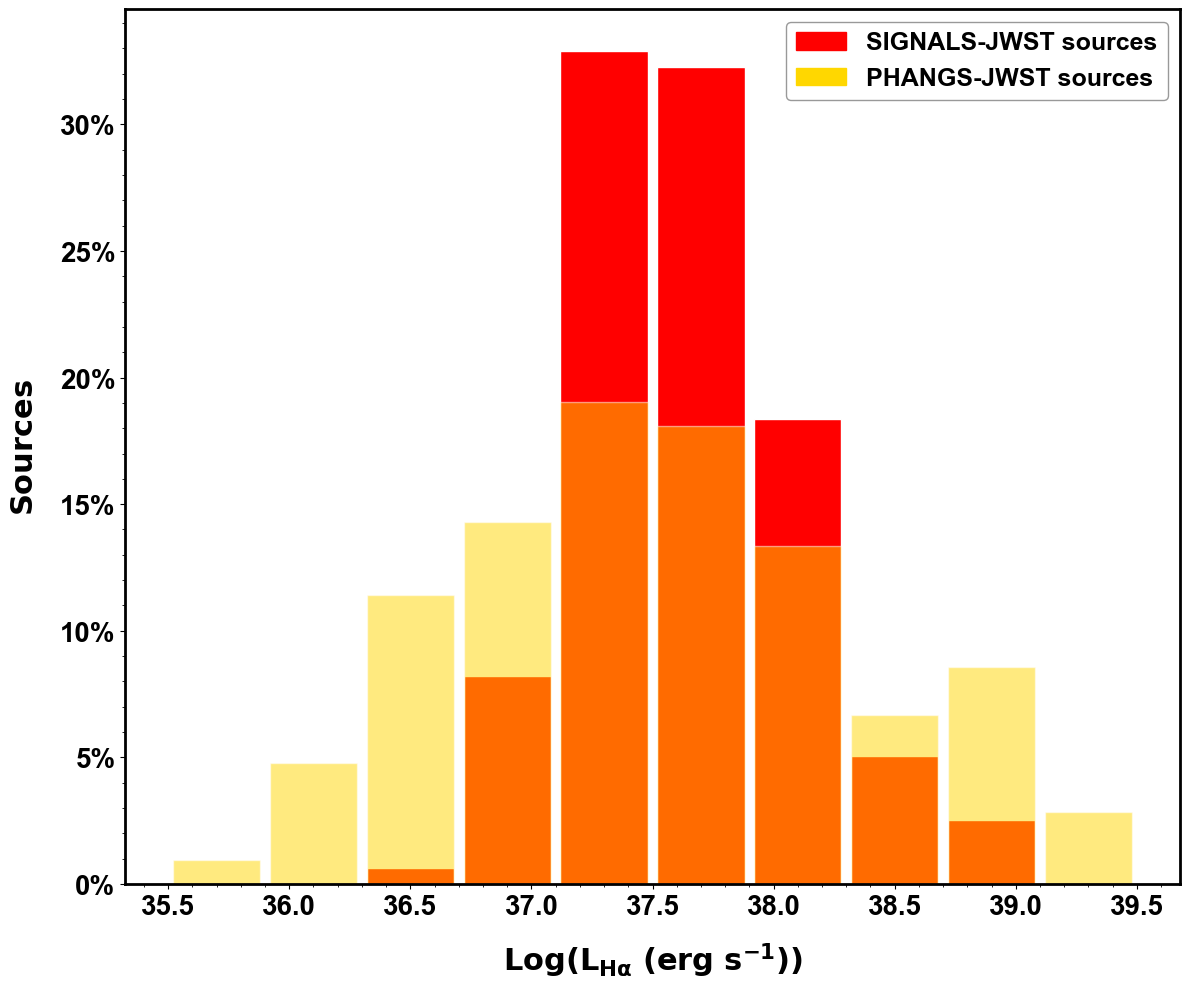}
	\includegraphics[width=0.32\linewidth]{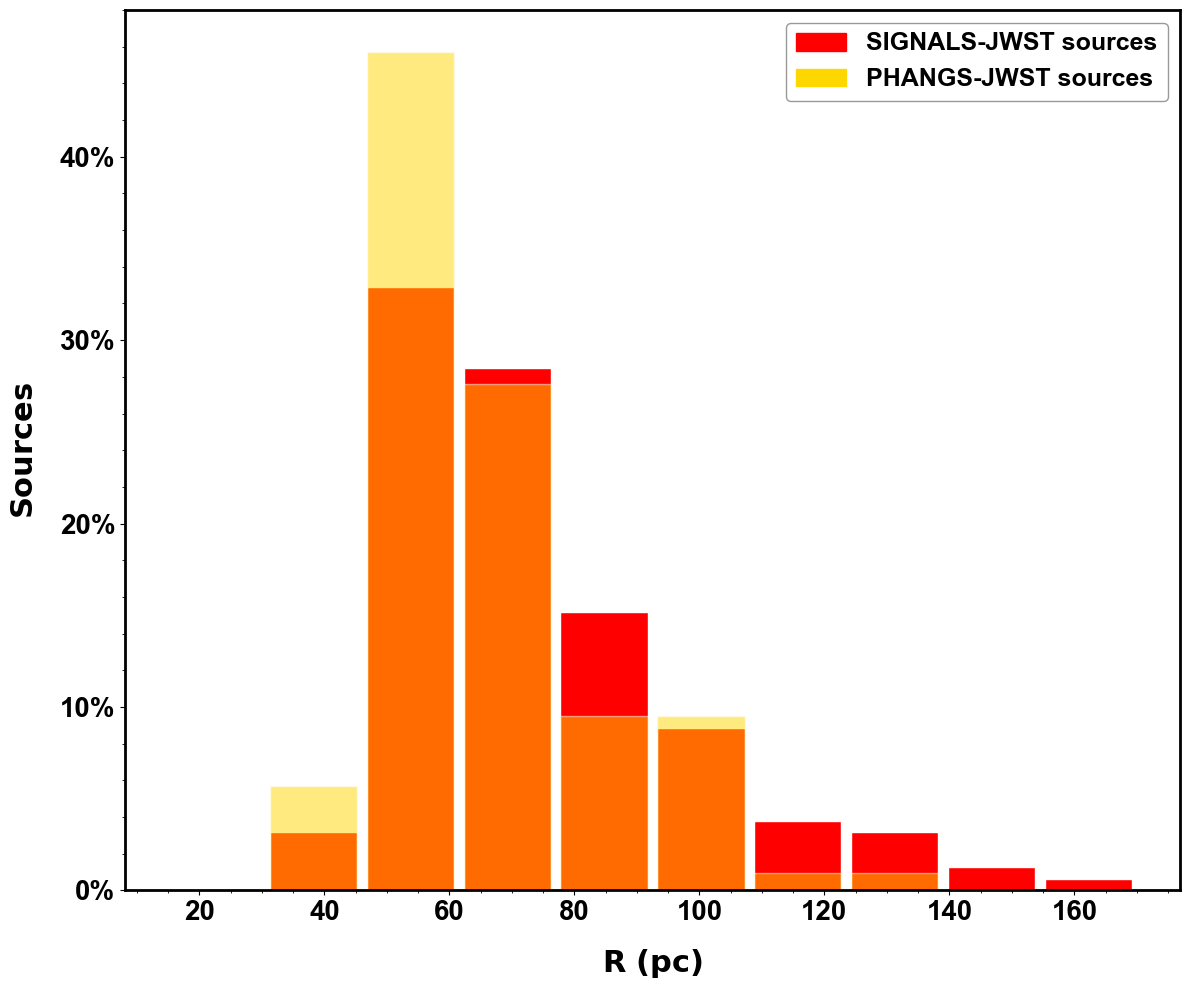}
	\includegraphics[width=0.32\linewidth]{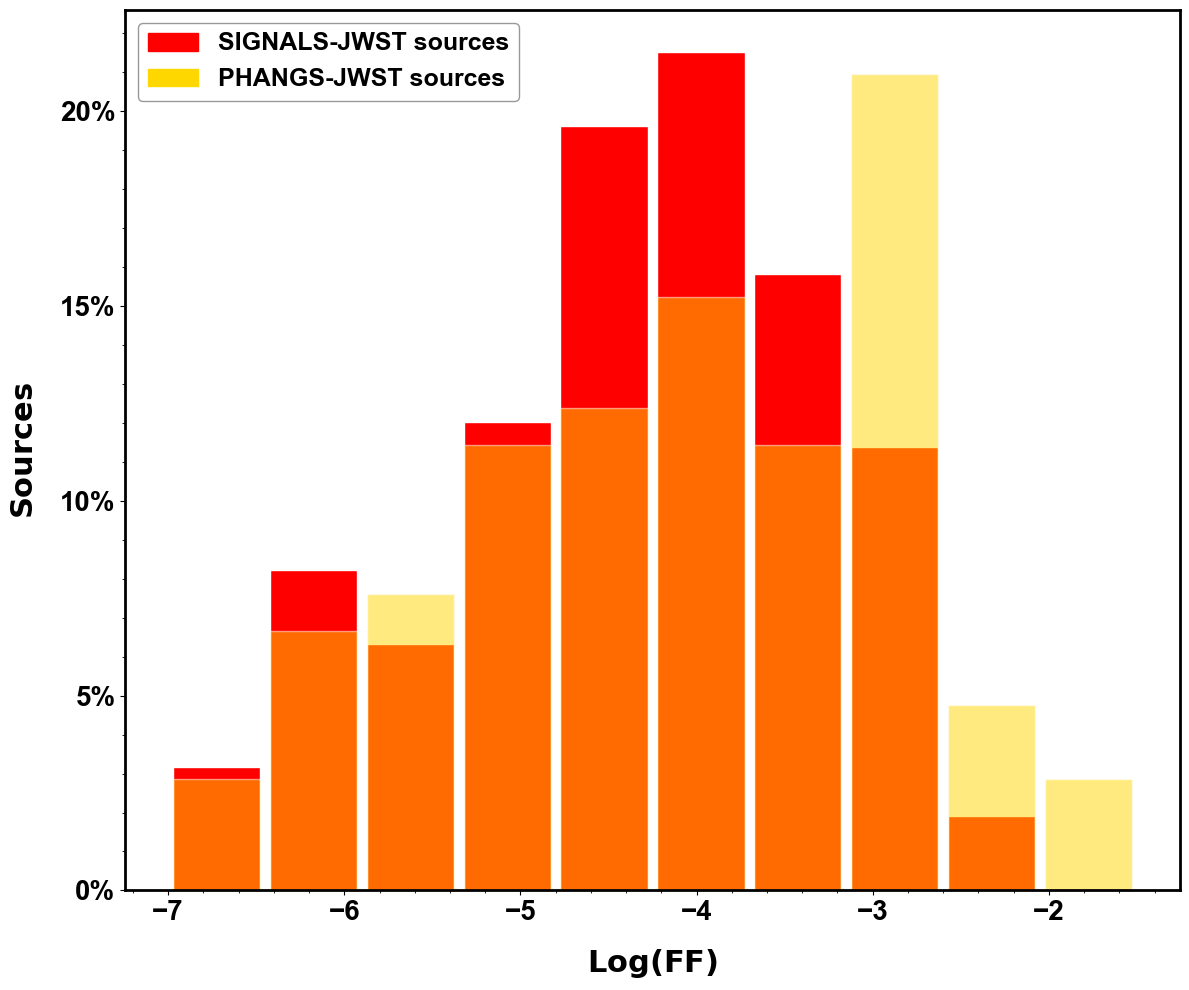}
	\caption{Histograms showing the H$\alpha$ luminosity distribution (left panel), the size distribution of the circularized radii (central panel), and the number of regions as a function of the FF (right panel) for SIGNALS-JWST sample (yellow) and PHANGS-H\,II sample (red) (Sect.~\ref{subsec:PAH}). The vertical axis represents the total number of regions in percentage.}
	\label{fig:LHaRFFJWST}
\end{figure*}
\end{appendix}
\end{document}